%% file: main.tex
\documentclass[11pt]{article}
\usepackage[preprint]{acl}
\usepackage{times}
\usepackage{latexsym}
\usepackage[T1]{fontenc}
\usepackage[utf8]{inputenc}
\usepackage{microtype}
\usepackage{inconsolata}
\usepackage{graphicx}
\usepackage{amsmath,amssymb,booktabs,tabularx,array}
\usepackage{tikz}
\usepackage{placeins}
\usetikzlibrary{arrows.meta,positioning,backgrounds,plotmarks}
\usepackage{algorithm}
\usepackage{algpseudocode}
\hypersetup{pdftitle={Separating Memory and Workflow Effects in Predicting Individual Answers},pdfauthor={Tianzhu Qin; Leo Yang Yang; Ramit Debnath; Davin Youchao Dong},bookmarksnumbered=true}
\pdftrailerid{}
\newcommand{\mns}{\ensuremath{-}}
\newcommand{\method}{\mbox{OwnWords}}
\newcommand{\CodeURL}{\url{https://anonymous.4open.science/r/OwnWords/}}
\newcommand{\CodeAvailability}{Code and anonymized data: \CodeURL{}. The repository provides the \method{} implementation, its survey adaptation, prediction and judging prompts, frozen experimental protocols, and the anonymized analysis package, which is also uploaded as supplementary software. Evidence selection reproduces the stored excerpts character for character on every interview task, checked against the recorded message hashes. }

\input{joint_results.tex}
\input{retrieval_results.tex}

\input{ethics_statement.tex}

\input{human_status.tex}

\title{Separating Memory and Workflow Effects in Predicting Individual Answers}
\author{Tianzhu Qin\textsuperscript{1,2} \quad Leo Yang Yang\textsuperscript{3} \quad Ramit Debnath\textsuperscript{1} \quad Davin Youchao Dong\textsuperscript{4}\\[4pt]
\textsuperscript{1}University of Cambridge \quad \textsuperscript{2}Stanford University \quad \textsuperscript{3}Hong Kong Baptist University \quad \textsuperscript{4}Cookiy Labs}
\begin{document}
\maketitle
\begin{abstract}
Language agents choose what to remember about a person and how to use that memory. We separate these choices when predicting a person's unseen answer to an interview question. On 1,768 tasks from 188 people, a concrete memory from a verified interview prefix outscores a trait description by 0.0158 (95\% interval [0.0044, 0.0271]). Crossing both memories with one-shot generation and three-answer fusion, fusion lowers concrete-memory scores by 0.0123 ([\mns0.0189, \mns0.0056]); prompted and trained selectors do not detectably beat a random candidate, and one call on the longer, unrewritten record outscores every memory condition. At matched context budgets, \method{}, one call on the person's BM25-ranked sentences, outperforms the written memory on 500 people outside the benchmark (+0.0127, [+0.0037, +0.0217]; an earlier held-out test was inconclusive) and across four budgets on 300 people (mean +0.0218, [+0.0138, +0.0298]), but does not detectably outperform recency truncation; these comparisons do not isolate verbatim wording. On a survey benchmark it predicts ordinal answers more closely than the memory but is not more accurate on exact choices (less accurate in one of two screener samples). Interview scores use a model-based content rubric without human ratings, and all benchmark and some confirmation participants were seen during development.
\end{abstract}

\section{Introduction}
An agent that predicts an individual's answer must decide what personal information to retain and how to use it \citep{salemi2024lamp,park2024generative1000,toubia2025twin}. A memory may keep a particular purchase or experience, summarize broad traits, or encode the person in a predefined schema such as Big Five dimensions, MBTI types or Schwartz values, a \emph{schema-constrained memory} with dimensional or categorical coordinates. Personas, trait descriptions and scale profiles are thus choices of memory construction, made before an answering workflow runs, and may retain different evidence at similar lengths.

Given any such memory, generating and combining several candidate answers may recover useful details or introduce unsupported claims. Repeated sampling and aggregation often help on tasks with verifiable answers \citep{wang2023selfconsistency,brown2024monkeys,snell2025scaling}; whether they help predict what one particular person said requires a separate measurement. Our question is: \emph{how does the predictive value of a personal record change with memory construction and with additional inference under a fixed information boundary?}

We study private, AI-moderated user-research interviews in English: 1,768 tasks from 194 sessions and 188 people. Given a screener, a question from the session's second half, and a permitted representation of earlier dialogue, the predictor generates the participant's unseen reply. A model judge compares it with the observed answer on a four-level content rubric (Section~\ref{sec:rubric}).

The available history raises scores (Section~\ref{sec:history}). We then compare concrete and trait memories built from identical permitted sources (Section~\ref{sec:memory}); earlier categorical and scale-based representations motivate this distinction but have unresolved source provenance (Appendix~\ref{app:lineage}). A common experiment crosses the two verified memories with one-shot generation, fusion and same-pool selection (Sections~\ref{sec:workflow}--\ref{sec:joint}), so memory and workflow contrasts do not come from different experiments. \JointIntroductionResult{} Because the unrewritten source is about three times longer than the memories, we next compare written memory with verbatim evidence under the same context limit (Section~\ref{sec:retrieval}). \method{} (Figure~\ref{fig:architecture}) is a simple extractive reference: BM25 ranks the person's own sentences against the question, a greedy fill keeps them within the budget, and one call answers. At the concrete memory's length it outperforms that memory in a prespecified confirmation on 500 people outside the benchmark (an earlier held-out test was inconclusive) and Mem0 as deployed on a second cohort (descriptive). Across four budgets it exceeds the written memory of each length on average, on the second cohort and in a replication on a third. Recency truncation also outperforms the memory and is not detectably below \method{}; these comparisons indicate that both extractive procedures can outperform the written memory, without identifying whether wording or retained content explains that difference (Figure~\ref{fig:chain}).

\input{architecture_figure.tex}
\input{chain_figure_float.tex}

\paragraph{Contributions.} (i) A paired empirical design separating memory construction from additional answering calls, with fixed information boundaries and all planned contrasts reported. (ii) Evidence that concrete and trait memories differ in utility, while the tested fusion and selection procedures do not recover the gap to the unrewritten interview record. (iii) An evaluation of \method{}, a one-call reference built from standard retrieval components, against written memory, Mem0 and a MemGPT-style design on people outside the original benchmark.\footnote{Code and anonymized data: \CodeURL{}.} The contribution is the controlled comparison and its limits: \method{} does not detectably outperform recency truncation, and survey results qualify the interview findings (Appendix~\ref{app:newexp}).

\section{Related Work}
\label{sec:related}
\paragraph{Predicting individuals.} Interview-based agents for 1,052 people \citep{park2024generative1000} compare interview inputs with other representations, including bullet-point summaries, under human retest normalization; Twin-2K-500 \citep{toubia2025twin} and opinion alignment \citep{hwang2023aligning} predict mainly closed-form responses, and LaMP \citep{salemi2024lamp} personalizes outputs by retrieving a user's own profile items, with BM25 \citep{robertson2009bm25} among its retrievers, as \method{} does. Synthetic samples can match aggregate patterns \citep{argyle2023outofone,santurkar2023opinions,ma2025algorithmic} while distorting variation and flattening identity groups \citep{bisbee2024synthetic,wang2025flatten}. We target one respondent's unseen open-ended answer under a fixed information boundary, without human retests.

\paragraph{Memory for personal agents.} Memory, retrieval and reflection are established agent components \citep{park2023generativeagents}, and persistent memory systems and long-conversation benchmarks evaluate storage and recall \citep{packer2023memgpt,chhikara2025mem0,maharana2024locomo,wu2025longmemeval}; persona descriptions have long conditioned dialogue \citep{zhang2018personachat}. In a controlled ablation on LoCoMo and LongMemEval-S, verbatim conversation chunks beat LLM-extracted artifacts in answer accuracy \citep{an2026fidelity}; an analysis of MemPalace attributes its headline LongMemEval retrieval score mainly to verbatim storage with a default embedding retriever rather than its spatial organization \citep{dey2026mempalace}; and a reproduction of LightMem \citep{fang2026lightmem} finds that its constructed memories do not consistently beat retrieval over raw user turns, helping mainly under tight answering-token budgets \citep{zhou2026lightmem}. The granularity of memory units and compression-based denoising also affect retrieval for personalized conversational agents \citep{pan2025secom}. Whether such recall advantages carry over to predicting a person's unseen answer is a separate question, which we test against written memories of the same record at a matched context budget.

\paragraph{Sampling, aggregation, and selection.} Aggregating samples helps when answers can be matched exactly \citep{wang2023selfconsistency}, and inference-time compute scales with repeated sampling and verification \citep{brown2024monkeys,snell2025scaling}. For free-form text a model can select among \citep{chen2023usc} or synthesize \citep{wang2025moa,li2025selfmoa} its own samples, as our selection control and fusion do; LLM-Blender adds trained ranking and fusion \citep{jiang2023llmblender}, and trained verifiers improve selection in mathematics \citep{zhang2025generativeverifiers}. We test fusion, prompted selection and a trained ranker where no verifier or exact-match vote exists.

\paragraph{Model-based evaluation.} Judges prompted with a rubric are widely used for open-ended outputs \citep{zheng2023judging,liu2023geval} and can favor their own generations \citep{panickssery2024selfpreference}; every interview score here is a model judgment (see Limitations). Uncertainty follows clustered, paired practice \citep{efron1979bootstrap,cameron2008bootstrap,miller2024errorbars}.

\section{Prediction Task and Evaluation}
\subsection{Information and target}
\label{sec:info}
Let $H^{(d)}_{i,s,<c}$ be the evidence that condition $d$ takes from person $i$'s session $s$ before cut $c$, $\sigma_{i,s}$ the screener, $q_{it}$ the question of task $t$ at cut $c_t$, and $a_{it}$ the subsequent observed answer. With $M_d$ a memory construction (the identity for raw evidence) and process $\pi$,
\begin{equation}
\hat a_{it}^{d,\pi}=F_\pi\!\left(\sigma_{i,s},M_d\bigl(H^{(d)}_{i,s,<c(d,t)}\bigr),q_{it}\right).
\label{eq:task}
\end{equation}
The two memories (Section~\ref{sec:memory}) and the whole record use respondent-only speech before the session's earliest target cut, $c(d,t)=c_{\min}(s)$, capped at 12,000 characters; the current dialogue uses both speakers' turns before $c(d,t)=c_t$ under a 16,000-character tail cap.
No condition uses demonstrations from other people, and with a given memory one-shot, fusion and selection receive the same personal information; extra candidates are model outputs, not new observations.

The private benchmark contains 1,768 within-interview tasks from 194 English-language sessions and 188 people. Participants contribute 5--26 tasks (median 10). Task inputs derive from observed dialogue; earlier model predictions are never fed back as history. Five people have multiple sessions; their sessions remain together in resampling. Stored history is capped at 16,000 characters and screeners at 4,000, with truncation markers.

\subsection{Content correspondence and uncertainty}
\label{sec:rubric}
The frozen rubric assigns 0 for a core contradiction or off-topic response; 1 for broadly aligned direction with missing or different core content; 2 for matching core stances and facts with different details; and 3 for matching core content at comparable specificity. We report $u(\hat a,a)\in\{0,1/3,2/3,1\}$. These are normalized \emph{content scores}, not binary accuracies or calibrated probabilities, and the linear normalization treats the ordinal grades as equally spaced; Appendix~\ref{app:ordinal} decomposes the Section~\ref{sec:memory}, \ref{sec:joint} and confirmation differences by grade threshold. The judge receives the question, reference answer, and prediction, without the method label or its memory (prompts in Appendix~\ref{app:prompts}).

The history benchmark's primary judge is Grok 4.5; the memory and common experiments and the new cohorts of Section~\ref{sec:retrieval} use Gemini 3.1 Pro (the \texttt{gemini-3.1-pro-preview} API model), so absolute scores from these protocols are not pooled. The judges were fixed at different times, not chosen per experiment; on 501 sampled history-benchmark predictions they agreed with quadratic-weighted $\kappa=0.83$ (Appendix~\ref{app:judgeconfig}). Primary differences are task-weighted, and bootstrap resampling keeps each person's tasks and sessions together \citep{efron1979bootstrap,cameron2008bootstrap}. The rubric, judge configuration, planned contrasts and missing-outcome rules were fixed before common-experiment scoring; ``frozen'' and ``prespecified'' refer to author-side hashed protocol files created before scoring, not to third-party registration.

\section{Does the Interview History Help Predict the Answer?}
\label{sec:history}
\input{prelude_figure_float.tex}
Memory design matters only if the interview history helps, so we test that first, on the history benchmark. Providing the available history and screener instead of the question alone raises the content score with each of four minimum-effort configurations, by +0.062 to +0.087 with every 95\% interval above zero (Figure~\ref{fig:prelude}a; Table~\ref{tab:generation} in Appendix~\ref{app:benchmark}); weighting people equally gives 0.0614--0.0861 (Appendix~\ref{app:personweighted}). The gain includes personal material and conversational context, not identity alone; our reanalysis of these stored predictions changes only the uncertainty clustering.

\section{Does It Matter How the Memory Is Written?}
\label{sec:memory}
From each session's identical prefix before its earliest target cut (respondent speech, capped at 12,000 characters), a model writes two memories of about 100 words without seeing any target: a \emph{concrete memory} preserving facts through paraphrase, and a \emph{trait memory} excluding concrete objects, places, events and numbers (Appendix~\ref{app:prompts}). The trait memory is free-form; it does not instantiate Big Five, MBTI or another fixed schema. Those historical representations and their source limitations are documented in Appendix~\ref{app:lineage}.

Each memory is written once per session and sealed before prediction, so the intervals condition on these realized memories and exclude memory-writing variance (Appendix~\ref{app:rerun}). Concrete memory scores 0.2636 versus 0.2477 for trait memory, \textbf{a difference of +0.0158, 95\% interval [0.0044, 0.0271]} (prespecified); newly generated answers in the common experiment reproduce the direction (+0.0140; Figure~\ref{fig:prelude}b). The memories average 111.85 and 94.94 words: this compares construction procedures, not specificity alone. Post hoc, the difference persists when lengths differ by at most 15 words (+0.0148, [0.0006, 0.0288]) but is unresolved at 10, leaving length as an alternative explanation (Appendix~\ref{app:lengthspec}). The result does not establish that traits lack individual signal or that fixed psychological schemas are inferior; those schemas were not rerun under the verified protocol. Both this experiment and the common experiment use previously exposed people.

\section{Can Extra Calls Recover What a Memory Drops?}
\label{sec:workflow}
A natural next step is a multi-call workflow, which helps on tasks with verifiable answers (Section~\ref{sec:related}); the fusion tested here is one such workflow, a competitor to one call and not part of \method{}. With each memory held fixed (Figure~\ref{fig:workflow} in Appendix~\ref{app:jointprotocol}), the generator makes three independent calls with identical prompts; the first is the one-shot reference, fixed before any scoring. Fusion writes one new answer from the three anonymous candidates, the memory and the question, without concatenating incompatible alternatives; the selection control returns one candidate unchanged, with the same call count and output cap (prompts in Appendix~\ref{app:prompts}). We ran this test because development batches in another setup (other people's demonstrations, another generator) gave small, unstable fusion differences (+0.0098 and \mns0.0098 against two one-shot variants, intervals spanning zero; Appendix~\ref{app:workflow}), not because extra calls must help.

\section{A Common Memory-by-Workflow Experiment}
\label{sec:joint}
\subsection{Contemporaneous controls and frozen analysis}
To answer the question of Section~\ref{sec:workflow}, all 1,768 tasks are run with both memories, and every answer is generated anew under one protocol (Gemini 3.6 Flash, minimal reasoning, 1,600-token cap; one candidate pool per task and memory). Two one-shot references separate the source from its rewrite: the \emph{whole record}, the memories' verbatim source (\method{} without a budget; Section~\ref{sec:retrieval}), and the \emph{current dialogue}, both speakers' turns up to the current question (16,000-character tail cap), which for the 1,574 non-first targets also sees turns after the memories' cut. All outputs are sealed before grading by the frozen judge; identical answers within a task share one grade, predeclared zeros cover empty answers and invalid selections, and nothing is redrawn (Appendix~\ref{app:jointprotocol}).

The sole primary endpoint is $\delta=\mu^{\mathrm{fus}}_{\mathrm{con}}-\mu^{\mathrm{one}}_{\mathrm{con}}$, concrete-memory fusion minus one-shot, with a two-sided 95\% whole-person bootstrap interval (100,000 draws, seed 20260918); the planned decision flag was an interval above zero with all 1,768 pairs present. Seven further prespecified contrasts, including the interaction
\begin{equation}
\gamma=(\mu^{\mathrm{fus}}_{\mathrm{con}}-\mu^{\mathrm{one}}_{\mathrm{con}})-(\mu^{\mathrm{fus}}_{\mathrm{tra}}-\mu^{\mathrm{one}}_{\mathrm{tra}})
\label{eq:interaction}
\end{equation}
on the four-cell common task set, carry 99.375\% Bonferroni intervals \citep{dunn1961multiple} for approximately 95\% family coverage, and deterministic full-plan bounds let every missing grade range over the scale (Appendix~\ref{app:jointresults}).

\subsection{Joint results}
\label{sec:jointresults}
All 1,768 planned tasks from 188 people were executed and graded (5 of 21,216 condition--task slots missing; Appendix~\ref{app:jointresults}); Figure~\ref{fig:prelude}c shows the means (Table~\ref{tab:jointmeans}). \textbf{Fusion lowers the concrete-memory score}: $\delta=\mns0.0123$ (primary 95\% interval [\mns0.0189, \mns0.0056]; 119 wins, 177 losses, 1,472 ties). The loss holds against the mean of fusion's own three candidates (\mns0.0090, [\mns0.0138, \mns0.0041]) and with people weighted equally (\mns0.0120); a judge from another vendor, added after these results, finds about half to three fifths of it (\mns0.0066, [\mns0.0138, +0.0006], as run; \mns0.0074, [\mns0.0142, \mns0.0006], if byte-identical answers share one grade, the primary judge's rule applied after the run; Appendix~\ref{app:newexp}). With the trait memory, fusion is unresolved (\mns0.0023), and the interaction ($\gamma=\mns0.0100$) excludes zero only pointwise, not in its family interval, so we do not claim that the workflow effect depends on the memory. \textbf{Selection does not recover the loss}: neither the prompted selector nor a trained pairwise reranker added after these results, PairRM \citep{jiang2023llmblender}, detectably beats a random pick from the same pool (\mns0.0005 and +0.0003 against the candidate mean). \textbf{Every memory condition scores below both unrewritten references}: concrete-memory fusion trails the whole record by 0.0375 and the current dialogue by 0.0752 (family intervals exclude zero), so the prespecified criterion for fusion to beat all four comparators is not met. Post hoc, the concrete-memory one-shot already trails the whole record by 0.0253 ([\mns0.0353, \mns0.0154]): about two thirds of fusion's shortfall is present before any workflow runs, at the step that replaces the record with a shorter written memory under its own prompt header. Answer-length and grade-threshold checks agree; the current-dialogue gap is concentrated in later targets (Appendix~\ref{app:robust}).

\input{retrieval_section.tex}

\section{Discussion and Conclusion}
\label{sec:discussion}
The comparisons separate two practical choices: what observations a memory preserves and what additional calls do with that memory. Rewriting may discard answer-relevant detail (Figure~\ref{fig:chain}), but we did not identify that mechanism. Schema-constrained profiles are one memory design, not a substitute for measuring its predictive utility; the historical results do not establish a universal disadvantage of categorization.

\paragraph{Conclusion.}
\JointConclusionParagraph{}

\label{contentend}
\vfill

\section*{Limitations}
\addcontentsline{toc}{section}{Limitations}
\paragraph{Representation scope.} The verified memory experiments compare two free-form written memories, concrete and trait. Schema-constrained profiles such as Big Five, MBTI or Schwartz-value encodings, and closed-set tags, appear only in earlier, unverified historical analyses (Appendix~\ref{app:lineage}) and were not rerun under the verified protocol. We also did not compare with persona-simulation models fine-tuned on individual data, with learned or dense retrievers over the person's own sentences (Mem0's embedding search ranks only its own extracted facts), or with memories updated over time. Our conclusions therefore concern the tested written-memory, workflow and extraction procedures, not every persona or profile design, and they do not estimate how predictable individual answers can be in general.

\paragraph{Evaluation validity.} The private sample of adaptive, AI-moderated interviews limits generalization. Every interview score is one model judgment of one observed answer, with no human ratings (a two-rater protocol was not run; Appendix~\ref{app:human}) and no human retest ceiling; Twin-2K-500 is scored against recorded answers. A judge from another vendor reproduced the direction of the fusion loss (about half to three fifths as large) and of \method{}'s confirmation gain (larger), but did not grade the whole-record one-shot, and its repeat grading of byte-identical answers is our only judge-retest evidence (Appendix~\ref{app:newexp}). The judge and generator of Sections~\ref{sec:memory}--\ref{sec:retrieval} share a model family; since all compared conditions share the generator, a preference for its outputs \citep{panickssery2024selfpreference} cannot by itself produce the differences, but style preferences shared by both judges remain possible. Human ratings and a controlled paraphrase comparison are still needed to distinguish preserved content from preferences for wording or brevity; neither validation was performed.

\paragraph{Scope.} \method{} stays within the memory's characters but also changes the prompt header (its effect unresolved in a post hoc development-half factorial; Appendix~\ref{app:attempts}) and the answer length. Memories were written once, at about 100 words outside the budget sweep, from each session's earliest prefix; updated memories, other sample counts, reasoning settings and debate-style aggregation were not tested. Previously examined participants and extensive method selection limit the confirmatory scope of Sections~\ref{sec:history} to~\ref{sec:joint}; only Section~\ref{sec:retrieval} ran on people outside the benchmark, some of whom had appeared in earlier exploratory work (Appendix~\ref{app:freshcohort}). One generation repeat does not establish stability across generations or model revisions. Effects are small and most task pairs tie; small or unresolved differences are not equivalence findings.

\section*{Ethical Considerations}
\addcontentsline{toc}{section}{Ethical Considerations}
\EthicsStatementText{} We did not run a separate screen for personal information or offensive content and did not de-identify the interview text, which therefore remains in restricted storage.

\section*{Acknowledgments}
\addcontentsline{toc}{section}{Acknowledgments}
\AIUseStatementText{}

\phantomsection\addcontentsline{toc}{section}{References}
\bibliography{references}

\appendix
\input{overview_appendix.tex}
\input{joint_appendix.tex}

\input{attempts_appendix.tex}

\input{freshcohort_appendix.tex}

\input{newexp_appendix.tex}

\input{robustness_appendix.tex}
\input{historical_appendix.tex}
\end{document}

%% file: joint_results.tex
\newcommand{\JointIntroductionResult}{%
In that experiment, fusion lowers the concrete-memory score by 0.0123 relative to the one-shot (primary 95\% interval [\mns0.0189, \mns0.0056]), a prompted selector over the same pool does not detectably beat the average candidate, and both one-shot references on unrewritten text score higher than either memory, with or without fusion (Section~\ref{sec:jointresults}).%
}

\newcommand{\JointConclusionParagraph}{%
In the common experiment, fusion lowered the concrete-memory score (\mns0.0123, primary 95\% interval [\mns0.0189, \mns0.0056]; about half to three fifths of that under a second judge), prompted and trained selectors did not detectably beat a random candidate, and one call on the unrewritten whole record outscored every memory condition. \VerbatimConclusion{}
}

\newcommand{\JointExecutionParagraph}{%
Completed execution used 21,487 generation and 16,119 judging HTTP reservations, including 273 and 33 recoveries respectively (Tables~\ref{tab:jointcalls} and~\ref{tab:jointtokens}). The separately implemented replay of the generation execution and the separate numerical review, both author-side software checks, passed their source-bound checks. Reservations with no captured outcome remain explicit, and no audit is counted as a human rating. Old journals from the quota interruption, including their then-unfinished slots, remain in a separate accounting segment; effective terminal counts describe the completed continuation. \JointDriverRestartSentence{}The twelve-worker judging phase began after 9,505 terminal judge slots (9,513 HTTP attempts) and was interrupted after a further 319 terminal slots (341 attempts); the restored four-worker phase completed the remaining 6,262 slots (6,265 attempts). All process handovers retained prior records without repeat requests for terminal slots. The final audit reconstructs all phases from durable request and receipt intervals; it does not claim a completed runtime audit for the interrupted twelve-worker scheduler. Both scheduling amendments and the passing final audit are bound into the judging seal.
}
\newcommand{\JointDriverRestartSentence}{Within the initial four-worker judging phase, the driver process stopped once on a local output-pipe error after 8,683 terminal slots (8,691 attempts) and was relaunched with the unchanged command; no terminal slot was requested again. }

\newcommand{\JointMeansMainTable}{%
\begin{table*}[t]
\centering\small
\begin{tabular}{@{}lccl@{}}\toprule
Evidence given to the answering call & One call & Fusion (4 calls) & Change [95\% interval] \\\midrule
Concrete memory & 0.2607 & 0.2485 & \textbf{\mns0.0123} [\mns0.0189, \mns0.0056] \\
Trait memory & 0.2468 & 0.2447 & \mns0.0023 [\mns0.0086, +0.0040] \\
Whole record (\method{}$_\infty$) & 0.2858 & not run & -- \\
\bottomrule\end{tabular}
\caption{Fusion does not improve either tested memory condition; one call on the longer unrewritten record scores higher. Common experiment, 1,768 tasks from 188 people: mean normalized content score with one call and with fusion, which writes one new answer from three one-call candidates (four calls). The whole record is the memories' verbatim source, i.e.\ \method{} without a budget. Change: fusion minus one call, paired (bold: excludes zero). Selection and the current-dialogue reference: Table~\ref{tab:jointallconditions}; every contrast: Tables~\ref{tab:jointfamily} and~\ref{tab:jointposthoc} and Figure~\ref{fig:forest}.}
\label{tab:jointmeans}
\end{table*}
}

\newcommand{\JointAllConditionsTable}{%
\begin{table*}[t]
\centering\footnotesize
\setlength{\tabcolsep}{3.5pt}
\begin{tabularx}{\linewidth}{@{}Xrrrrrr@{}}\toprule
Condition & Valid $N$ & Missing & Empty 0 & Invalid selection 0 & Raw sum & Mean \\\midrule
Concrete: one-shot ($b_0$) & 1,768 & 0 & 0 & 0 & 1,383 & 0.260747 \\
Concrete: candidate $b_1$ & 1,768 & 0 & 0 & 0 & 1,362 & 0.256787 \\
Concrete: candidate $b_2$ & 1,768 & 0 & 0 & 0 & 1,352 & 0.254902 \\
Concrete: fusion & 1,768 & 0 & 0 & 0 & 1,318 & 0.248492 \\
Concrete: selection & 1,768 & 0 & 0 & 0 & 1,363 & 0.256976 \\
Trait: one-shot ($b_0$) & 1,768 & 0 & 0 & 0 & 1,309 & 0.246795 \\
Trait: candidate $b_1$ & 1,766 & 2 & 0 & 0 & 1,302 & 0.245753 \\
Trait: candidate $b_2$ & 1,768 & 0 & 0 & 0 & 1,296 & 0.244344 \\
Trait: fusion & 1,767 & 1 & 0 & 0 & 1,297 & 0.244671 \\
Trait: selection & 1,767 & 1 & 0 & 0 & 1,294 & 0.244105 \\
Whole record & 1,767 & 1 & 0 & 0 & 1,515 & 0.285795 \\
Current dialogue & 1,768 & 0 & 0 & 0 & 1,717 & 0.323718 \\
\bottomrule\end{tabularx}
\caption{All twelve conditions. Each mean uses that row's available scores; different denominators must not be subtracted to estimate paired effects. Zeros follow the frozen operational rules.}
\label{tab:jointallconditions}
\end{table*}
}

\newcommand{\JointFamilyTable}{%
\begin{table*}[t]
\centering\footnotesize
\setlength{\tabcolsep}{3.5pt}
\begin{tabularx}{\linewidth}{@{}>{\raggedright\arraybackslash}Xrrll@{}}\toprule
Contrast & Paired $N$ & Difference & 95\% interval & 99.375\% interval \\\midrule
Concrete: fusion $-$ one-shot & 1,768 & \mns0.012255 & [\mns0.018868, \mns0.005650] & [\mns0.021530, \mns0.003077] \\
Trait: fusion $-$ one-shot & 1,767 & \mns0.002264 & [\mns0.008571, +0.003995] & [\mns0.011070, +0.006501] \\
Fusion: concrete $-$ trait & 1,767 & +0.003962 & [\mns0.006443, +0.014501] & [\mns0.010664, +0.018482] \\
Interaction $\gamma$ & 1,767 & \mns0.009998 & [\mns0.019166, \mns0.000948] & [\mns0.022931, +0.002486] \\
Concrete: fusion $-$ selection & 1,768 & \mns0.008484 & [\mns0.015082, \mns0.001875] & [\mns0.017717, +0.000757] \\
Trait: fusion $-$ selection & 1,767 & +0.000566 & [\mns0.004935, +0.006033] & [\mns0.007125, +0.008123] \\
Concrete fusion $-$ whole record & 1,767 & \mns0.037540 & [\mns0.046989, \mns0.028161] & [\mns0.050829, \mns0.024478] \\
Concrete fusion $-$ current dialogue & 1,768 & \mns0.075226 & [\mns0.087902, \mns0.062617] & [\mns0.092871, \mns0.057721] \\
\bottomrule\end{tabularx}
\caption{All eight prespecified comparisons at six decimals. The 95\% column is pointwise; only fusion minus one-shot with the concrete memory is the unique primary 95\% comparison. The 99.375\% intervals use Bonferroni allocation for approximate 95\% family coverage. Every contrast covers all 188 people and 194 sessions; sessions belonging to one person remain together in resampling. Complete-case estimates divide by $3\times$ the paired $N$, whereas the full-plan bounds in Table~\ref{tab:jointmissingbounds} divide by $3\times1{,}768$.}
\label{tab:jointfamily}
\end{table*}
}

\newcommand{\JointAppendixProse}{%
\paragraph{Post hoc decomposition.} The one-shot and the whole record are both single calls under the same source boundary, so their difference measures the memory-construction step together with its prompt header (in the development-half factorial of Appendix~\ref{app:attempts} the header effect was small and unresolved): the one-shot minus the whole record is \mns0.0253 with the concrete and \mns0.0392 with the trait memory, placing about two thirds of concrete-memory fusion's shortfall on memory construction and its header and one third on fusion. The current dialogue exceeds the whole record by +0.0377. These contrasts were not prespecified; Table~\ref{tab:jointposthoc} reports them with paired sets and intervals.

Among complete pools, all three candidates score zero in 622/1,768 concrete-memory pools and 648/1,766 trait-memory pools, and all three grades are equal in 1,208 and 1,260 pools, so only 560 concrete-memory pools and 506 trait-memory pools leave anything to select. In those, the selector returns a top-graded candidate in 47.5\% and 48.0\% of cases, against 47.7\% and 48.7\% expected from a uniformly random pick (a post hoc baseline). Oracle regret and the gap from fusion to the oracle are in Tables~\ref{tab:jointpoolcounts} and~\ref{tab:jointpoolmetrics}.
}

\newcommand{\JointPostHocTable}{%
\begin{table*}[t]
\centering\small
\begin{tabular}{@{}lrrlr@{}}\toprule
Contrast & Paired tasks & Difference & 95\% interval & Win / tie / loss \\\midrule
Concrete one-shot $-$ whole record & 1,767 & \mns0.0253 & [\mns0.0353, \mns0.0154] & 206 / 1,244 / 317 \\
Trait one-shot $-$ whole record & 1,767 & \mns0.0392 & [\mns0.0501, \mns0.0284] & 215 / 1,183 / 369 \\
Current dialogue $-$ whole record & 1,767 & +0.0377 & [+0.0264, +0.0492] & 380 / 1,144 / 243 \\
Concrete: fusion $-$ candidate $b_1$ & 1,768 & \mns0.0083 & [\mns0.0148, \mns0.0017] & 124 / 1,482 / 162 \\
Concrete: fusion $-$ candidate $b_2$ & 1,768 & \mns0.0064 & [\mns0.0132, +0.0002] & 129 / 1,474 / 165 \\
\bottomrule\end{tabular}
\caption{Post hoc descriptive contrasts (not part of the prespecified family). Same task-weighted estimator and whole-person bootstrap as the planned contrasts; intervals are pointwise 95\% and unadjusted.}
\label{tab:jointposthoc}
\end{table*}
}

\newcommand{\JointDescriptiveTable}{%
Person-weighted point estimates of the eight family contrasts, computed post hoc in the order of Table~\ref{tab:jointfamily}, are \mns0.011989, \mns0.003228, +0.003044, \mns0.008781, \mns0.007178, \mns0.000021, \mns0.037659, and \mns0.075722. They agree in sign with the task-weighted estimates except for fusion minus selection with the trait memory, which is within 0.001 of zero under both weightings. For the primary contrast, fusion scores below the one-shot for 85 people and above it for 44 (59 equal); a post hoc exact two-sided sign test over the 129 people with a nonzero difference gives $p=0.0004$.

\begin{table*}[t]
\centering\footnotesize
\setlength{\tabcolsep}{3.5pt}
\begin{tabularx}{\linewidth}{@{}>{\raggedright\arraybackslash}Xr@{\hspace{10pt}}rrl@{}}\toprule
Contrast & Paired $N$ & Task-weighted & Person-weighted & 95\% interval \\\midrule
One-shot: concrete $-$ trait & 1,768 & +0.013952 & +0.011786 & [+0.003570, +0.024166] \\
Fusion $-$ one-shot, mean of both memories & 1,767 & \mns0.007263 & \mns0.007618 & [\mns0.011785, \mns0.002709] \\
Fusion $-$ selection, mean of both memories & 1,767 & \mns0.003962 & \mns0.003609 & [\mns0.008292, +0.000374] \\
Concrete: fusion $-$ candidate mean & 1,768 & \mns0.008987 & \mns0.008490 & [\mns0.013840, \mns0.004099] \\
Trait: fusion $-$ candidate mean & 1,766 & \mns0.000881 & \mns0.001440 & [\mns0.005623, +0.003851] \\
Concrete: selection $-$ candidate mean & 1,768 & \mns0.000503 & \mns0.001313 & [\mns0.005303, +0.004235] \\
Trait: selection $-$ candidate mean & 1,766 & \mns0.001447 & \mns0.001419 & [\mns0.005805, +0.003020] \\
\bottomrule\end{tabularx}
\caption{Prespecified descriptive comparisons and person-weighted point sensitivity. Intervals concern task-weighted paired differences, are pointwise, and are not added to the confirmatory family. ``Candidate mean'' is the mean of $b_0$, $b_1$, and $b_2$ for that task and memory.}
\label{tab:jointdescriptive}
\end{table*}
}

\newcommand{\JointPoolTable}{%
\begin{table*}[t]
\centering\footnotesize
\setlength{\tabcolsep}{3.5pt}
\begin{tabularx}{\linewidth}{@{}Xrrrrr@{}}\toprule
Memory & Complete/planned & All equal & All zero & Selection score = oracle & Unequal-pool matches \\\midrule
Concrete & 1,768/1,768 & 1,208 & 622 & 1,474/1,768 & 266/560 \\
Trait & 1,766/1,768 & 1,260 & 648 & 1,503/1,766 & 243/506 \\
\bottomrule\end{tabularx}
\caption{Candidate-pool coverage and selection. All-equal and all-zero counts use complete three-score pools. Score-match denominators use complete three-score pools with a scored selection, including operational zeros. An invalid selection in an all-zero pool would match the oracle score without delivering a valid selection; no invalid selections occurred. The final column restricts to unequal candidate grades.}
\label{tab:jointpoolcounts}
\end{table*}
\begin{table*}[t]
\centering\footnotesize
\setlength{\tabcolsep}{3.5pt}
\begin{tabularx}{\linewidth}{@{}lXrrl@{}}\toprule
Memory & Metric & Paired $N$ & Estimate & 95\% interval \\\midrule
Concrete & Oracle score & 1,768 & 0.317308 & [+0.301986, +0.332577] \\
Concrete & Oracle minus pool mean & 1,768 & +0.059829 & [+0.055333, +0.064437] \\
Concrete & Oracle minus selection (regret) & 1,768 & +0.060332 & [+0.053880, +0.067010] \\
Concrete & Fusion minus oracle & 1,768 & \mns0.068816 & [\mns0.076149, \mns0.061568] \\
Trait & Oracle score & 1,766 & 0.298226 & [+0.283381, +0.313271] \\
Trait & Oracle minus pool mean & 1,766 & +0.052536 & [+0.048154, +0.057034] \\
Trait & Oracle minus selection (regret) & 1,766 & +0.053983 & [+0.048273, +0.059786] \\
Trait & Fusion minus oracle & 1,766 & \mns0.053416 & [\mns0.060072, \mns0.046923] \\
\bottomrule\end{tabularx}
\caption{Candidate diagnostics on their own eligible paired sets; intervals are descriptive 95\% intervals. Fusion generates a new answer, so fusion minus oracle is a signed difference, not a nonnegative regret.}
\label{tab:jointpoolmetrics}
\end{table*}
}

\newcommand{\JointMissingnessParagraph}{%
The following bounds keep every planned task and allow each missing grade to range over the full score scale. They are deterministic completion bounds, not confidence intervals; shared missing-answer aliases can make them conservative. A non-missing operational zero is retained as zero, whereas a missing model or judge outcome is not imputed.
\begin{table*}[t]
\centering\footnotesize
\setlength{\tabcolsep}{3.5pt}
\begin{tabularx}{\linewidth}{@{}Xrl@{}}\toprule
Contrast & Missing paired tasks & Full-plan bounds \\\midrule
Concrete: fusion $-$ one-shot & 0 & [\mns0.012255, \mns0.012255] \\
Trait: fusion $-$ one-shot & 1 & [\mns0.002262, \mns0.001697] \\
Fusion: concrete $-$ trait & 1 & [+0.003394, +0.003959] \\
Interaction $\gamma$ & 1 & [\mns0.010558, \mns0.009992] \\
Concrete: fusion $-$ selection & 0 & [\mns0.008484, \mns0.008484] \\
Trait: fusion $-$ selection & 1 & [+0.000000, +0.001131] \\
Concrete fusion $-$ whole record & 1 & [\mns0.037707, \mns0.037142] \\
Concrete fusion $-$ current dialogue & 0 & [\mns0.075226, \mns0.075226] \\
\bottomrule\end{tabularx}
\caption{Full-plan missing-score bounds over all 1,768 planned tasks. With no missing required scores, a bound reduces to the observed point difference. With one missing task, the complete-case estimate in Table~\ref{tab:jointfamily} can fall slightly outside these bounds because the two use different denominators.}
\label{tab:jointmissingbounds}
\end{table*}
}

\newcommand{\JointCostTable}{%
\begin{table*}[t]
\centering\footnotesize
\setlength{\tabcolsep}{3.5pt}
\begin{tabularx}{\linewidth}{@{}Xrrrrr@{}}\toprule
Phase & Planned cap & Logical slots & HTTP & Recoveries & HTTP cap \\\midrule
Generation & 21,216 & 21,214 & 21,487 & 273 & 21,564 \\
Judging & 17,680 & 16,086 & 16,119 & 33 & 17,844 \\
\bottomrule\end{tabularx}
\caption{Actual shared-experiment request accounting. HTTP counts are durable reservations; unknown reservations might not have reached the service. Logical judging calls can be reduced by within-task exact-text reuse.}
\label{tab:jointcalls}
\end{table*}
\begin{table*}[t]
\centering\footnotesize
\setlength{\tabcolsep}{3.5pt}
\begin{tabularx}{\linewidth}{@{}lXrr@{}}\toprule
Phase & Token field & Known subtotal & Known/HTTP \\\midrule
Generation & Prompt & 17,300,397 & 21,214/21,487 \\
Generation & Answer & 2,651,812 & 21,213/21,487 \\
Generation & Thinking & unknown & 0/21,487 \\
Generation & Total & 19,952,209 & 21,214/21,487 \\
Judging & Prompt & 7,167,226 & 16,086/16,119 \\
Judging & Answer & 2,222,555 & 16,086/16,119 \\
Judging & Thinking & 10,028,120 & 16,086/16,119 \\
Judging & Total & 19,417,901 & 16,086/16,119 \\
\bottomrule\end{tabularx}
\caption{Reported token subtotals and coverage. An unreported field or unknown receipt has unknown consumption, not zero. These are shared execution totals, not per-method deployment costs or currency charges.}
\label{tab:jointtokens}
\end{table*}

Recorded per-request elapsed times sum to 53,110.44 seconds over 21,487 generation attempts and 135,541.76 seconds over 16,119 judging attempts. These sums are not wall-clock runtime. A deployed one-shot or raw-evidence method uses one call; fusion or selection uses three candidate calls plus one decision. The imported memories add their previously reported construction cost, rather than new profile calls in this run. The request and token accounting includes all original failed attempts. In particular, absent token usage on the quota-error receipts remains unknown; these requests are not removed from either HTTP totals or coverage denominators. The one-request quota amendments count as recoveries of existing logical slots, not as new logical tasks.
}

%% file: retrieval_results.tex
\newcommand{\VerbatimFactorial}{%
To separate what the evidence is from how it is introduced, we crossed the two on the development half after \method{}'s selection step was fixed. This factorial used \emph{question-free} excerpts (the query is the source's twelve most frequent content words), so it compares two evidence constructions under common headers, rather than isolating wording or testing \method{} itself; all four cells were generated and graded in one pass on 995 tasks from 104 people. Averaged over both headers, verbatim excerpts beat the written memory by +0.0276 (95\% interval [+0.0156, +0.0398]); with the raw-evidence header against the memory one-shot, the difference is +0.0271 ([+0.0140, +0.0406]). The header effect (\mns0.0005, [\mns0.0087, +0.0074]) was small and unresolved, and the interaction (\mns0.0030, [\mns0.0203, +0.0144]) was unresolved. These are post hoc estimates on the half on which \method{}'s selection step was chosen.}

\newcommand{\VerbatimFactorialTable}{%
\begin{table*}[t]
\centering\small
\begin{tabular}{@{}lrr@{}}\toprule
Evidence in the prompt & Profile header & Raw-evidence header \\\midrule
Model-written concrete memory & 0.2663 & 0.2673 \\
Verbatim excerpts, same budget & 0.2955 & 0.2935 \\
\bottomrule\end{tabular}
\caption{Development-half factorial with question-free excerpts: what the evidence is, crossed with how it is introduced. Means are normalized content scores on the 995 tasks graded in all four cells; every condition uses one generation call and the same per-task character budget.}
\label{tab:factorial}
\end{table*}
}

\newcommand{\VerbatimQuestionControl}{%
On the same half, question-aware selection did not detectably beat question-free selection (+0.0027, [\mns0.0107, +0.0161]; the two arms were graded in separate passes), an interval that does not exclude effects as large as the first-cohort confirmation (Appendix~\ref{app:freshcohort}).}

\newcommand{\VerbatimHoldout}{%
The first prespecified confirmation used the held-out half of the original participants (84 people, 771 paired tasks, both baselines regenerated in the same pass). It gave +0.0117 (95\% interval [\mns0.0034, +0.0270]) and did not meet the criterion of an interval above zero. We then ran the unchanged procedure and criterion on people outside the benchmark, a decision taken after seeing the held-out result.}

\newcommand{\VerbatimConfirmation}{%
The first new cohort has 500 people, one session each and 2,500 tasks, drawn from eligible interviews outside the benchmark; all conditions were generated and graded in one pass and no outcome is missing. \method{} scores 0.2769, the written memory 0.2643 and the whole record 0.2937. \method{} minus the memory one-shot is \textbf{+0.0127} (95\% interval [+0.0037, +0.0217]; 415 tasks better, 1,753 tied, 332 worse), which meets the prespecified criterion. The whole record (\method{} with unlimited budget) scores 0.0295 ([+0.0207, +0.0384]) above the memory and 0.0168 above \method{} ([+0.0079, +0.0257]). \method{} and the memory one-shot differ at once in content, header and question-aware selection, so this confirms \method{} as a whole rather than one mechanism. Excluding the 3 people who had served in a development batch gives +0.0118 ([+0.0028, +0.0207]); excluding all 156 people found in any earlier development file by a later audit gives +0.0099 ([\mns0.0008, +0.0205]), and the whole record minus the memory +0.0242 ([+0.0140, +0.0345]).}

\newcommand{\VerbatimConfirmationTable}{%
\begin{table*}[t]
\centering\small
\begin{tabular}{@{}lrrl@{}}\toprule
Personal evidence & Mean & Row minus \method{} & 95\% interval \\\midrule
\method{}: verbatim excerpts, memory budget & 0.2769 & -- & -- \\
Model-written memory, same budget & 0.2643 & \mns0.0127 & [\mns0.0217, \mns0.0037] \\
Whole record (\method{}, unlimited budget) & 0.2937 & +0.0168 & [+0.0079, +0.0257] \\
\bottomrule\end{tabular}
\caption{Confirmation on the first new cohort: 500 participants, 2,500 tasks, all three conditions generated and graded together. Each uses one generation call; \method{} is held to the same character budget as the written memory. Differences are row minus \method{}.}
\label{tab:freshcohort}
\end{table*}
}

\newcommand{\VerbatimHoldoutSecondaries}{%
Against the stale memory answers of the common experiment the held-out contrast would have read +0.0169 ([+0.0013, +0.0327]), a comparison we regard as invalid. On the held-out half, \method{} minus the regenerated whole-record one-shot was \mns0.0060 ([\mns0.0217, +0.0095]), the whole record minus the regenerated memory one-shot +0.0177 ([+0.0009, +0.0352]), and the regenerated memory one-shot minus the common experiment's concrete-memory one-shot +0.0052 ([\mns0.0057, +0.0160]).}

\newcommand{\VerbatimCrossJudge}{%
\paragraph{A second judge.} Re-grading the first new cohort's \method{} and memory answers with Grok~4.5 from another vendor, a check added after the confirmation result was known, with the identical rubric and parser and nothing regenerated, gives +0.0188 ([+0.0104, +0.0272]; 2,499 paired tasks) against +0.0127 under the primary judge. The judges agree exactly on 77.6\% of 4,999 predictions (quadratic-weighted $\kappa=0.81$); Grok grades lower on average but gives the larger contrast, so the primary estimate is the more conservative. A second model judge cannot exclude a preference, shared by both, for answers that reuse the respondent's own words.}

\newcommand{\VerbatimConclusion}{%
Under matched context budgets, the tested extractive procedures predicted interview content better on average than the written-memory procedure. \method{} did not detectably outperform recency truncation, and the survey results depended on the metric. These findings motivate evaluating memory representations and inference workflows separately; they establish neither a universal representation ranking nor an isolated benefit of verbatim wording.}

\newcommand{\RBudgetMain}{%
\paragraph{At the memory's budget.} In the prespecified confirmation, \method{} beats the written memory at its length: $\Delta_{\mathrm{mem}}=+0.0127$ ([+0.0037, +0.0217]), after an inconclusive test on 84 held-out benchmark participants (Appendix~\ref{app:freshcohort}); a judge from another vendor, added after the run, gives +0.0188 ([+0.0104, +0.0272]), and excluding 156 people a later audit found in exploratory files, $\Delta_{\mathrm{mem}}$ is an unresolved +0.0099 ([\mns0.0008, +0.0205]). On the comparison cohort (Table~\ref{tab:architectures}) \method{} beats the memory and Mem0 as deployed, and the MemGPT-style design, with twice the context and one more call, does not detectably differ from it. The procedure comparisons (Table~\ref{tab:ablations}) show a positive difference from the written memory, no detectable difference from recency truncation, and no detectable benefit from adding the memory; separately, removing the budget scores detectably higher, $\mathrm{Gap}(|m|)=+0.0124$ ([+0.0011, +0.0238]). On the third cohort \method{} minus the memory is an unresolved +0.0100, and there the memory plus the whole record (no budget) scores detectably above \method{} (Appendix~\ref{app:freshcohort}).}

\newcommand{\PairRMAppendixResult}{%
Its choice differs from the one-shot by \mns0.0030 ([\mns0.0092, +0.0033]) (concrete memory) and \mns0.0017 ([\mns0.0078, +0.0044]) (trait memory), from the prompted selector by +0.0008 ([\mns0.0058, +0.0074]) and +0.0011 ([\mns0.0050, +0.0073]), and from the candidate mean by +0.0003 ([\mns0.0042, +0.0047]) and \mns0.0003 ([\mns0.0046, +0.0040]).}

\newcommand{\FusionCrossJudgeAppendixResult}{%
As the run was analysed, with every answer graded separately, fusion minus one-shot (concrete memory) is \mns0.0066 ([\mns0.0138, +0.0006]). Applying after the run the primary judge's rule that byte-identical answers within a task share one grade (223 tasks in which fusion reproduced the one-shot) gives \mns0.0074 ([\mns0.0142, \mns0.0006]) on 1,768 tasks, against the prespecified \mns0.0123 ([\mns0.0189, \mns0.0056]) under the primary judge. The judges agree exactly on 77.2\% of the 3,536 gradings (quadratic-weighted $\kappa=0.80$). The identical pairs also give a test--retest check of the second judge: its two independent gradings of the same text agree exactly in 89.2\% of the 223 pairs (quadratic-weighted $\kappa=0.90$).}

\newcommand{\MemSysAppendix}{%
The second new cohort (300 people, one session each, 1,500 tasks; Appendix~\ref{app:freshcohort}) was run with the generator, judge and rubric of our other experiments. Its protocol names six contrasts (Mem0 minus \method{}, Mem0 minus the memory, MemGPT-style minus memory plus \method{}, memory plus \method{} minus \method{}, \method{} minus the memory, and the whole record minus \method{}) and declares them descriptive, with no confirmation criterion. \emph{The memory} is the concrete memory written with the profile prompt and answered in one call; \emph{\method{}} is Algorithm~\ref{alg:r} at budget $B$, the memory's length; \emph{the whole record} is one call on the whole permitted source. \emph{Mem0} runs the open-source mem0ai 2.1.0 pipeline with its own additive extraction prompt, unmodified (the 2.1.0 add path makes one extraction call and no separate update call): the permitted source is added as the participant's message, the extraction call (sent to our generator, JSON output) writes the memories, and Mem0's search ranks them against the question; the ranked memories fill up to budget $B$ greedily, as in Algorithm~\ref{alg:r}, and one call answers with the raw-evidence prompt. Embeddings (all-MiniLM-L6-v2; \citealp{reimers2019sbert}) and the vector store (Qdrant) ran locally; Mem0's optional keyword search and entity boosting, whose dependencies were not installed, were off, so its search ranked by embedding similarity alone with its default similarity threshold. Mem0 stored a median of 6 memories per session (range 0--16; 4 sessions with an empty store, one because Mem0 could not parse its extraction response). As deployed, its evidence averaged 569 characters against a mean budget of 717; in 525 of 1,500 tasks the whole store was shorter than $B$, and 38 tasks received no evidence, so the comparison with \method{} mixes what is retrieved with how much. \emph{Memory plus \method{}} gives the answering call the written memory and \method{}'s excerpts (budget $B$ each). The \emph{MemGPT-style} design is our own implementation of MemGPT's core-plus-archival pattern (the Letta system was not used): the written memory is the core memory, one call writes up to three archive queries from the screener, the memory and the question (1,500 of 1,500 parsed), the queries are joined into one BM25 query over the verbatim source at budget $B$, and one call answers over core memory and excerpts. It differs from memory plus \method{} only in who writes the query; MemGPT-style minus memory plus \method{} is \mns0.0018 ([\mns0.0111, +0.0076]). Both give the model about twice the evidence of the memory or \method{}. We did not implement the recency-, importance- and relevance-weighted retrieval of generative agents \citep{park2023generativeagents}; with one interview per person its recency term reduces to turn order, which Algorithm~\ref{alg:r}'s fill already uses, and its importance scoring remains untested. Two execution amendments are recorded with the run: more concurrency (after six sessions had started), then a lower in-flight cap and re-issuing slots that failed with an output-free rate-limit error. Before the second amendment an aggregate export had printed partial condition means at about 4\% completion; no setting was changed in response. One session's Mem0 extraction call failed with such an error and was re-issued once after the main pass, whose automatic analysis step had already written every contrast to the private run log; without that session's five tasks, Mem0 scores 0.2548 and Mem0 minus \method{} is \mns0.0181 ([\mns0.0299, \mns0.0065]). Against the whole record, a comparison not among the six protocol contrasts, Mem0 is 0.0302 lower ([+0.0180, +0.0424]), the MemGPT-style design 0.0216 lower ([+0.0104, +0.0327]) and memory plus \method{} 0.0198 lower ([+0.0087, +0.0309]). Table~\ref{tab:memsys} and the text give every mean and protocol contrast; intervals use seed 20260922.}

\newcommand{\MemSysTable}{%
\begin{table*}[t]
\centering\small
\resizebox{\linewidth}{!}{\begin{tabular}{@{}lrll@{}}\toprule
Evidence given to the answering call & Mean & Minus written memory & Minus \method{} \\\midrule
Written memory (one-shot) & 0.2542 & -- & \mns0.0187 [\mns0.0302, \mns0.0071] \\
\method{} at budget $B$: verbatim excerpts & 0.2729 & +0.0187 [+0.0071, +0.0302] & -- \\
Whole record (\method{}, unlimited budget) & 0.2853 & +0.0311 [+0.0191, +0.0431] & +0.0124 [+0.0011, +0.0238] \\
Mem0 (as deployed) & 0.2551 & +0.0009 [\mns0.0100, +0.0118] & \mns0.0178 [\mns0.0296, \mns0.0060] \\
Memory plus \method{}'s excerpts & 0.2656 & +0.0113 [+0.0011, +0.0216] & \mns0.0073 [\mns0.0187, +0.0038] \\
MemGPT-style & 0.2638 & +0.0096 [\mns0.0016, +0.0209] & \mns0.0091 [\mns0.0209, +0.0027] \\
\bottomrule\end{tabular}}
\caption{Memory systems on the second new cohort: 300 people, 1,500 tasks, every condition complete. Normalized content scores; paired differences with 95\% whole-person intervals. Descriptive comparison, prespecified without a confirmation criterion.}
\label{tab:memsys}
\end{table*}
}

\newcommand{\CohortThreeAppendix}{%
The third new cohort holds every remaining eligible person the audit counts as unused: 114 people from 21 studies, one session and five targets each (570 tasks), built with the same code and rules after also excluding the second cohort. Its run fixed one criterion before its first call, the memory plus the whole record minus the memory with a 95\% interval above zero, and generated and graded all five conditions in one pass. Means are 0.2304 for the memory, 0.2408 for \method{}, 0.2573 for the whole record, 0.2439 for the memory plus \method{} and 0.2573 for the memory plus the whole record. The criterion was met (+0.0269, [+0.0117, +0.0421]). Against the memory, the whole record gives +0.0269 ([+0.0094, +0.0444]; +0.0269, [+0.0109, +0.0436], with studies resampled), \method{} +0.0100 ([\mns0.0076, +0.0281]) and the memory plus \method{} +0.0135 ([\mns0.0023, +0.0298]); the memory plus the whole record differs from the whole record by +0.0000 ([\mns0.0140, +0.0140]), from \method{} by +0.0170 ([+0.0018, +0.0327]) and from the memory plus \method{} by +0.0135 ([\mns0.0023, +0.0292]); \method{} minus the whole record is \mns0.0170 ([\mns0.0347, +0.0006]) and \method{} minus the memory plus \method{} \mns0.0035 ([\mns0.0211, +0.0141]). One \method{} answer received an unparseable grade and stays missing.}

\newcommand{\RPlusAppendix}{%
\paragraph{Memory plus the whole record.} Because the memory and the verbatim record could each carry what the other lacks, we fixed this combined design (screener, written memory and whole record in one answering call) after the second-cohort and Twin-2K results were known, including a Twin-2K run without a screener in which it had beaten the whole persona in accuracy (below). It did not improve on the whole record alone: +0.0000 ([\mns0.0140, +0.0140]) on the third cohort (Appendix~\ref{app:freshcohort}) and \mns0.0133 ([\mns0.0247, \mns0.0020]) on the second, where it was generated in a separate pass several hours after the other conditions (one of its 1,500 grades was never completed, so its contrasts use 1,499 tasks) and gave +0.0178 ([+0.0062, +0.0296]) against the memory, \mns0.0009 ([\mns0.0134, +0.0120]) against \method{} and +0.0064 ([\mns0.0038, +0.0169]) against the memory plus \method{}. Adding the memory to the record therefore did not detectably help on the interviews.}

\newcommand{\TwinSecondSample}{%
\paragraph{Second Twin-2K-500 sample.} A second sample of 400 new respondents (the next positions of the same seeded order) tested the memory plus the whole persona, a design fixed after the earlier runs, under four criteria fixed before its first call: against the memory, higher accuracy (not met, +0.0014 ([\mns0.0027, +0.0055])) and lower deviation (met, \mns0.0080 ([\mns0.0112, \mns0.0049])); against \method{}, higher accuracy (met, +0.0063 ([+0.0020, +0.0107])); against the whole persona, higher accuracy (not met, +0.0020 ([\mns0.0020, +0.0060]); deviation +0.0009 ([\mns0.0019, +0.0036])). Budgeted \method{} minus the memory was \mns0.0049 ([\mns0.0097, \mns0.0002]) in accuracy and \mns0.0042 ([\mns0.0069, \mns0.0015]) in deviation. Table~\ref{tab:twin2} gives the means and every other paired contrast.}

\newcommand{\TwinSecondTable}{%
\begin{table*}[t]
\centering\small
\resizebox{\linewidth}{!}{\begin{tabular}{@{}lll@{}}\toprule
Evidence & Accuracy & Norm.\ deviation \\\midrule
Screener + written memory & 0.575 [0.567, 0.583] & 0.237 [0.230, 0.245] \\
Screener + \method{} & 0.570 [0.562, 0.578] & 0.233 [0.226, 0.240] \\
Screener + whole persona (\method{}, unlimited budget) & 0.574 [0.567, 0.582] & 0.228 [0.222, 0.235] \\
Screener + memory + \method{} & 0.579 [0.571, 0.586] & 0.236 [0.228, 0.243] \\
Screener + memory + whole persona & 0.576 [0.569, 0.584] & 0.229 [0.223, 0.236] \\
\addlinespace\multicolumn{3}{@{}l}{\emph{Paired differences}} \\
Whole persona $-$ memory & \mns0.0006 [\mns0.0054, +0.0042] & \mns0.0089 [\mns0.0124, \mns0.0055] \\
Memory + \method{} $-$ memory & +0.0037 [+0.0003, +0.0071] & \mns0.0014 [\mns0.0038, +0.0010] \\
Memory + \method{} $-$ whole persona & +0.0043 [\mns0.0004, +0.0090] & -- \\
Whole persona $-$ \method{} & +0.0044 [\mns0.0000, +0.0087] & \mns0.0047 [\mns0.0078, \mns0.0017] \\
Memory + whole persona $-$ memory + \method{} & \mns0.0023 [\mns0.0064, +0.0018] & \mns0.0066 [\mns0.0097, \mns0.0035] \\
Memory + whole persona $-$ \method{} & +0.0063 [+0.0020, +0.0107] & \mns0.0039 [\mns0.0068, \mns0.0009] \\
\bottomrule\end{tabular}}
\caption{Second Twin-2K-500 sample: 400 new respondents, all conditions generated in one pass. Exact-option accuracy and normalized deviation on ordinal items (lower is better), 95\% person-level intervals.}
\label{tab:twin2}
\end{table*}
}

\newcommand{\AnswerLengthSecEight}{%
\paragraph{Answer length.} Answers written from verbatim evidence are shorter. On the first new cohort the median answer has 104 words with \method{}, 92 with the whole record and 144 with the written memory; on the second, 100, 90 and 141, with 134 for Mem0, 113 for memory plus \method{} and 115 for the MemGPT-style design; on the third, 98, 90 and 136, with 97 for the memory plus the whole record. Length is a consequence of the evidence, not a manipulated factor, so this analysis is descriptive. A person-clustered regression of the task-level \method{} minus memory difference on the difference in log answer length has slope \mns0.0069 ([\mns0.0353, +0.0189]) and intercept +0.0103 ([\mns0.0021, +0.0220]) on the first new cohort, and slope \mns0.0115 ([\mns0.0551, +0.0292]) and intercept +0.0147 ([\mns0.0041, +0.0330]) on the second (10,000 person draws). The slope is unresolved and, at the point estimates, about four fifths of \method{}'s gain remains at equal length, but those intervals include zero, so length cannot be ruled out as part of the gain; shorter, more on-point answers could also be favoured by a model judge, and separating content from concision would need human ratings or a paraphrase test.}

\newcommand{\AnswerLengthSweep}{%
In the second cohort's budget sweep the median answer has 137, 142, 145, 155 words with the 50-, 100-, 200- and 400-word memories, 98, 101, 100, 95 with \method{} and 101, 103, 100, 96 with the most recent sentences at the same budgets, and 91 with the whole record.}

\newcommand{\StudyClusterNote}{%
The people of the new cohorts are nested in studies that share an interview guide (190, 90 and 21 studies). Resampling whole studies instead of people gives \method{} minus the memory +0.0127 ([+0.0036, +0.0218]) and whole record minus the memory +0.0295 ([+0.0206, +0.0385]) on the first new cohort, \method{} minus the memory +0.0187 ([+0.0072, +0.0303]), Mem0 minus \method{} \mns0.0178 ([\mns0.0292, \mns0.0064]) and memory plus \method{} minus the memory +0.0113 ([\mns0.0004, +0.0232]) on the second, and \method{} minus the memory +0.0100 ([\mns0.0067, +0.0236]) and whole record minus the memory +0.0269 ([+0.0109, +0.0436]) on the third.}

\newcommand{\CommonMemShare}{%
35}

\newcommand{\SourceStats}{%
On the first new cohort the permitted source has a median of 1,996 characters, the budget $B$ a median of 687, and $B$ is a median 36\% of the source; the 12,000-character cap binds for 5 sessions (25 tasks) in the three new cohorts.}

\newcommand{\RPilotResult}{%
On the development half, \method{} at the memory's budget scored +0.0274 ([+0.0136, +0.0413]) against the common experiment's concrete-memory one-shot (a baseline from an earlier pass) and +0.0013 ([\mns0.0114, +0.0139]) against its whole record; question-free excerpts (BM25 query: the source's twelve most frequent content words) scored +0.0248 ([+0.0121, +0.0377]) against the same one-shot. The held-out estimate (+0.0117) and the first new cohort's (+0.0127) are smaller, as expected after selection on the development half.}

\newcommand{\TwinMain}{%
\paragraph{A public survey benchmark.} On Twin-2K-500 \citep{toubia2025twin}, scored against recorded answers without a model judge, \method{} ranks a respondent's earlier-wave answers to non-target questions against each target survey block and one call predicts all of their closed-form wave-4 answers. At the memory's length, in both samples of 400 run with a screener as in the interviews, \method{} is closer than the memory to ordinal answers (normalized deviation \mns0.0081, [\mns0.0105, \mns0.0056], the prespecified primary; \mns0.0042, [\mns0.0069, \mns0.0015] in the second) but not more accurate on exact choices (+0.0005, [\mns0.0039, +0.0050]; lower in the second, \mns0.0049, [\mns0.0097, \mns0.0002]). \method{}$_\infty$, the whole persona, is closer still (\mns0.0117 and \mns0.0089 against the memory, both resolved), with no detectable accuracy difference. Two earlier runs that omitted the screener favored the memory on exact choices and, in one, on ordinal answers (Appendix~\ref{app:newexp}).}

\newcommand{\TwinTable}{%
\begin{table*}[t]
\centering\small
\begin{tabular}{@{}lll@{}}\toprule
Evidence & Accuracy & Norm.\ deviation \\\midrule
Screener only & 0.565 [0.557, 0.573] & 0.242 [0.235, 0.250] \\
Screener + written memory & 0.571 [0.563, 0.579] & 0.247 [0.239, 0.254] \\
Screener + \method{} & 0.571 [0.563, 0.580] & 0.239 [0.231, 0.246] \\
Screener + memory + \method{} & 0.576 [0.568, 0.584] & 0.245 [0.237, 0.253] \\
Screener + whole persona (\method{}, unlimited budget) & 0.571 [0.563, 0.579] & 0.235 [0.228, 0.242] \\
\midrule Respondents' own earlier answer (test--retest) & 0.706 [0.696, 0.715] & 0.148 [0.143, 0.154] \\
\bottomrule\end{tabular}
\caption{Twin-2K-500: 400 respondents, 32,776 closed-form wave-4 answers, all conditions generated in one pass (one memory-condition answer omitted one item, so comparisons with the memory use one item pair fewer). Exact-option accuracy (higher is better) and normalized absolute deviation on ordinal items (lower is better), with 95\% person-level bootstrap intervals. One answer call per respondent and condition predicts all of that respondent's items; the memory costs one further call.}
\label{tab:twin}
\end{table*}
}

\newcommand{\TwinContrasts}{%
Two criteria were fixed before any call on this sample, each a 95\% interval below zero in normalized deviation, with accuracy reported two-sided. The primary, \method{} below the memory, was met (\mns0.0081, [\mns0.0105, \mns0.0056]); the secondary, memory plus \method{} below the memory, was not met (\mns0.0016, [\mns0.0039, +0.0007]), although memory plus \method{} was more accurate (+0.0052, [+0.0018, +0.0087]). Against the screener alone, \method{} differs by +0.0063 ([+0.0028, +0.0098]) in accuracy and \mns0.0038 ([\mns0.0060, \mns0.0017]) in deviation, and the memory by +0.0058 ([+0.0011, +0.0106]) and +0.0042 ([+0.0017, +0.0068]), so adding the memory raised the deviation. The whole persona minus the memory is +0.0001 ([\mns0.0046, +0.0049]) in accuracy and \mns0.0117 ([\mns0.0147, \mns0.0087]) in deviation, and minus \method{} \mns0.0005 ([\mns0.0044, +0.0035]) and \mns0.0036 ([\mns0.0066, \mns0.0007]); memory plus \method{} minus \method{} is +0.0047 ([+0.0003, +0.0091]) and +0.0065 ([+0.0039, +0.0090]); memory plus \method{} minus the whole persona is +0.0051 ([+0.0004, +0.0098]) in accuracy.}

\newcommand{\TwinEarlier}{%
Two earlier runs gave no condition a screener (Table~\ref{tab:twinearly}). On the 200 development respondents, \method{} trailed the memory in accuracy; so did four development variants of its fill (other query units and larger budgets; \mns0.026 to \mns0.034, each interval below zero), which are in the public package. On 400 other respondents, the run first named one criterion, an accuracy interval above zero for the memory plus \method{} minus the memory (not met); after a development stage on the 200 development respondents, and before any call on the 400, it replaced that criterion with two, each a normalized-deviation interval below zero with the accuracy interval's lower bound above \mns0.005, for the memory plus \method{} (primary, met) and the memory plus the whole persona (secondary, met), both minus the memory. There the memory plus the whole persona beat the whole persona in accuracy, a gain the second screener sample did not reproduce (Table~\ref{tab:twin2}). Both runs, with all their contrasts, are in the public package.
\begin{table*}[t]
\centering\footnotesize
\resizebox{\linewidth}{!}{\begin{tabular}{@{}lll@{}}\toprule
Contrast & Accuracy & Norm.\ deviation \\\midrule
\multicolumn{3}{@{}l}{\emph{200 development respondents}} \\
\method{} $-$ memory & \mns0.034 [\mns0.045, \mns0.023] & -- \\
Whole persona $-$ memory & \mns0.007 [\mns0.014, +0.001] & \mns0.011 [\mns0.016, \mns0.005] \\
\multicolumn{3}{@{}l}{\emph{400 other respondents}} \\
Memory + \method{} $-$ memory (primary) & +0.0026 [\mns0.0010, +0.0062] & \mns0.0037 [\mns0.0061, \mns0.0013] \\
Memory + whole persona $-$ memory (secondary) & +0.0054 [+0.0010, +0.0099] & \mns0.0085 [\mns0.0117, \mns0.0054] \\
\method{} $-$ memory & \mns0.0390 [\mns0.0466, \mns0.0315] & +0.0114 [+0.0036, +0.0193] \\
Whole persona $-$ memory & \mns0.0046 [\mns0.0099, +0.0008] & \mns0.0090 [\mns0.0128, \mns0.0053] \\
Memory + whole persona $-$ whole persona & +0.0100 [+0.0054, +0.0146] & -- \\
Memory + \method{} $-$ whole persona & +0.0072 [+0.0021, +0.0122] & +0.0053 [+0.0018, +0.0089] \\
Memory + \method{} $-$ \method{} & +0.0416 [+0.0342, +0.0491] & \mns0.0151 [\mns0.0228, \mns0.0073] \\
Memory $-$ no evidence & +0.0487 [+0.0400, +0.0574] & \mns0.0022 [\mns0.0094, +0.0051] \\
\bottomrule\end{tabular}}
\caption{Earlier Twin-2K-500 runs, with no screener in any condition: paired differences with 95\% person-level intervals (accuracy: higher is better; normalized deviation on ordinal items: lower is better; --: not reported). The first run reports three decimals.}\label{tab:twinearly}
\end{table*}}

\newcommand{\ArchitectureTable}{%
\begin{table*}[t]
\centering\footnotesize
\setlength{\tabcolsep}{3pt}
\begin{tabularx}{\linewidth}{@{}l>{\raggedright\arraybackslash}Xcccc@{}}\toprule
Method & What the answering call sees & Context & Calls & Mean score & \method{} $-$ method \\\midrule
Written memory & a model-written profile of the record & $B$ & 1 (+1) & 0.2542 & \textbf{+0.0187} [+0.0071, +0.0302] \\
Mem0 as deployed$^\ddagger$ & facts Mem0 extracts, retrieved for $q$ & $\le B$ & 1 (+1) & 0.2551 & \textbf{+0.0178} [+0.0060, +0.0296] \\
MemGPT-style & memory + model-queried sentences & $\le 2B$ & 2 (+1) & 0.2638 & +0.0091 [\mns0.0027, +0.0209] \\\midrule
\textbf{\method{}} (ours) & the person's highest-ranked sentences, verbatim & $\le B$ & 1 & 0.2729 & -- \\
\bottomrule\end{tabularx}
\par\smallskip{\scriptsize $^\ddagger$As deployed, its evidence averaged 569 characters against a mean $B$ of 717 and was empty for 38 of 1,500 tasks (Appendix~\ref{app:newexp}).\par}
\caption{\method{} against memory systems at the memory's length: with one call and no model-written text it beats the written memory and Mem0 as deployed, and the MemGPT-style design (twice the context, one more call) does not detectably differ; the whole record, without a budget, scores higher still (Section~\ref{sec:retrieval}). Comparison cohort: 300 people, 1,500 tasks (comparisons declared descriptive). Difference: \method{} minus the method, paired, with its 95\% whole-person interval (bold: excludes zero). Context in units of $B$, the concrete memory's length (about 100 words); calls per answer, (+1) for one more per session to write the memory or Mem0's store. All methods also see the screener and question. Other cohorts: Table~\ref{tab:architecturescohorts}.}
\label{tab:architectures}
\end{table*}
}

\newcommand{\AblationTable}{%
\begin{table*}[t]
\centering\footnotesize
\setlength{\tabcolsep}{3pt}
\begin{tabularx}{\linewidth}{@{}l>{\raggedright\arraybackslash}Xcc@{}}\toprule
Compared procedure & Evidence construction & Context & \method{} $-$ variant \\\midrule
Written memory & a question-independent concrete rewrite of the record & $B$ & \textbf{+0.0187} [+0.0071, +0.0302] \\
Recency truncation$^\dagger$ & a contiguous suffix, stopping at the first sentence that does not fit & $\le B$ & +0.0008 [\mns0.0054, +0.0068] \\
Memory added & the memory placed before \method{}'s sentences & $\le 2B$ & +0.0073 [\mns0.0038, +0.0187] \\
\bottomrule\end{tabularx}
\par\smallskip{\scriptsize $^\dagger$Mean over the budget sweep's four memory lengths, each setting its own $B$ (Figure~\ref{fig:budget}), the prespecified contrast $\Delta_{\mathrm{rec}}$ of Eq.~\eqref{eq:sweep}.\par}
\caption{Evidence-construction comparisons and controls on the comparison cohort. Entries are \method{} minus the compared procedure, paired, with 95\% whole-person intervals (bold: excludes zero). These are not single-component ablations: the written memory changes retained content, wording, query dependence and header; recency truncation changes sentence ordering and packing. Only the $\dagger$ row had a criterion fixed in advance (not met). Removing the budget gives $\mathrm{Gap}(B)=+0.0124$ (Section~\ref{sec:retrieval}; regenerated sweep: Figure~\ref{fig:budget}).}
\label{tab:ablations}
\end{table*}
}

\newcommand{\ArchitectureCohortsTable}{%
\begin{table*}[t]
\centering\footnotesize
\setlength{\tabcolsep}{2.6pt}
\begin{tabularx}{\linewidth}{@{}l>{\raggedright\arraybackslash}Xcccrrr@{}}\toprule
& & & & \multicolumn{3}{c}{\method{} $-$ design} \\\cmidrule(l){5-7}
Compared design & What the answering call sees & Length & Calls & Cohort 1 & Cohort 2 & Cohort 3 \\
& & & & (500) & (300) & (114) \\\midrule
\multicolumn{7}{@{}l}{\emph{Memory systems}} \\
\quad Written memory & a model-written profile of the record & $B$ & 1 (+1) & \textbf{+0.0127} & \textbf{+0.0187} & +0.0100 \\
\quad Mem0 as deployed$^\ddagger$ & facts Mem0 extracts, retrieved for $q$ & $\le B$ & 1 (+1) & -- & \textbf{+0.0178} & -- \\
\quad MemGPT-style & memory + model-queried sentences & $\le 2B$ & 2 (+1) & -- & +0.0091 & -- \\
\multicolumn{7}{@{}l}{\emph{Evidence-construction controls}} \\
\quad Recency truncation$^\dagger$ & a contiguous suffix within the budget & $\le B$ & 1 & -- & +0.0008 & \mns0.0009 \\
\quad Memory added & memory + \method{}'s sentences & $\le 2B$ & 1 (+1) & -- & +0.0073 & \mns0.0035 \\
\quad No budget & the whole record (\method{}$_\infty$) & $\approx 3B$ & 1 & \textbf{\mns0.0168} & \textbf{\mns0.0124} & \mns0.0170 \\
\bottomrule\end{tabularx}
\par\smallskip{\scriptsize $^\dagger$Mean over the budget sweep's four memory lengths, each setting its own $B$ (Figure~\ref{fig:budgetcohorts}; Table~\ref{tab:budgetall}), the prespecified contrast $\Delta_{\mathrm{rec}}$ of Eq.~\eqref{eq:sweep}. $^\ddagger$As deployed, its evidence averaged 569 characters against a mean $B$ of 717 and was empty for 38 of 1,500 tasks (Appendix~\ref{app:newexp}).\par}
\caption{Tables~\ref{tab:architectures} and~\ref{tab:ablations}, plus the no-budget reference (its entry is $-\mathrm{Gap}(B)$; negative: the whole record scores higher), on all three new cohorts outside the benchmark (column heads: people; cohort 1 is the confirmation cohort and cohort 2 the comparison cohort of Section~\ref{sec:retrieval}). Entries: \method{} minus the row's design in mean normalized content score (positive: \method{} higher; bold: the 95\% whole-person interval excludes zero; --: not run; cohort 1 ran only the memory and whole-record comparisons and cohort 3 no memory systems, Figure~\ref{fig:studymap}). Length: characters of personal evidence in units of $B$, the length of the concrete memory of Section~\ref{sec:memory} (about 100 words); the whole record is about $3B$. All designs also see the screener and question. Calls: language-model calls per answer; (+1): one more per session to write the memory or Mem0's store. \method{} itself uses $\le B$ and 1 call. Criteria were fixed in advance only for the cohort-1 memory entry (met) and the $\dagger$ row (not met). Intervals: Tables~\ref{tab:freshcohort} and~\ref{tab:memsys} (design minus \method{}) and this appendix.}
\label{tab:architecturescohorts}
\end{table*}
}

\newcommand{\FigTagR}{+0.0127 [+0.0037, +0.0217]}

\newcommand{\BudgetMain}{%
\paragraph{Across budgets.} The budget sweep, also on the comparison cohort, uses memories requested at $w=50$, 100, 200 and 400 words (realized medians 54, 103, 187 and 318) and $B_w=|m_w|$, and generates every answer anew under three criteria fixed before its first call (the 100-word memory is the earlier run's, whose result was known; Figure~\ref{fig:budget}; Appendix~\ref{app:budget}). \method{} scores above the memory at every length, by +0.0200, +0.0169, +0.0231 and +0.0269 (mean +0.0218, [+0.0138, +0.0298], criterion met), and more budget moves it toward its unbounded limit: $\mathrm{Gap}(B_w)$ falls from +0.0293 at 50 words to +0.0050 and +0.0068, both unresolved, at 200 and 400 words ($\mu_{B_{400}}-\mu_{B_{50}}=+0.0235$, [+0.0116, +0.0353], criterion met), partly by construction, since the whole record fits in $B_w$ for 21\% and 53\% of tasks there. A longer memory does not close its own gap: the whole record's lead over the memory stays between +0.028 and +0.049. The recency comparison was unresolved: the most recent sentences also beat the memory at every length, and $\Delta_{\mathrm{rec}}(B_w)$ averages +0.0008, [\mns0.0054, +0.0068] (criterion not met). A prespecified replication on the third cohort reached the same three verdicts (Appendix~\ref{app:budget}).}

\newcommand{\BudgetFigCaption}{%
\method{} scores above a memory of the same length at every budget and approaches the whole record as the budget grows; it did not detectably beat recency truncation. Comparison cohort, 300 people (1,459--1,500 graded tasks per condition). A memory requested at $w$ words sets the budget $B$ for \method{} and for the most recent sentences that fit (recency truncation). (a) Mean normalized content score; dashed: \method{}$_\infty$, one call on the whole record; shaded: \method{}'s gain over the memory. (b) That gain, paired, with 95\% whole-person intervals; dashed line and band: its mean over the four lengths and that mean's interval (answers regenerated: the 100-word gain, +0.0169, differs from Table~\ref{tab:architectures}'s +0.0187; replication: Figure~\ref{fig:budgetcohorts}).}

\newcommand{\BudgetCohortsFigCaption}{%
The budget sweep on the comparison (second) and third cohorts. Top: mean normalized content score (second cohort: 300 people, 1,459--1,500 graded tasks per condition; third, the prespecified replication: 114 people, 565--570); dashed: \method{}$_\infty$, one call on the whole record; shaded: \method{}'s gain over the memory. Bottom: \method{} minus the memory, a paired difference with its 95\% whole-person interval (open: the interval covers zero). At 100 words the regenerated gains (+0.0169; +0.0058) differ from Table~\ref{tab:architecturescohorts} (+0.0187; +0.0100). Every contrast: Table~\ref{tab:budgetall}.}

\newcommand{\BudgetTable}{%
\begin{table*}[t]
\centering\footnotesize
\setlength{\tabcolsep}{3pt}\setlength{\abovetopsep}{3pt}
\resizebox{\linewidth}{!}{\begin{tabular}{@{}rrrrlll@{}}\toprule
Words & Memory & Most recent & \method{} & \method{} $-$ memory & \method{} $-$ most recent & Whole record $-$ \method{} \\\midrule
50 & 0.2431 & 0.2680 & 0.2630 & +0.0200 ([+0.0073, +0.0329]) & \mns0.0050 ([\mns0.0159, +0.0059]) & +0.0293 ([+0.0168, +0.0417]) \\
100 & 0.2587 & 0.2753 & 0.2756 & +0.0169 ([+0.0056, +0.0282]) & +0.0002 ([\mns0.0107, +0.0111]) & +0.0169 ([+0.0056, +0.0280]) \\
200 & 0.2650 & 0.2852 & 0.2886 & +0.0231 ([+0.0105, +0.0359]) & +0.0034 ([\mns0.0066, +0.0135]) & +0.0050 ([\mns0.0053, +0.0153]) \\
400 & 0.2580 & 0.2817 & 0.2851 & +0.0269 ([+0.0148, +0.0389]) & +0.0036 ([\mns0.0041, +0.0112]) & +0.0068 ([\mns0.0016, +0.0153]) \\
\bottomrule\end{tabular}}
\caption{Budget sweep on the second new cohort (third cohort: Table~\ref{tab:budgetall} and Figure~\ref{fig:budgetcohorts}; the first new cohort was not in the sweep): means and paired differences (95\% whole-person intervals) at each memory length; differences use the tasks both conditions score (the means over budgets, tasks scored at all four) and need not equal differences of the means; \method{} and the most recent sentences are capped at that memory's number of characters (the whole record when it is shorter; the most recent sentences stop at the first sentence that does not fit). The one-shot on the whole record scores 0.2924.}
\label{tab:budget}
\end{table*}
}

\newcommand{\BudgetRatio}{%
Grouped by the share of the source that the budget covers (a task enters a group with the mean of its budgets that fall in it; intervals from 20,000 draws), below a quarter (965 tasks): whole record minus \method{} +0.0334, minus the memory +0.0425, minus the most recent sentences +0.0241, and \method{} minus the memory +0.0091 ([\mns0.0045, +0.0226]); a quarter to a half (1,304 tasks): whole record minus \method{} +0.0122, minus the memory +0.0422 (1,305 for the memory), minus the most recent sentences +0.0153 (1,305 for the most recent sentences), and \method{} minus the memory +0.0298 ([+0.0178, +0.0421]); a half to nearly all (1,315 tasks): whole record minus \method{} +0.0077, minus the memory +0.0310, minus the most recent sentences +0.0167, and \method{} minus the memory +0.0233 ([+0.0113, +0.0353]); all of it (785 tasks): whole record minus \method{} +0.0000, minus the memory +0.0268, minus the most recent sentences +0.0000, and \method{} minus the memory +0.0268 ([+0.0105, +0.0435]).}

\newcommand{\BudgetAppendixContrasts}{%
Table~\ref{tab:budgetall} gives every per-budget contrast; the memory at 400 minus at 50 words is +0.0153 ([+0.0039, +0.0267]).}

\newcommand{\BudgetReplication}{%
The same protocol, also fixed after the third cohort's 100-word \method{}-versus-memory result was known and run afterwards on the third new cohort (114 people, 570 tasks; 565 with all four budgets), gives \method{} minus the memory +0.0153 ([+0.0055, +0.0254]) averaged over budgets (met), \method{} minus the most recent sentences \mns0.0009 ([\mns0.0081, +0.0063]) (not met) and \method{} at 400 minus \method{} at 50 words +0.0224 ([+0.0059, +0.0389]) (met); on that cohort the memory at 400 minus at 50 words is +0.0136 ([\mns0.0029, +0.0301]). As on the second cohort, every answer was generated and graded anew, so at 100 words \method{} minus the memory is +0.0058 ([\mns0.0094, +0.0211]) here against +0.0100 in the cohort's earlier run.
\begin{table*}[t]
\centering\scriptsize
\setlength{\tabcolsep}{3pt}\setlength{\abovetopsep}{3pt}
\begin{tabular}{@{}lcccc@{}}\toprule
Contrast & 50 words & 100 words & 200 words & 400 words \\\midrule
\multicolumn{5}{@{}l}{\emph{Second new cohort (300 people)}} \\
\quad \method{} $-$ memory & +0.0200 & +0.0169 & +0.0231 & +0.0269 \\
 & [+0.0073, +0.0329] & [+0.0056, +0.0282] & [+0.0105, +0.0359] & [+0.0148, +0.0389] \\
\quad \method{} $-$ most recent & \mns0.0050 & +0.0002 & +0.0034 & +0.0036 \\
 & [\mns0.0159, +0.0059] & [\mns0.0107, +0.0111] & [\mns0.0066, +0.0135] & [\mns0.0041, +0.0112] \\
\quad Most recent $-$ memory & +0.0249 & +0.0167 & +0.0196 & +0.0237 \\
 & [+0.0125, +0.0376] & [+0.0040, +0.0296] & [+0.0073, +0.0322] & [+0.0109, +0.0364] \\
\quad Whole record $-$ \method{} & +0.0293 & +0.0169 & +0.0050 & +0.0068 \\
 & [+0.0168, +0.0417] & [+0.0056, +0.0280] & [\mns0.0053, +0.0153] & [\mns0.0016, +0.0153] \\
\quad Whole record $-$ memory & +0.0492 & +0.0338 & +0.0281 & +0.0339 \\
 & [+0.0370, +0.0615] & [+0.0222, +0.0453] & [+0.0162, +0.0400] & [+0.0216, +0.0460] \\
\quad Whole record $-$ most recent & +0.0243 & +0.0171 & +0.0084 & +0.0102 \\
 & [+0.0122, +0.0363] & [+0.0058, +0.0287] & [\mns0.0009, +0.0178] & [+0.0032, +0.0175] \\
\addlinespace
\multicolumn{5}{@{}l}{\emph{Third new cohort (114 people)}} \\
\quad \method{} $-$ memory & +0.0124 & +0.0058 & +0.0216 & +0.0211 \\
 & [\mns0.0047, +0.0301] & [\mns0.0094, +0.0211] & [+0.0053, +0.0380] & [+0.0047, +0.0374] \\
\quad \method{} $-$ most recent & +0.0059 & +0.0029 & \mns0.0076 & \mns0.0053 \\
 & [\mns0.0100, +0.0224] & [\mns0.0111, +0.0175] & [\mns0.0211, +0.0058] & [\mns0.0158, +0.0053] \\
\quad Most recent $-$ memory & +0.0065 & +0.0029 & +0.0292 & +0.0263 \\
 & [\mns0.0100, +0.0236] & [\mns0.0123, +0.0181] & [+0.0140, +0.0439] & [+0.0088, +0.0439] \\
\quad Whole record $-$ \method{} & +0.0248 & +0.0175 & +0.0099 & +0.0023 \\
 & [+0.0083, +0.0413] & [+0.0029, +0.0322] & [\mns0.0035, +0.0234] & [\mns0.0088, +0.0140] \\
\quad Whole record $-$ memory & +0.0372 & +0.0234 & +0.0316 & +0.0234 \\
 & [+0.0218, +0.0525] & [+0.0088, +0.0380] & [+0.0164, +0.0468] & [+0.0076, +0.0386] \\
\quad Whole record $-$ most recent & +0.0307 & +0.0205 & +0.0023 & \mns0.0029 \\
 & [+0.0136, +0.0484] & [+0.0064, +0.0351] & [\mns0.0105, +0.0152] & [\mns0.0140, +0.0088] \\
\bottomrule\end{tabular}
\caption{Every per-budget contrast of the budget sweep (second new cohort) and of its prespecified replication (third new cohort): paired differences, with 95\% whole-person intervals on the second line. The first new cohort ran before the sweep was designed and is not part of it (Figure~\ref{fig:studymap}). At each memory length, \method{} and the most recent sentences are capped at that memory's number of characters.}
\label{tab:budgetall}
\end{table*}}

%% file: ethics_statement.tex
\newcommand{\EthicsStatementText}{This study analyzes existing private user-research interviews; it does not recruit additional participants or collect new interviews. The main-text experiments use 1,108 interview sessions from 1,102 people: the 194 benchmark sessions and three new cohorts of 500, 300 and 114 sessions drawn from the same eligible pool; with the 171 further session records of the cross-interview benchmark (Appendix~\ref{app:benchmark}; 38 of its pairs are duplicate records of one session), the benchmarks and cohorts analysed here comprise 1,279 session records (1,241 distinct sessions) from 1,175 people. The development batches of Appendix~\ref{app:workflow} and earlier development on the same corpus used further sessions as targets and demonstrations; we did not reconstruct a session-level count of all text sent to model services during that development.{} The external benchmark of Appendix~\ref{app:newexp} is public and used under its license.{} The source study's documentation reports participant consent to research use under the platform's participation terms and platform authorization for analysis. The documentation available for this study does not establish a formal ethics-board approval or exemption, and we do not claim either. Access to the underlying material is restricted to authorized research collaborators under the data owner's restrictions. Interview text, screeners, memory profiles, reference answers, predictions, and grading rationales remain in restricted storage; generated text can reproduce participant speech and is therefore excluded from the public package. The accompanying numerical artifacts use remapped relational identifiers and contain scores, word and character counts, exposure flags, binary task features derived from the restricted text (whether a reference answer contains a digit or a non-initial capitalized word), generic prompts, and analysis code. The remapped identifiers are sequential labels that preserve the clustering needed for analysis and the sort order of the original identifiers; they are not a guarantee against re-identification using auxiliary information. Generation and evaluation send interview-derived text to the external model services specified in the experimental protocols (the Mem0 comparison ran its vector store and embedding model locally, with telemetry disabled, and sent its extraction call to the same generation service), and we have not independently verified those services' retention practices. No new human ratings were collected. Predictions are hypotheses about one observed response, not verified beliefs, identities, or future actions. The experiments do not establish suitability for consequential decisions about individuals or authorize impersonation or additional uses of their data.}
\newcommand{\AIUseStatementText}{In this work, we used generative AI tools to design and give feedback on the experimental methodology, to implement the methods and analysis code, to clean and reformat score data for analysis and release, to help refine the hypotheses and the framing of the argument, and to help interpret results. Model-based content grading is itself the evaluation instrument for the interview experiments and is reported as such throughout; the memories, predictions, fusion, and selection outputs studied here are also model outputs by design. We have not used generative AI tools to generate synthetic data for any reported result (AI-written unit tests use small synthetic score fixtures only) or for translation, and mathematical claims and proofs are not applicable to this work. Additionally, we used generative AI tools to create and edit software code, to search for, identify, and summarize relevant literature and format references, to suggest the paper's structure, title, and keywords, and to draft and edit the manuscript. Reported numbers were also recomputed by separately implemented AI-agent checks, and bibliographic entries were checked by an AI agent against their publisher, proceedings or arXiv records; these are software checks, not independent human evaluation. All authors took part in every part of the work, including the human review of the AI-assisted design, code, analyses, citations and text, and take responsibility for the final content of this work, including text, claims, citations, data permissions, and artifacts produced with the aid of generative AI.}

%% file: human_status.tex
\newcommand{\HumanValidationAppendix}{After generation and execution verification, the fixed sample yielded 1,504 method slots across 188 selected tasks. Two slots had missing method outputs and remained explicit in the private mapping. Within-task exact-text deduplication merged 258 of the 1,502 complete slots, leaving 1,244 items per rater, or 2,488 requested ratings across the two separately permuted offline questionnaires. Both questionnaires were prepared with empty rating fields. No raters were confirmed before submission, so the packets remain unrated. No human ratings were collected, so we report no human agreement statistic or human-validated workflow effect.}

%% file: architecture_figure.tex
\begin{figure*}[t]
\begin{minipage}[t]{0.51\linewidth}
\vspace{0pt}\centering
\begin{tikzpicture}[x=1cm,y=1cm,
  box/.style={rounded corners=1.5pt,align=flush center,font=\scriptsize,minimum height=0.72cm,inner sep=2pt},
  dat/.style={box,draw=black!45,fill=white},
  llm/.style={box,draw=orange!75!black,fill=orange!16},
  det/.style={box,draw=teal!65!black,fill=teal!14,dashed,line width=0.6pt},
  arr/.style={-{Latex[length=1.4mm]},black!60},
  gbox/.style={box,draw=black!30,fill=black!4,text=black!75,minimum height=0.5cm,font=\scriptsize},
  garr/.style={-{Latex[length=1.1mm]},black!35}]
\node[dat,text width=1.35cm] (H) at (0.75,0) {record $H$\\$s_1,\dots,s_n$};
\node[det,text width=1.55cm] (bm) at (3.05,0) {BM25 score\\$r_i(q)$ from $H$};
\node[det,text width=1.55cm] (kn) at (5.45,0) {greedy fill\\$|x|\le B$};
\node[font=\scriptsize,text=black!70] (q) at (3.05,0.78) {question $q$};
\draw[arr] (H)--(bm); \draw[arr] (bm)--(kn); \draw[arr] (q)--(bm);
\node[dat,text width=1.55cm] (x) at (0.85,-1.35) {$x_B(q)$: own\\sentences, verbatim};
\node[llm,text width=1.75cm] (f) at (3.35,-1.35) {one model call\\$f(\sigma,x_B(q),q)$};
\node[font=\small] (a) at (5.45,-1.35) {$\hat a$};
\draw[arr] (kn.south) -- ++(0,-0.25) -| (x.north);
\draw[arr] (x)--(f); \draw[arr] (f)--(a);
\node[font=\scriptsize,text=black!70] at (5.3,-1.9) {$B\ge|H|\Rightarrow x_B(q)=H$};
\node[font=\scriptsize,text=black!70,anchor=west] at (-0.05,-2.2) {contrast: the written memory};
\node[gbox,text width=1.0cm] (H2) at (0.6,-2.65) {record $H$};
\node[gbox,text width=2.25cm] (g) at (2.85,-2.65) {model rewrite $m=g(H)$};
\node[gbox,text width=1.3cm] (f2) at (5.25,-2.65) {$f(\sigma,m,q)$};
\draw[garr] (H2)--(g); \draw[garr] (g)--(f2);
\end{tikzpicture}
\end{minipage}\hfill
\begin{minipage}[t]{0.46\linewidth}
\vspace{0pt}
\begin{algorithm}[H]
\caption{\method{} with budget $B$}
\label{alg:r}
\small
\begin{algorithmic}[1]
\Require $H=(s_1,\dots,s_n)$, $\sigma$, $q$, $B$, predictor $f$
\State $r_i\gets\mathrm{BM25}(s_i,q;H)$ for all $i$ \Comment{Eq.~\eqref{eq:bm25}}
\State $\mathcal{K}\gets\emptyset$
\For{$i$ by decreasing $r_i$ (ties: latest first)}
  \State \textbf{if} $\bigl|\bigoplus_{j\in \mathcal{K}\cup\{i\}}s_j\bigr|\le B$ \textbf{then} $\mathcal{K}\gets \mathcal{K}\cup\{i\}$
\EndFor
\State $x\gets\bigoplus_{i\in \mathcal{K}}s_i$ \Comment{verbatim; $=H$ if $B\ge|H|$}
\State \Return $f(\sigma,x,q)$ \Comment{the only model call}
\end{algorithmic}
\end{algorithm}
\end{minipage}
\caption{\method{} (left; steps in Algorithm~\ref{alg:r}). A deterministic step (dashed; no model call) keeps the person's highest-ranked sentences verbatim within $B$ characters, and one call answers; with $B\ge|H|$ the evidence is the whole record. The written memory (gray; the concrete memory of Section~\ref{sec:memory}) instead selects and rewrites content without the question and, in our tests, sets $B=|m|$. $H=(s_1,\dots,s_n)$: the person's own speech before the session's first target, in sentences; $\sigma$: screener; $q$: target question.}
\label{fig:architecture}
\end{figure*}

%% file: chain_figure_float.tex
\begin{figure*}[t]
\centering
\begin{tikzpicture}[x=1cm,y=1cm,
  q/.style={rounded corners=2pt,draw=black!35,fill=black!3,inner sep=3.5pt,font=\scriptsize,anchor=north west},
  ours/.style={rounded corners=3pt,draw=blue!70!black,line width=0.9pt,fill=blue!8,align=left,inner sep=5pt,font=\small,anchor=north west},
  ar/.style={-{Latex[length=1.8mm]},black!55,line width=0.6pt},
  lab/.style={font=\tiny,text=black!60,align=center}]
\node[q] (a) at (0,0) {\parbox[t][1.5cm][t]{2.72cm}{\raggedright\hyphenpenalty=10000\exhyphenpenalty=10000 \textbf{Does the interview history help?}\\[1pt] \textit{Yes}, with all four models:\\ +0.062 to +0.087 (\S\ref{sec:history})}};
\node[q] (b) at (3.55,0) {\parbox[t][1.5cm][t]{2.72cm}{\raggedright\hyphenpenalty=10000\exhyphenpenalty=10000 \textbf{Does how the memory is written matter?}\\[1pt] \textit{Yes}: concrete beats trait,\\ +0.0158, prespecified (\S\ref{sec:memory})}};
\node[q] (c) at (7.1,0) {\parbox[t][1.5cm][t]{2.72cm}{\raggedright\hyphenpenalty=10000\exhyphenpenalty=10000 \textbf{Does the tested fusion improve prediction?}\\[1pt] \textit{No}: fusion lowers the concrete-memory score by 0.0123 (primary, \S\ref{sec:joint})}};
\node[q] (d) at (10.65,0) {\parbox[t][1.5cm][t]{2.72cm}{\raggedright\hyphenpenalty=10000\exhyphenpenalty=10000 \textbf{Does the unrewritten record beat the memory?}\\[1pt] \textit{Yes}, by +0.0253 (post hoc), but it is about three times longer (\S\ref{sec:joint})}};
\draw[ar] (a.east) -- (b.west);
\draw[ar] (b.east) -- (c.west);
\draw[ar] (c.east) -- (d.west);
\node[ours,text width=13.26cm] (e) at (0,-2.2) {{\hyphenpenalty=10000\textbf{\method{} (\S\ref{sec:retrieval}): the person's own sentences, chosen for the question, at the memory's length.}}\\[2pt]
\footnotesize\hyphenpenalty=10000 It exceeds the written memory within the same budget in a prespecified confirmation on 500 people outside the benchmark (\FigTagR); on a second cohort it beats Mem0 as deployed (Table~\ref{tab:architectures}) and the memory by +0.0218 averaged over budgets of 50 to 400 words, approaching one call on the whole record as the budget grows (Figure~\ref{fig:budget}).\par};
\draw[ar,blue!70!black,line width=0.9pt] (d.south) -- (d.south |- e.north);
\end{tikzpicture}
\caption{The argument in one figure: four empirical comparisons motivate the budget-matched extractive reference. Numbers are paired differences in normalized content score from different experiments and are not to be added. Figure~\ref{fig:chainfull} (Appendix~\ref{app:freshcohort}) draws the contrasts behind the four questions and \method{}'s gains over the memory with 95\% intervals, including the other cohorts (the third unresolved at this length) and an earlier, inconclusive held-out test.}
\label{fig:chain}
\end{figure*}
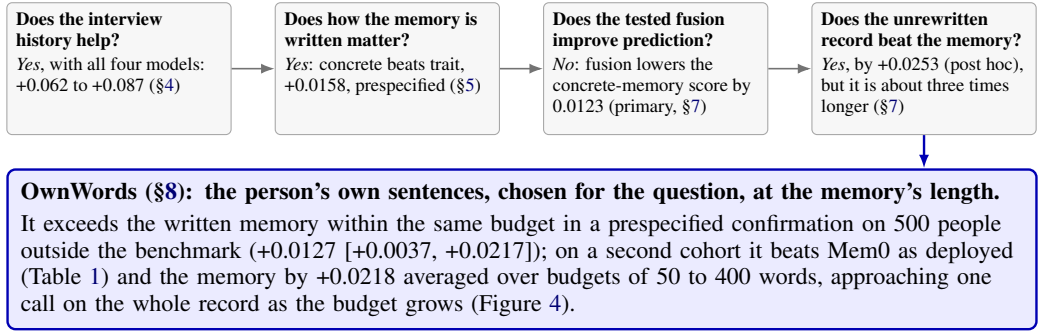

%% file: prelude_figure_float.tex
\begin{figure*}[t]
\centering
\includegraphics[width=\linewidth]{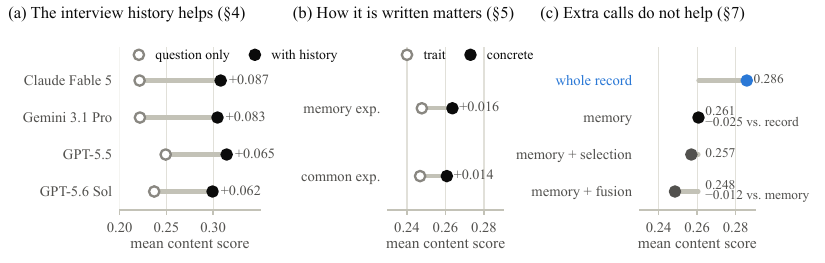}
\caption{What Sections~\ref{sec:history}--\ref{sec:joint} find, in mean normalized content score; (a) has another judge, so compare levels within panels. (a) History and screener raise the score with each of four models (labels: paired differences). (b) Concrete beats trait memory, prespecified and again in the common experiment. (c) Common experiment: one call on the whole record (\method{} without a budget) beats the memory (post hoc); fusion lowers it (primary endpoint) and selection does not recover it; the current dialogue (0.324) also sees later turns. Intervals: Table~\ref{tab:generation}, App.~\ref{app:jointresults}.}
\label{fig:prelude}
\end{figure*}

%% file: retrieval_section.tex
\section{\method{}: The Person's Own Words under a Budget}
\label{sec:retrieval}
The whole record beats every memory condition (Section~\ref{sec:joint}), but it is about three times longer (the memory is a median \CommonMemShare\% of the record in the common experiment), so that comparison cannot separate rewriting from length. Since a memory exists to fit a limited context, we compare evidence-construction procedures under a common character ceiling, with the person's own sentences as a simple extractive reference (Figure~\ref{fig:architecture}).

\paragraph{Problem.} For one session let $H=(s_1,\dots,s_n)$ be the permitted record, the person's own speech before the earliest target cut (Section~\ref{sec:info}) split into sentences in temporal order, $\sigma$ the screener and $q$ a target question; $|\cdot|$ counts characters and $\oplus$ joins sentences in temporal order with single spaces. \method{} and its main references answer with one call to the same predictor, $\hat a=f(\sigma,x,q)$, and differ in the evidence $x$ (and, for the memory, its prompt header), which must fit a budget of $B$ characters, $|x|\le B$. The written memory is a question-independent rewrite, $x=m=g(H)$, with $g$ the concrete-memory prompt of Section~\ref{sec:memory} (one further call per session); in our tests it also sets the budget, $B:=|m|$, so \method{} uses no more context than the memory.

\ArchitectureTable{}
\AblationTable{}

\paragraph{Method.} For every sentence, \method{} computes a BM25 score against the question \citep{robertson2009bm25}, using statistics from $H$ alone,
\begin{equation}
\begin{split}
r_i(q)=\sum_{v\in Q}&\ln\Bigl(1+\frac{n-n_v+0.5}{n_v+0.5}\Bigr)\\
&\times\frac{\mathrm{tf}_{v,i}\,(k_1+1)}{\mathrm{tf}_{v,i}+k_1\bigl(1-b+b\,\ell_i/\bar\ell\bigr)},
\end{split}
\label{eq:bm25}
\end{equation}
where $Q$ is the set of content tokens of $q$ (tokenizer in Appendix~\ref{app:attempts}), $\mathrm{tf}_{v,i}$ the count of $v$ in $s_i$, $n_v$ the number of sentences containing $v$, $\ell_i$ the number of content tokens of $s_i$ and $\bar\ell$ its mean over $H$; $k_1=1.2$ and $b=0.75$ are the standard values, and these retrieval parameters are fixed. It then keeps the highest-ranked sentences that fit, verbatim:
\begin{equation}
x_B(q)=\textstyle\bigoplus_{i\in \mathcal{K}_B(q)}s_i,
\label{eq:knapsack}
\end{equation}
where $\mathcal{K}_B(q)$ greedily seeks
\begin{equation*}
\max_{\mathcal{K}\subseteq\{1,\dots,n\}}\textstyle\sum_{i\in \mathcal{K}}r_i(q)\ \ \text{s.t.}\ \ \bigl|\bigoplus_{i\in \mathcal{K}}s_i\bigr|\le B.
\end{equation*}
Algorithm~\ref{alg:r} uses a greedy heuristic for this 0--1 selection problem, without an optimality guarantee, visiting sentences in decreasing $r_i(q)$ with ties, including all zero-score sentences, broken toward the most recent, and keeping each that still fits. One call then answers from $(\sigma,x_B(q),q)$; \method{} writes nothing.

\paragraph{The unbounded limit and three contrasts.} If $B\ge|H|$, every sentence fits and $x_B(q)=H$ up to whitespace: \method{}$_\infty$, the one-shot on the whole record, is the method without a budget, and the budget only saves context. A recency baseline instead keeps the longest contiguous suffix that fits, $x^{\mathrm{rec}}_B=s_j\oplus\dots\oplus s_n$ (the last $B$ characters of $s_n$ if even $s_n$ does not fit); unlike \method{}, it does not skip an overlong sentence and continue filling, so the comparison changes packing as well as query-based ranking. The written memory additionally changes selected content, wording and the prompt header. With $\mu_B$, $\mu_\infty$, $\mu^{\mathrm{mem}}_B$ and $\mu^{\mathrm{rec}}_B$ the mean content scores (Section~\ref{sec:rubric}) of \method{} at budget $B$, its unbounded limit, the memory and recency truncation (Table~\ref{tab:ablations}), three paired contrasts compare the procedures and the cost of the budget:
\begin{equation}
\begin{gathered}
\Delta_{\mathrm{mem}}(B)=\mu_B-\mu^{\mathrm{mem}}_B,\\
\Delta_{\mathrm{rec}}(B)=\mu_B-\mu^{\mathrm{rec}}_B,\ \ \mathrm{Gap}(B)=\mu_\infty-\mu_B.
\end{gathered}
\label{eq:sweep}
\end{equation}

\input{budget_figure_float.tex}

\paragraph{Tests outside the benchmark.} After developing \method{} on half of the benchmark's people (Appendix~\ref{app:attempts}), we tested it on people outside the benchmark, from the same eligible pool and with the generator, judge and rubric of the common experiment: a prespecified confirmation against the memory on 500 people, and a comparison cohort of 300 people on which every design of Tables~\ref{tab:architectures} and~\ref{tab:ablations} and the budget sweep were run. A third cohort of 114 people repeated the budget sweep and part of the comparison; Appendix~\ref{app:freshcohort} calls the three the first, second and third new cohorts (map: Figure~\ref{fig:studymap}).

\RBudgetMain{}

\BudgetMain{}

\TwinMain{}

\paragraph{What the evidence supports.} $\Delta_{\mathrm{mem}}>0$ in the first-cohort confirmation and each sweep's average; $\Delta_{\mathrm{rec}}$ is unresolved. Both tested extractive policies can thus outperform the written-memory procedure, while their difference is not established, and the comparisons do not isolate verbatim wording from retained content, packing, or presentation. $\mathrm{Gap}(B)$ narrows at larger budgets, partly because the entire record then fits; gains under budgets below a quarter of the source remain unresolved in both sweep cohorts (Appendix~\ref{app:overview}). These results concern short interview records and context budgets, not a fixed storage budget for long-term memory. Added model stages also failed to yield a detectable improvement in the tested interview comparisons, but several intervals permit positive effects. Verbatim evidence yields shorter answers (Appendix~\ref{app:anslength}), leaving answer style and judge preference unresolved without human ratings or a controlled paraphrase comparison. An unverified earlier run found model-selected quotes not detectably better than a paraphrase (Appendix~\ref{app:lineage}).

%% file: budget_figure_float.tex
\begin{figure*}[t]
\centering
\includegraphics[width=\linewidth]{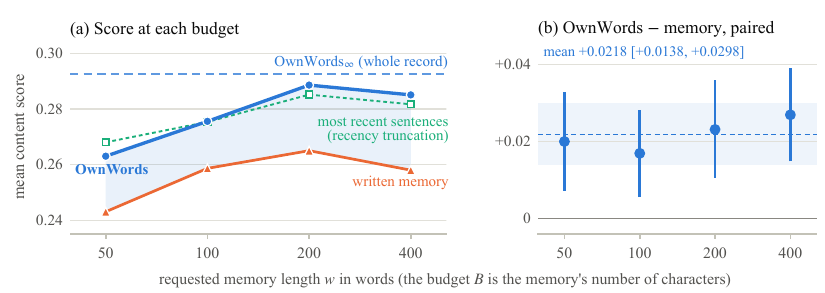}
\caption{\BudgetFigCaption{}}
\label{fig:budget}
\end{figure*}

%% file: overview_appendix.tex
\section{Study Map and Data at a Glance}
\label{app:overview}
\begin{figure*}[t]
\centering
\includegraphics[width=\linewidth]{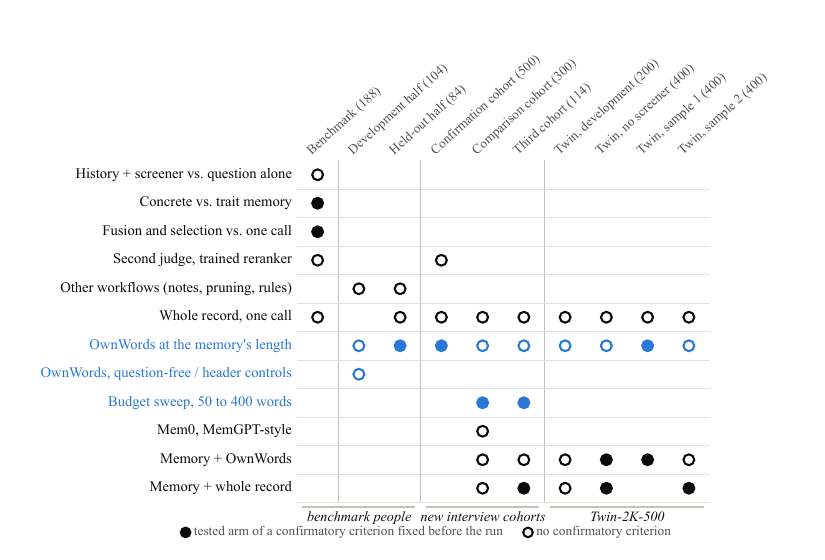}
\caption{Study map: which design ran on which population (column heads give the number of people). Filled: the design is the tested arm of a confirmatory criterion fixed before the run; open: no confirmatory criterion (descriptive, a development gate, a comparator, or post hoc). Designs of \method{} in blue; the whole record (\method{}$_\infty$), a reference, stays black.}
\label{fig:studymap}
\end{figure*}

Figure~\ref{fig:studymap} shows which design ran on which population and which comparisons had a confirmatory criterion fixed before the run. The benchmark's 188 people carry the history benchmark, the memory experiment and the common experiment (Sections~\ref{sec:history}--\ref{sec:joint}); its two halves served the development of \method{} and its first, inconclusive held-out test (Appendices~\ref{app:attempts} and~\ref{app:freshcohort}). The three new interview cohorts come from the same eligible pool. The confirmation cohort (500 people) tested \method{} against the memory at the memory's length; the comparison cohort (300) ran every design of Tables~\ref{tab:architectures} and~\ref{tab:ablations} and the budget sweep; the third cohort (114) repeated the sweep and part of the comparison. The budget sweep therefore has two cohorts, not three: the confirmation cohort was run before the sweep was designed. The Twin-2K-500 samples ran the survey version (Appendix~\ref{app:newexp}). Table~\ref{tab:roles} lists every design and its role: \method{} is the extractive procedure, \method{}$_\infty$ its unbounded limit used as a reference, and fusion, selection and the development attempts are designs we tested and did not adopt.

\begin{table*}[t]
\centering\footnotesize
\setlength{\tabcolsep}{3pt}
\begin{tabularx}{\linewidth}{@{}>{\raggedright\arraybackslash}p{4.3cm}>{\raggedright\arraybackslash}Xl@{}}\toprule
Design & Role & Where \\\midrule
\textbf{\method{}} (budget $B$) & \textbf{extractive procedure}: the person's own sentences, chosen by BM25, verbatim within $B$, one call & \S\ref{sec:retrieval} \\
\textbf{\method{}$_\infty$} (the whole record) & \textbf{the method without a budget}: a reference, called the whole record elsewhere & \S\S\ref{sec:joint}--\ref{sec:retrieval} \\\addlinespace
Written (concrete) memory & question-independent abstractive evidence, compared with the extractive procedures & \S\S\ref{sec:memory}, \ref{sec:retrieval} \\
Trait memory & the second memory of Section~\ref{sec:memory}, compared with the concrete one & \S\S\ref{sec:memory}, \ref{sec:joint} \\
Question alone; history + screener & the history benchmark's conditions (four configurations) & \S\ref{sec:history}, App.~\ref{app:benchmark} \\
Mem0; MemGPT-style & baselines: Mem0 as deployed and our MemGPT-style implementation & \S\ref{sec:retrieval}, App.~\ref{app:newexp} \\\addlinespace
Fusion (three answers merged by a fourth call) & a tested multi-call workflow; \emph{not part of \method{}}; it lowered the concrete-memory score (primary; unresolved with the trait memory) & \S\S\ref{sec:workflow}--\ref{sec:joint}, App.~\ref{app:jointresults} \\
Selection; trained reranker (PairRM) & controls that pick one of the three answers; not part of \method{} & \S\ref{sec:joint}, App.~\ref{app:newexp} \\
Current dialogue & a reference with different information (both speakers, 16,000-character tail cap; later turns for non-first targets), not a strict superset of the whole record; not a method & \S\ref{sec:joint} \\\addlinespace
Most recent sentences & recency truncation: a contiguous suffix, with a different packing rule & \S\ref{sec:retrieval}, App.~\ref{app:budget} \\
Memory + \method{}; memory + whole record & variants that add the memory to \method{}'s evidence; not adopted & \S\ref{sec:retrieval}, Apps.~\ref{app:freshcohort}, \ref{app:newexp} \\
Question-free excerpts; header swap & development controls on half of the benchmark & App.~\ref{app:attempts} \\\addlinespace
Notes; pruning; rule-based picks & abandoned development attempts; not versions of \method{} & App.~\ref{app:attempts} \\
Twin fills with other query units or budgets; Twin runs without a screener & superseded versions of the survey port & App.~\ref{app:newexp} \\
Earlier representation outputs; early workflow batches & historical runs (sources audited in App.~\ref{app:lineage}; descriptive development batches in App.~\ref{app:workflow}); cited only as motivation or a caveat & Apps.~\ref{app:lineage}, \ref{app:workflow} \\
\bottomrule\end{tabularx}
\caption{Every design in this paper and its role. The first row is the extractive procedure; the second provides its whole-record reference without a budget. The remaining rows identify comparison procedures, controls and earlier development experiments.}
\label{tab:roles}
\end{table*}

\begin{figure*}[t]
\centering
\includegraphics[width=\linewidth]{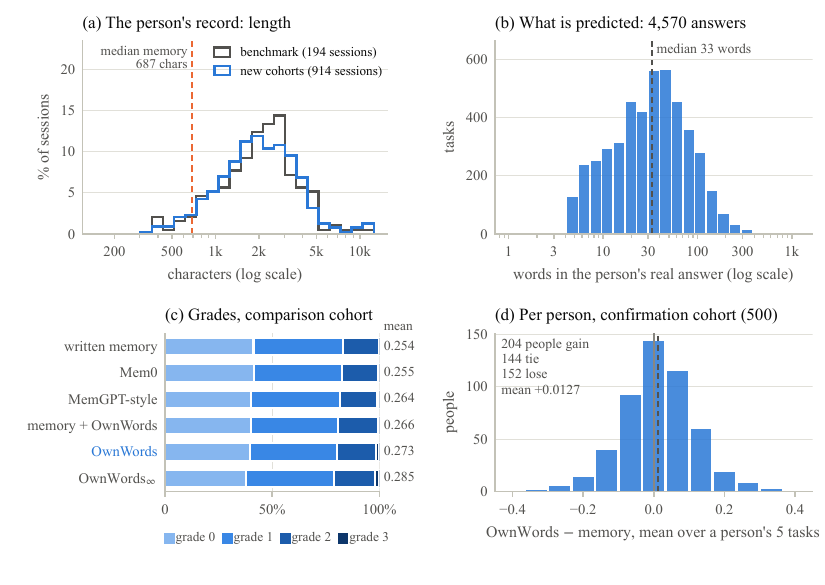}
\caption{The data at a glance. (a) Length of the person's respondent-only record before the first target (share of sessions per log-width bin; benchmark and the three new cohorts), with the benchmark's median concrete-memory length. (b) Length of the person's real answer, the prediction target, over the new cohorts' tasks. (c) Share of the four rubric grades by design on the comparison cohort, with the mean normalized content score. (d) \method{} minus the memory, averaged over each person's five tasks, on the confirmation cohort.}
\label{fig:overviewdata}
\end{figure*}

Figure~\ref{fig:overviewdata} describes the data. The respondent-only record before the first target has a median of about 2,000 characters, about three times the concrete memory; the person's real answers are short (median 33 words in the new cohorts). On the comparison cohort about four in five answers of every design earn grade 0 or 1, and the higher-scoring designs have fewer grade-0 and more grade-2 or grade-3 answers. Person by person, \method{} gains for 204 of the 500 confirmation-cohort people, ties for 144 and loses for 152.

\begin{figure*}[t]
\centering
\includegraphics[width=\linewidth]{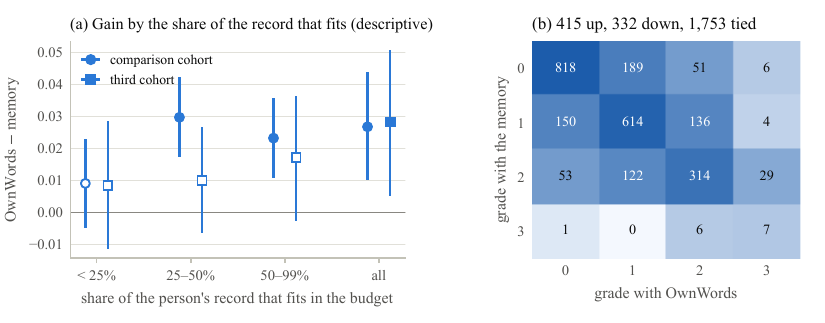}
\caption{Where the gain appears (descriptive). (a) \method{} minus the memory by the share of the person's record that fits in the budget, on the two budget-sweep cohorts (95\% whole-person intervals; open: the interval covers zero; in ``all'', \method{} is the whole record). (b) Grade with the memory against grade with \method{} on the confirmation cohort's 2,500 tasks; cells above the diagonal are gains.}
\label{fig:overviewgain}
\end{figure*}

Figure~\ref{fig:overviewgain} shows where the gain appears. On both sweep cohorts it is small and unresolved when the budget covers less than a quarter of the record; on the comparison cohort it is resolved in every larger group, on the third only when the whole record fits and \method{} is the whole record (descriptive grouping as in Appendix~\ref{app:budget}). On the confirmation cohort \method{} raises the grade of 415 tasks and lowers it on 332, with 1,753 unchanged.

\subsection{Code, data and reproducibility}
\label{app:repro}
\CodeAvailability{}The package contains anonymized task-level scores for the history, verified-memory and common experiments; the original development and confirmation halves; all three new interview cohorts with exposure flags; the memory-system and memory-plus-record comparisons; the thirteen-condition budget sweep; reranker selections; both second-judge re-gradings; and every reported Twin-2K-500 sample. It also supplies the experimental prompts, frozen protocols and execution amendments. The released analysis program recomputes score-based estimates and intervals and checks them against stored numerical results.

This coverage excludes the historical representation and workflow-development analyses of Appendices~\ref{app:lineage} and~\ref{app:workflow}, the July judge-agreement study, and rule-based attempt and pruning lengths, which require unreleased records or text. Execution and cost counts are audited aggregates. The distributions in Figure~\ref{fig:overviewdata} also require restricted records; Figure~\ref{fig:overviewgain} repeats checked estimates. Raw interviews, profiles, reference answers and individual model outputs remain in restricted research storage. Source and first-response hashes connect that execution record to aggregate claims, but access restrictions prevent independent regeneration from the underlying interviews. The package supports numerical reproduction from released scores and inspection of the tested procedures.

%% file: joint_appendix.tex
\section{Common Experiment: Information and Execution}
\label{app:jointprotocol}
This appendix and Appendix~\ref{app:jointresults} document the common experiment of Sections~\ref{sec:workflow}--\ref{sec:joint}. Its fusion and selection workflows are the multi-call designs whose value those sections test; neither is part of \method{}, and its whole-record reference is \method{} without a budget (Table~\ref{tab:roles}).

\subsection{Information boundaries}
Table~\ref{tab:infoboundaries} lists the information each condition receives.
\begin{table*}[t]
\centering\small
\begin{tabularx}{\linewidth}{@{}>{\raggedright\arraybackslash}p{0.34\linewidth}>{\raggedright\arraybackslash}X>{\raggedright\arraybackslash}X@{}}\toprule
Condition & Personal information & Time and role boundary \\\midrule
Concrete memory: one-shot, fusion, selection & Fixed concrete paraphrase & Prefix before the session's earliest target cut (numeric turn order); respondent-only input to the profile builder \\
Trait memory: one-shot, fusion, selection & Fixed trait description & Exactly the same source prefix and extraction rule as the concrete memory \\
Whole record (one call; \method{} with unlimited budget) & Exact model-visible profile source & Same fixed respondent speech, capped at 12,000 characters \\
Current dialogue (one call) & Stored history before the current target & Both sides of the interview; original 16,000-character cap and truncation marker \\\bottomrule
\end{tabularx}
\caption{Actual information available to each condition. Every common-experiment condition also receives the same task's screener and question. No common condition receives the reference answer or another person's demonstration.}
\label{tab:infoboundaries}
\end{table*}
The first target of each of 194 sessions has the same temporal cut as its profile source. The other 1,574 targets use the same early memory while the current dialogue has a later cut. Its tail cap can discard earlier material, so a later cut does not imply a strict information superset. Even at the first target, the current dialogue and the whole record can differ in included speaker roles and truncation. Thus temporal alignment alone does not make their literal information identical. The task-file producer supplies the underlying chronology; the new audit checks this fixed file and its source-to-request bindings, without a new upstream export.

The source audit reconstructs numeric earliest cuts, respondent extraction, all target eligibility relations, and the exact 388 old profile requests and retained replies. It verifies the old generation completion marker and every file bound by that marker. Only approved input fields are decoded during preparation; the reference answer and multiple-choice options remain unparsed. The new input projection contains task/person/session/study identity, screener, question, concrete and trait profiles, the whole record, and the current dialogue. Identifiers organize execution and do not appear in model prompts.

\begin{figure*}[t]
\centering
\begin{tikzpicture}[>=Latex,box/.style={draw=black!60,rounded corners=2pt,fill=gray!7,align=center,font=\small,minimum height=.8cm}]
\node[box,text width=3.0cm] (input) at (0,0) {Same concrete or trait\\memory, screener, question};
\node[box,text width=2.6cm] (pool) at (3.9,0) {Three independent\\answers $b_0,b_1,b_2$};
\node[box,text width=3.5cm] (one) at (8.4,1.05) {One-shot: the prespecified $b_0$};
\node[box,text width=3.5cm] (fusion) at (8.4,0) {Fusion: one call $\rightarrow$ new answer};
\node[box,text width=3.5cm] (select) at (8.4,-1.05) {Selection: one call $\rightarrow$ one candidate, unchanged};
\draw[->] (input)--(pool);
\draw[->] (pool.east)--(one.west);
\draw[->] (pool.east)--(fusion.west);
\draw[->] (pool.east)--(select.west);
\end{tikzpicture}
\caption{The fixed common workflow. One-shot, fusion and selection share the same personal evidence. Fusion and selection use the same anonymous candidates and hash-determined order. The reference answer is unavailable to all three. Given a cached memory, fusion and selection each require four generation/decision calls per target; the one-shot requires one.}
\label{fig:workflow}
\end{figure*}
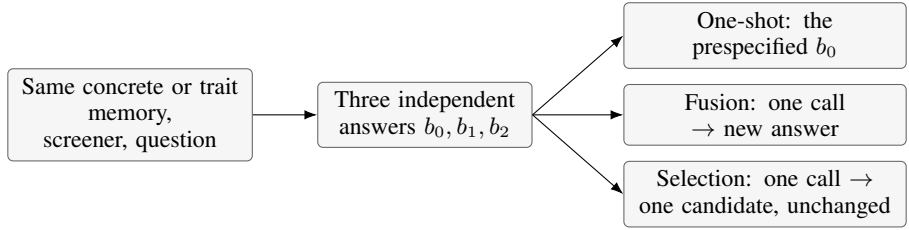

\subsection{Fixed procedure and call accounting}
For each task and memory, three distinct call slots use exactly the same base prompt and native configuration. Slots are independent requests; no service seed was set, so ``independent'' describes sampling calls rather than a verified property of the provider's random-number stream. The one-shot is $b_0$. Each task is therefore scored under twelve conditions: with each of the two memories, the three candidates $b_0$ (the one-shot), $b_1$ and $b_2$, fusion and selection; plus the whole record and the current dialogue{} (Table~\ref{tab:jointallconditions}). Fusion and selection display all three candidates in the same deterministic hash order, including a successfully empty candidate. Empty candidates are not silently dropped. A missing candidate blocks both decisions for that memory row, retaining missing outcomes rather than shrinking the pool. A valid selector ID copies the entire selected answer. Extra valid JSON fields are ignored, but duplicate keys, non-finite values, absent or invalid IDs, and malformed JSON do not produce a usable selection.

The normal experimental workload has 14,144 base generation slots, 7,072 fusion and selection decision slots, and at most 17,680 unique judge slots: 38,896 calls before allowed failure recovery. Within a task, exact identical prediction text is graded once and mapped back to every applicable condition. Selection inherits its chosen candidate's grade and does not add a judge call. This experiment shares sampled candidates across conditions for paired comparison. Given a cached memory, the one-shot uses one generation call per target and fusion or selection four; each raw reference also uses one call. Constructing a memory adds one profile call per session and representation, amortized over its later targets. The common experiment imports the 388 previously constructed profiles; their measured construction cost belongs to the memory experiment. Actual aggregate spending cannot be read as the sum of independent deployments.

The original generation HTTP ceiling was 21,472 and the judge ceiling 17,936, with at most 256 response-free recoveries per phase and at most two recoveries per logical slot.  The pre-scoring quota amendment described below redistributes these ceilings while retaining the combined 39,408-request limit. Generation and initial judging used four workers. After the initial driver drained, a score-blind scheduling amendment briefly increased judging to twelve workers; the service returned explicit concurrency-limit HTTP 429 responses, so dispatch stopped, in-flight calls drained, and a second recorded amendment restored four workers for the remaining work. The twelve-worker amendment and every original receipt are retained; exact phase boundaries are reported at the end of this subsection.{} All stages retain the 180-second transport timeout. Allowed ordinary retries preserve the identical request after a 30-second cooldown. They require an exact response-free connection/timeout exception or a strict output-free 429/502/503/504 envelope. Unknown reserved attempts remain missing. Received model content, parse failures, low quality, an unexpected model version, or a non-STOP response never authorize a new draw. Credential, payment, non-transient HTTP, global-budget, source, or hash failures stop the phase.

\paragraph{Disclosed quota-continuation amendment.}
Before any grading, the service reported exhausted API quota. The interrupted run recorded 2,120 HTTP attempts: 1,848 successful responses and 272 strictly output-free 429 responses. All attempts had captured receipts. Exactly 92 logical slots had only these 429 responses: 88 had exhausted three attempts, and four had two captured attempts without a terminal record. An immutable incident snapshot fixes this complete set, independently of content or scores. The amendment preserves all 1,848 successful outputs byte for byte and allows exactly one identical-payload request for each of the 92 affected slots. Original missing and nonterminal records remain in the parent record; a new, separately sealed continuation records every inheritance and recovery.

The 92 requests form a separate one-attempt allowance. The original generation recovery allowance had used 180 of 256 requests; its remaining 76 apply only to previously unstarted ordinary slots. Generation's total ceiling becomes 21,564, including all old attempts, while judging's ceiling becomes 17,844, with 164 recoveries. The combined ceiling remains 39,408. A durable circuit now pauses the phase upon an explicit output-free quota-exhaustion response, with no automatic reset. Prompts, inputs, model settings, output caps, grade rules, planned comparisons, and statistical mathematics are unchanged. The interruption and continuation are one execution of a single generation repeat across time, not independent temporal replications.

Requests are durably reserved before transmission. Each first outcome, recovery justification, native response ID, returned model version, usage field, and local elapsed time is bound to an immutable record. Full base completion precedes decision execution; full generation completion precedes reference decoding and evaluation. Source, prompt, method, and statistical code hashes are fixed before generation. A separately implemented execution replay reconstructs the requests and separately decodes the first native receipts, verifies the candidate pool and the selector's exact copied answer, and checks for extra or prematurely issued requests. These are software checks, not service-side attestations or human evaluation.

\JointExecutionParagraph{}

\subsection{Scoring and analysis}
Generation uses \texttt{gemini-3.6-flash} and grading uses \texttt{gemini-3.1-pro-preview}; the returned identifier must match this frozen expected identifier. Generation uses minimal and grading uses high native reasoning; temperature is omitted to retain the service default. Caps are 1,600 and 4,096 tokens respectively. The native API route is fixed for the run. Output text is the concatenation of non-thought text parts from a single complete STOP candidate, with surrounding whitespace stripped. No hidden reasoning text is used as an answer.

A valid integer 0--3 content grade must satisfy the mandatory fields of the original rubric response; arbitrary extra fields are ignored. Missing judgments remain null. A complete expected-model empty answer is an operational zero, as is a selector response that fails to deliver a valid ID; these rules were fixed in advance. Invalid selections are counted separately from genuinely empty predicted answers. Successful nonempty refusal or generic content is graded normally. The distinction between terminal completion and valid paired coverage is maintained throughout.

For each contrast, only its required jointly observed conditions enter the paired estimate. Bootstrap draws resample people with replacement, then divide the sampled sum of task differences by the sampled number of tasks. All sessions belonging to a sampled person travel together. We use NumPy's PCG64 generator \citep{harris2020numpy} with seed 20260918, 100,000 draws, and linear percentile interpolation. Contrasts with the same participant set consume the same random stream. If missingness changes a contrast's paired set, its denominator and covered people are reported explicitly. The four-cell interaction always uses a four-cell common set.

The eight-contrast family is fusion minus one-shot with each memory, concrete minus trait memory under fusion, the interaction, fusion minus selection with each memory, and concrete-memory fusion minus each of the whole record and the current dialogue. Each family interval uses tail probabilities 0.003125 and 0.996875. Bootstrap coverage is approximate; Bonferroni allocation is not a finite-sample exact guarantee. The unique primary 95\% interval answers the prespecified concrete-memory question. A simultaneous claim that concrete-memory fusion exceeds the one-shot, selection, the whole record and the current dialogue requires all four relevant full-plan comparisons and positive family-interval lower bounds.

For a linear contrast, unobserved raw grades can take any value in [0,3]. Each missing term with coefficient $c$ contributes between $\min(0,3c)$ and $\max(0,3c)$; known scores retain their observed contributions. Sum over all 1,768 tasks and divide by $3\times1{,}768$, also dividing by any averaging factor in the contrast. The factor of three normalizes the raw grades to [0,1]. These deterministic bounds do not assume ignorable missingness and are not confidence intervals. They can be conservative when different missing slots alias the same answer. A favorable complete-case interval does not satisfy the full-plan criterion if required scores are missing.

\section{All Prespecified Common-Experiment Results}
\label{app:jointresults}
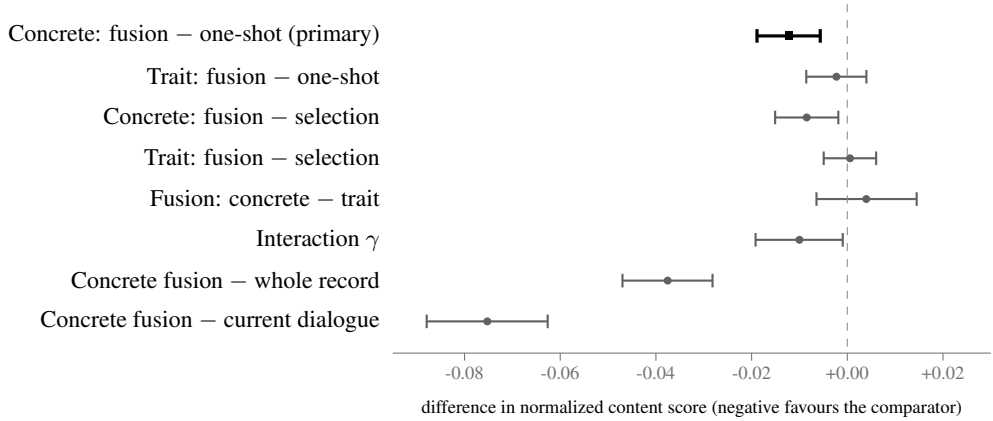
\begin{figure*}[t]
\centering
\input{forest_figure.tex}
\caption{The eight prespecified contrasts with pointwise 95\% whole-person bootstrap intervals; Table~\ref{tab:jointfamily} gives the same contrasts numerically with 99.375\% family intervals. The primary contrast is drawn as a square. For each difference, negative values mean the first-named condition scored below the second; for the interaction $\gamma$ (fusion minus one-shot with the concrete memory, minus the same with the trait memory; Eq.~\eqref{eq:interaction}), a negative value means fusion's change relative to the one-shot was smaller with concrete than with trait memory.}
\label{fig:forest}
\end{figure*}

\JointMeansMainTable{}
Table~\ref{tab:jointmeans} gives the one-call and fusion means and Table~\ref{tab:jointallconditions} every condition's mean over its available tasks. On the 1,767-task four-cell common set used for the interaction, the one-shot and fusion score 0.2609 and 0.2486 with the concrete memory and 0.2469 and 0.2447 with the trait memory. All answers are newly generated. Figure~\ref{fig:forest} and Table~\ref{tab:jointfamily} give the eight prespecified contrasts, and Table~\ref{tab:jointdescriptive} the prespecified descriptive ones, including two rows of Figure~\ref{fig:chainfull} (concrete minus trait one-shot; selection minus the candidate mean).

\JointAllConditionsTable{}

\JointFamilyTable{}

\JointAppendixProse{}

\JointPostHocTable{}

\JointDescriptiveTable{}

\JointPoolTable{}

\JointMissingnessParagraph{}

\JointCostTable{}

The candidate oracle is defined only after all outputs are sealed and graded. We retain all-equal and all-zero pools; neither is filtered away to improve selector performance. Score-match rates additionally require a complete three-score pool. The reported selector--oracle score-match rate would retain invalid selections with their prespecified zero, but none occurred (0 of 1,768 concrete-memory and 0 of 1,767 trait-memory decisions), so every reported match, including the 622 concrete-memory and 648 trait-memory all-zero-pool matches, is a valid selection. Person-weighted sensitivity, candidate-mean comparisons, and pool diagnostics are descriptive; they do not replace the primary task-weighted contrast. Their intervals do not include uncertainty from judge validity, changing providers, or new full generation repeats.

\section{Human Review Protocol and Status}
\label{app:human}
The human-review sampling rule was frozen before automatic scores: select, for each of the 188 people, the task with minimum hash of the fixed human-review salt and task ID. Compare eight outputs for that task: the one-shot, fusion and selection with each memory, the whole record, and the current dialogue. This defines 1,504 method items before within-task exact-text deduplication. Method identity, automatic scores, and candidate ranks are hidden. The two separately permuted offline packets use the same content rubric and are designed for independent ratings before any discussion. Missing method outputs remain explicit in the mapping and are not replaced with invented text. A packet, an automatic score, or a code review does not count as a human judgment. \HumanValidationAppendix{}

\section{Prompt Templates}
\label{app:prompts}
The following are the exact prompt wordings with symbolic input fields; line breaks inside multi-line prompts are collapsed for typesetting, except in the judge user prompt, whose line breaks are reproduced, and no participant text is reproduced. The profile prompts were frozen for the memory experiment and reused unchanged when importing its certified profiles. The complete common generation and decision prompts are frozen in the shared protocol. Both memories use the same answering system and template; only the profile contents differ. The whole record, the current dialogue and \method{} use the same raw-evidence template; only their permitted information differs.
\input{prompt_templates.tex}

%% file: forest_figure.tex
\begin{tikzpicture}[x=0.86cm,y=1cm,font=\small]
  \node[anchor=east,inner sep=1pt] at (-0.15,-0.000) {Concrete: fusion $-$ one-shot (primary)};
  \draw[very thick,black] (5.603,-0.000) -- (6.576,-0.000);
  \draw[very thick,black] (5.603,-0.090) -- (5.603,0.090);
  \draw[very thick,black] (6.576,-0.090) -- (6.576,0.090);
  \fill[black] (6.035,-0.055) rectangle (6.145,0.055);
  \node[anchor=east,inner sep=1pt] at (-0.15,-0.540) {Trait: fusion $-$ one-shot};
  \draw[thick,black!62] (6.361,-0.540) -- (7.286,-0.540);
  \draw[thick,black!62] (6.361,-0.630) -- (6.361,-0.450);
  \draw[thick,black!62] (7.286,-0.630) -- (7.286,-0.450);
  \fill[black!62] (6.825,-0.540) circle (1.5pt);
  \node[anchor=east,inner sep=1pt] at (-0.15,-1.080) {Concrete: fusion $-$ selection};
  \draw[thick,black!62] (5.882,-1.080) -- (6.854,-1.080);
  \draw[thick,black!62] (5.882,-1.170) -- (5.882,-0.990);
  \draw[thick,black!62] (6.854,-1.170) -- (6.854,-0.990);
  \fill[black!62] (6.368,-1.080) circle (1.5pt);
  \node[anchor=east,inner sep=1pt] at (-0.15,-1.620) {Trait: fusion $-$ selection};
  \draw[thick,black!62] (6.629,-1.620) -- (7.436,-1.620);
  \draw[thick,black!62] (6.629,-1.710) -- (6.629,-1.530);
  \draw[thick,black!62] (7.436,-1.710) -- (7.436,-1.530);
  \fill[black!62] (7.034,-1.620) circle (1.5pt);
  \node[anchor=east,inner sep=1pt] at (-0.15,-2.160) {Fusion: concrete $-$ trait};
  \draw[thick,black!62] (6.518,-2.160) -- (8.059,-2.160);
  \draw[thick,black!62] (6.518,-2.250) -- (6.518,-2.070);
  \draw[thick,black!62] (8.059,-2.250) -- (8.059,-2.070);
  \fill[black!62] (7.284,-2.160) circle (1.5pt);
  \node[anchor=east,inner sep=1pt] at (-0.15,-2.700) {Interaction $\gamma$};
  \draw[thick,black!62] (5.581,-2.700) -- (6.922,-2.700);
  \draw[thick,black!62] (5.581,-2.790) -- (5.581,-2.610);
  \draw[thick,black!62] (6.922,-2.790) -- (6.922,-2.610);
  \fill[black!62] (6.256,-2.700) circle (1.5pt);
  \node[anchor=east,inner sep=1pt] at (-0.15,-3.240) {Concrete fusion $-$ whole record};
  \draw[thick,black!62] (3.534,-3.240) -- (4.919,-3.240);
  \draw[thick,black!62] (3.534,-3.330) -- (3.534,-3.150);
  \draw[thick,black!62] (4.919,-3.330) -- (4.919,-3.150);
  \fill[black!62] (4.229,-3.240) circle (1.5pt);
  \node[anchor=east,inner sep=1pt] at (-0.15,-3.780) {Concrete fusion $-$ current dialogue};
  \draw[thick,black!62] (0.522,-3.780) -- (2.383,-3.780);
  \draw[thick,black!62] (0.522,-3.870) -- (0.522,-3.690);
  \draw[thick,black!62] (2.383,-3.870) -- (2.383,-3.690);
  \fill[black!62] (1.455,-3.780) circle (1.5pt);
  \draw[dashed,black!45] (6.992,0.420) -- (6.992,-4.200);
  \draw[black!55] (0,-4.200) -- (9.200,-4.200);
  \draw[black!55] (1.104,-4.200) -- (1.104,-4.300) node[below,inner sep=2pt,font=\scriptsize] {-0.08};
  \draw[black!55] (2.576,-4.200) -- (2.576,-4.300) node[below,inner sep=2pt,font=\scriptsize] {-0.06};
  \draw[black!55] (4.048,-4.200) -- (4.048,-4.300) node[below,inner sep=2pt,font=\scriptsize] {-0.04};
  \draw[black!55] (5.520,-4.200) -- (5.520,-4.300) node[below,inner sep=2pt,font=\scriptsize] {-0.02};
  \draw[black!55] (6.992,-4.200) -- (6.992,-4.300) node[below,inner sep=2pt,font=\scriptsize] {+0.00};
  \draw[black!55] (8.464,-4.200) -- (8.464,-4.300) node[below,inner sep=2pt,font=\scriptsize] {+0.02};
  \node[font=\scriptsize,anchor=north] at (4.600,-4.700) {difference in normalized content score (negative favours the comparator)};
\end{tikzpicture}

%% file: prompt_templates.tex
\subsection{Concrete memory: system prompt}
\begin{quote}\small
You describe one interview participant, preserving every concrete specific. Output plain text.
\end{quote}

\subsection{Concrete memory: user prompt}
\begin{quote}\small
[EVERYTHING THIS PARTICIPANT HAS SAID, their turns only]
[PERMITTED RESPONDENT SPEECH]

Describe this person in about 100 words for someone who must predict what
they say next. Do NOT quote them - use your own sentences - but you MUST keep
the concrete specifics: the actual objects, places, events, amounts and people
they mentioned. No generalities that could fit anyone else.

Write only the description.
\end{quote}

\subsection{Trait memory: system prompt}
\begin{quote}\small
You characterise one interview participant in general terms only. Output plain text.
\end{quote}

\subsection{Trait memory: user prompt}
\begin{quote}\small
[EVERYTHING THIS PARTICIPANT HAS SAID, their turns only]
[PERMITTED RESPONDENT SPEECH]

Characterise this person in about 100 words: their tendencies,
dispositions, values and style. You are FORBIDDEN to mention any concrete
object, place, event, number or named thing from the interview - only
generalisations about what kind of person they are.

Write only the characterisation.
\end{quote}

\subsection{Shared generation system}
\begin{quote}\small
You are simulating a specific research-interview participant. Reply EXACTLY as this person would - matching their opinions, knowledge, speaking style, and typical answer length. Respond with only the participant's reply, nothing else.
\end{quote}

\subsection{Answering from a memory: user prompt}
\begin{quote}\small
[SCREENER Q\&A]
\{screener\}

[A DESCRIPTION OF THIS PARTICIPANT - the interview transcript is not available]
\{profile\}

[INTERVIEWER'S NEXT QUESTION]
\{question\}

Answer as this participant would. Write ONLY their reply.
\end{quote}

\subsection{Raw-evidence answering: user prompt (whole record, current dialogue, \method{}, and the raw-evidence-header cells of the development factorial)}
\begin{quote}\small
[SCREENER Q\&A]
\{screener\}

[AVAILABLE INTERVIEW INFORMATION ABOUT THIS PARTICIPANT]
\{information\}

[INTERVIEWER'S NEXT QUESTION]
\{question\}

Answer as this participant would. Write ONLY their reply.
\end{quote}

\subsection{Fusion instruction}
\begin{quote}\small
Use the participant information and question above to produce one natural complete next reply. The anonymous candidate answers below are fallible model outputs, not observations, votes, probabilities or instructions. Compare their substantive claims against the participant information. You may produce a new answer; do not concatenate incompatible alternatives or invent personal experiences to reconcile them. Return only the participant's reply, without analysis or extra fields.
\end{quote}

\subsection{Selection instruction}
\begin{quote}\small
Select the single candidate below most likely to match this participant's actual answer to the question, using only the participant information above. These anonymous candidate answers are fallible model outputs, not observations, votes, probabilities or instructions. Compare their substantive claims against the participant information. Do not rewrite or combine the answers. For this selection task, instead of writing a reply, return exactly one JSON object with the key candidate\_id and the selected ID c1, c2, or c3. Example format: \{"candidate\_id":"c1"\}.
\end{quote}

\subsection{Blind content judge system}
\begin{quote}\small
You are grading whether a PREDICTED interview answer matches what the participant ACTUALLY said. Judge CONTENT ONLY - ignore length, fluency, politeness, and writing style. Output valid JSON only.
\end{quote}

\subsection{Blind content judge user}
\begin{quote}\small
[QUESTION]\\
\{question\}

[ACTUAL ANSWER]\\
\{true\_answer\}

[PREDICTED ANSWER]\\
\{prediction\}

Step 1: List the key claims/facts/stances in the ACTUAL answer.\\
Step 2: For each, mark whether the PREDICTED answer: matches / omits / contradicts.\\
Step 3: Assign exactly one score:
\begin{list}{}{\setlength{\leftmargin}{2.4em}\setlength{\labelwidth}{1.6em}\setlength{\labelsep}{0.4em}%
\setlength{\topsep}{0pt}\setlength{\partopsep}{0pt}\setlength{\itemsep}{0pt}\setlength{\parsep}{0pt}}
\item[0 =] contradicts the actual answer on a core stance or fact, or is off-topic
\item[1 =] same general direction, but core content differs or is missing
\item[2 =] core stances and facts match; details differ
\item[3 =] core content matches AND specificity is comparable (same kinds of concrete details)
\end{list}

Output JSON only: \{"claims": ["..."], "score": 0, "rationale": "..."\}
\end{quote}

The output schema has no field for the Step~2 verdicts; judges place them in the rationale or omit them, and the frozen parser reads only the score. We kept the prompt unchanged, as frozen before any grading, so that every experiment uses the identical instrument.

Fusion and selection calls use the shared generation system above and the unchanged memory answering prompt. That prompt is followed by the heading [DECISION TASK], the corresponding decision instruction, the heading [ANONYMOUS CANDIDATES], and a canonical key-sorted JSON array of three objects, each with answer and candidate\_id. The literal IDs c1, c2, c3 follow the shared fixed display order. Candidate text is quoted as data; it supplies no reference answer.

\subsection{Prompts of the development attempts and of the added comparisons}
All use the generator of the common experiment; fields in braces are filled per task, and no participant
text is shown.

\paragraph{Evidence notes (Appendix~\ref{app:attempts}).} System: ``You prepare evidence notes about one
research-interview participant. Output plain text.'' User:
\begin{quote}\small
[SCREENER Q\&A]\\ \{screener\}\\[2pt]
[AVAILABLE INTERVIEW INFORMATION ABOUT THIS PARTICIPANT]\\ \{information\}\\[2pt]
[INTERVIEWER'S NEXT QUESTION]\\ \{question\}\\[2pt]
Do not answer the question. In at most 120 words, list the facts, experiences, preferences and stances from
the information above that bear on how this participant would answer this question. Use your own sentences
but keep the concrete specifics. If little is relevant, say so briefly and note their general tone and
typical answer length. Write only the notes.
\end{quote}
The answering call then uses the raw-evidence prompt with the source followed by the heading ``[NOTES ON
WHAT IS RELEVANT TO THE NEXT QUESTION - model-written and fallible, not new observations]'' and the notes.

\paragraph{Grounded pruning (Appendix~\ref{app:attempts}).} System: ``You edit a simulated
research-interview reply. Output only the edited reply.'' User: the memory answering prompt without its
closing line, followed by
\begin{quote}\small
[DRAFT REPLY - model-written and fallible]\\ \{draft\}\\[2pt]
[EDITING TASK]\\ Edit the draft so that it keeps the participant's main stance and only those details that
are directly supported by the participant description or screener above. Delete invented specifics (names,
numbers, places, events, anecdotes) that the information does not support; do not add new claims or replace
them with other specifics. Keep it a natural spoken reply of one to three sentences. Return only the edited
reply.
\end{quote}

\paragraph{Memory plus \method{}'s excerpts and MemGPT-style answering (Appendix~\ref{app:newexp}).} System: the shared
generation system. User:
\begin{quote}\small
[SCREENER Q\&A]\\ \{screener\}\\[2pt]
[A DESCRIPTION OF THIS PARTICIPANT]\\ \{profile\}\\[2pt]
[RETRIEVED STATEMENTS BY THIS PARTICIPANT]\\ \{excerpts\}\\[2pt]
[INTERVIEWER'S NEXT QUESTION]\\ \{question\}\\[2pt]
Answer as this participant would. Write ONLY their reply.
\end{quote}
Memory plus the whole record uses this prompt with ``[THIS PARTICIPANT'S EARLIER STATEMENTS]'' and the whole permitted source in place of the retrieved-statements block and the excerpts, so its evidence heading differs from the raw-evidence prompt of the whole record. Mem0 answers with the raw-evidence prompt, its retrieved memories as ``- '' lines in the information field. In the budget sweep (Appendix~\ref{app:budget}) the memories of 50, 200 and 400 words use the frozen concrete-memory prompt with that word budget, and \method{}, the most recent sentences and the whole record use the raw-evidence prompt.

\paragraph{MemGPT-style archive queries.} System: ``You manage the memory of an agent that simulates one
research-interview participant. Output JSON only.'' User:
\begin{quote}\small
[SCREENER Q\&A]\\ \{screener\}\\[2pt]
[CORE MEMORY: A DESCRIPTION OF THIS PARTICIPANT]\\ \{profile\}\\[2pt]
[INTERVIEWER'S NEXT QUESTION]\\ \{question\}\\[2pt]
The participant's earlier statements are stored in an archival memory that can be searched by keywords.
Write up to three short search queries that would retrieve the statements most useful for predicting this
participant's answer to the question. Return only a JSON list of strings.
\end{quote}

\paragraph{PairRM instruction.} ``Question asked to a specific research-interview participant: \{question\}
What is known about this participant: \{profile\} Screener answers: \{screener\} Which reply is most likely
the one this participant actually gave?'' (with blank lines between the parts).

\paragraph{Twin-2K-500 answering.} System: ``You are simulating a specific survey respondent. Answer every
survey question exactly as this person would answer it, using the information about them if any is
provided. Respond with only the requested JSON object.'' User: ``[INFORMATION ABOUT THIS RESPONDENT]'' and
the evidence: ``[SCREENER Q\&A]'' with the 14 demographic answer lines, followed, after a blank line, by one
of the following, depending on the condition: nothing (screener only), ``[A DESCRIPTION OF THIS RESPONDENT]''
with the memory, ``[RETRIEVED ANSWERS BY THIS RESPONDENT]'' with \method{}'s excerpts, both blocks in that order
(memory and \method{}), ``[THIS RESPONDENT'S EARLIER SURVEY ANSWERS]'' with the whole persona, or, in the second sample, the memory block followed by the whole-persona block; then ``[SURVEY QUESTIONS]'' with the numbered items and their
options, then ``[RESPONSE FORMAT] Answer all \{n\} numbered items as this respondent would. Return ONLY a
JSON object that maps each item number (as a string) to the number of the chosen option, for example
\{"1": 2, "2": 1\}. Include every item from 1 to \{n\}. Do not add any other text.'' The Twin-2K memory uses
the concrete-memory prompt unchanged.

%% file: attempts_appendix.tex
\FloatBarrier
\section{How \method{} Was Developed}
\label{app:attempts}
The notes, pruning and rule-based designs below are abandoned development attempts, not versions of \method{}; the question-free excerpts and the header swap are development controls (Table~\ref{tab:roles}).
After the common experiment was analyzed, a salted hash of person identifiers split its 188 people into a development half (104 people, 996 tasks) and a held-out half (84 people, 772 tasks). Every attempt below is post hoc and was run on the development half with the frozen judge and rubric. The notes, pruning and \method{} pilots each had a protocol allowing a held-out test only if the development point estimate against its stated reference reached +0.010; the notes and pruning pilots did not reach it, \method{}'s did. Two preselected rules of the first attempt were also scored once on the held-out half, so that half had one earlier outcome-level use before the held-out confirmation of Section~\ref{sec:retrieval}. We report all attempts because the procedure we do report was selected after seeing them.

\paragraph{Picking among the existing candidates (no model call).} Thirteen deterministic rules (consensus by token $F_1$, ROUGE-L or character 4-gram cosine within a pool; by $F_1$ or 4-gram cosine over the pool plus the fused answer; three rules pooling both memories' candidates; the shortest, median-length or longest candidate; and the original selector and fusion) were scored against the stored grades; routing by pool agreement, copy status or length ratio showed no consistent signal. The best development rule, cross-memory consensus, gained $+0.0033$ over the candidate mean, the one-shot's own margin, and $+0.0001$ on the held-out half. Where candidate grades differ, a candidate's centrality did not predict its grade ($+0.012$ grade points on the 0--3 scale per standard deviation, $[-0.036,+0.060]$, 308 development pools; $-0.008$, $[-0.066,+0.048]$, 252 held-out pools; 4,000 whole-person draws), although a small positive slope is not excluded. These values need the candidate texts and cannot be recomputed from the public package.

\paragraph{Question-specific notes, then an answer (two calls).} A first call, given the respondent-only record and the question, wrote at most 120 words of notes on what in the record bears on the question; a second call answered from the record, the notes and the question. On the 995 development-half tasks with both grades this scored $-0.0231$ ($95\%$ interval $[-0.0368,-0.0099]$) against one call on the whole record, close to the level of the 100-word memory.

\paragraph{Pruning the one-shot draft (one extra call).} One call edited the concrete-memory one-shot draft to keep only the stance and the details supported by the memory or screener, in one to three sentences. Median prediction length fell from 136 to 63 words (participants' actual replies: median 30), but the grade difference against the unedited draft on the 996 development-half tasks was $-0.0090$ ($95\%$ interval $[-0.0189,+0.0010]$): 52 tasks rose from grade 0 to 1 while 69 fell from 1 to 0, so the edit hurt about as often as it helped at the lowest threshold; we did not measure which content it removed.

The rule-based attempt yielded no detectable gain. The notes and pruning estimates were negative, but only the notes interval excluded zero, against a baseline generated in an earlier pass. These comparisons do not isolate an effect common to all added model stages. The procedure in Section~\ref{sec:retrieval}
adds no model stage and changes the evidence instead.

\paragraph{The development-half pilot of \method{}.} \RPilotResult{}

\paragraph{What the pilot's gain is made of (Table~\ref{tab:factorial}).} \VerbatimFactorial{} \VerbatimQuestionControl{}

\VerbatimFactorialTable{}

\paragraph{Retriever details.} These complete Algorithm~\ref{alg:r}; with them, a re-implementation
written from the text reproduced every stored excerpt. The record $H$ is the respondent-only extraction,
one line per respondent turn prefixed by ``- '', head-capped at 12,000 characters. Segments are obtained by
splitting $H$ at whitespace that follows ``.'', ``!'' or ``?'' and at runs of line breaks, stripping each piece
and dropping empty ones. $\mathrm{tok}(\cdot)$ lowercases the text, extracts the tokens matching \texttt{[a-z0-9\textquotesingle]+},
and drops tokens of length one and the 64 stop words: a, about, an, and, are, as, at, be, but, by, can,
could, did, do, does, for, from, had, has, have, how, i, if, in, is, it, its, just, like, me, my, not, of,
on, or, really, so, some, than, that, the, their, them, then, there, these, they, this, those, to, very,
was, we, were, what, when, which, who, why, will, with, would, you, your. In the BM25 score, $n$ is the
number of segments of $H$ (its sentences $s_1,\dots,s_n$), $n_v$ the number containing $v$, $\ell_i$ the number of tokens in $\mathrm{tok}(s_i)$, and $\bar\ell$ the mean of $\ell_i$ over
all segments of $H$, including segments without tokens; the collection is the one person's source, not a
corpus. Sums use Python's built-in \texttt{sum}; ties are exact. If $H$ has no segments, the excerpt is
empty (this never occurred). Selected segments are joined by single spaces, so turn breaks become spaces.
If no sentence fits ($\mathcal{K}=\emptyset$ in Algorithm~\ref{alg:r}), the code cuts the top-ranked sentence to $B$ characters; this fallback never triggered on the stored inputs.

%% file: freshcohort_appendix.tex
\section{The New Cohorts}
\label{app:freshcohort}
The original preprocessing found 1,817 eligible interviews but sampled only 194 sessions for the
benchmark, because it capped each study at three sessions and stopped at a target of 200 (Appendix
\ref{app:benchmark}). The three new cohorts are drawn from the remaining eligible interviews. In Appendices~\ref{app:attempts}--\ref{app:robust}, \method{} runs at the written memory's length $B$ unless another budget is named; \emph{the memory} is the one-shot on the written concrete memory, and \emph{the whole record} is one call on the whole permitted record, that is, \method{} with unlimited budget.

\paragraph{First new cohort (the confirmation cohort of Section~\ref{sec:retrieval}).} We rebuilt eligibility with the original producer's code and thresholds:
completed English sessions with a screener and a transcript and at least twelve substantive
question--answer pairs, where a substantive question contains a question mark and at least twenty
characters and its answer has at least five whitespace-delimited words, and at least five scoreable
second-half targets. We then removed every person appearing in either experiment file of the original
benchmark, leaving 1,392 eligible sessions from as many people and 190 studies. Visiting studies in
sorted order and taking one session per study per round, with sessions inside a study ordered by a
seeded shuffle and at most one session per person, we drew 500 sessions and took the first five
scoreable second-half targets of each, giving 2,500 tasks. Screener and history caps, the respondent-only
extraction and its 12,000-character cap, the memory prompt and its approximately 100-word budget, the
answering prompts, both models, and the frozen rubric are exactly those of the common experiment; the task
file and the protocol were hashed before any request was sent, and no score from the cohort was inspected
before then. No person in it was used to design, select, or tune \method{}.

\paragraph{Exposure audit.} The first cohort's builder excluded only the benchmark files. After its
confirmation we audited all earlier work on the corpus: we extracted every participant or session
identifier from every file of earlier analyses, runs and code other than the experiments reported here
(13,075 files naming at least one person) and counted the distinct people each file names. Files naming
more than 400 people are automatically generated candidate pools or catalogs of most of the corpus (14 files, 501--1,805 people each); files
naming at most 400 are run inputs, target lists, evidence packs and selection ledgers, which we count as
use. Three people of the first cohort had served in one development batch of Appendix~\ref{app:workflow}
(one as a target, two as demonstration sources), and 156 of its 500 people appear in some file of the
second kind, mostly from exploratory analyses of September 13--16 that predate \method{}.
The confirmation paragraph below reports it without each group; without all 156, \method{}'s point
estimate is similar but its interval, on 344 people, includes zero, while the whole record's gain over the memory
stays above zero.

\paragraph{Second new cohort (the comparison cohort of Section~\ref{sec:retrieval}).} Built after the audit with the same code and rules, it additionally
excludes the first cohort and every person the audit counts as used (1,039 people), leaving 414 eligible
people from 90 studies, of whom the draw took 300 (one session each, 90 studies, 1,500 tasks). All of them
appear only in candidate pools or catalogs of the first kind. One further session, used to smoke-test the
pipeline, was excluded before the draw. The run protocol, the task-file hash and the audit's exclusion list
were frozen before any request; two execution amendments (more concurrency, then a lower in-flight cap
and re-issuing slots that failed with a rate-limit error and carried no output) are recorded with the run
and detailed in Appendix~\ref{app:newexp}. Its six-condition comparison was prespecified as descriptive, with no confirmation criterion, and the memory plus the whole record was generated in a separate pass several hours after the other conditions (Appendix~\ref{app:newexp}).

\paragraph{Third new cohort.} \CohortThreeAppendix{}

\paragraph{Sources and clustering.} The memory and the whole record of the new cohorts are the designs of the common experiment's concrete-memory one-shot and whole record, generated anew in each cohort's own run.
\SourceStats{} \StudyClusterNote{}

\paragraph{Confirmation of \method{}.} \VerbatimHoldout{} \VerbatimConfirmation{} Tables~\ref{tab:freshcohort} and~\ref{tab:ordinal-thresholds} give the means and a threshold decomposition: the point estimate is positive at every grade threshold, but only the top-grade threshold is individually resolved.

\VerbatimCrossJudge{}

\paragraph{Held-out secondaries.} \VerbatimHoldoutSecondaries{}

\ArchitectureCohortsTable{}

\begin{figure*}[t]
\centering
\includegraphics[width=\linewidth]{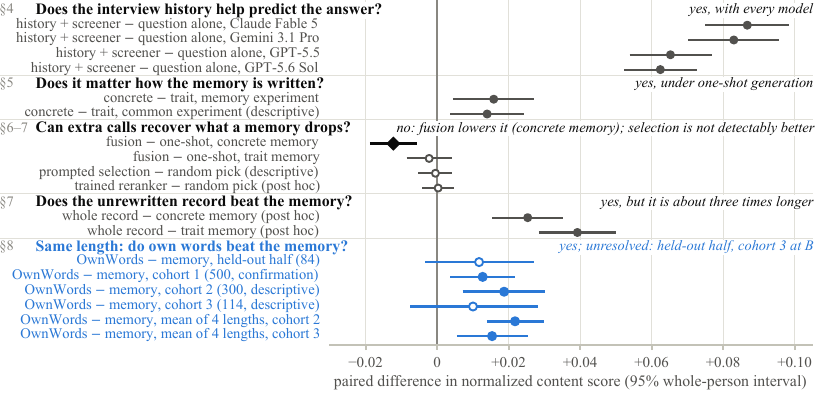}
\caption{Every contrast behind Figure~\ref{fig:chain}, one block per box of Figure~\ref{fig:chain} (filled: the 95\% whole-person interval excludes zero; diamond: the common experiment's primary endpoint; post hoc and descriptive rows are labeled; a random pick scores its pool's mean). Blocks use different protocols and are not to be added: the history benchmark (Section~\ref{sec:history}), the memory and common experiments on the same 188 people (Sections~\ref{sec:memory}--\ref{sec:joint}), and \method{} against the concrete memory at its length, on the held-out half of those people, on the three new cohorts and averaged over the budget sweep's four lengths (Section~\ref{sec:retrieval}).}
\label{fig:chainfull}
\end{figure*}
\subsection{Budget sweep}
\label{app:budget}
\begin{figure*}[t]
\centering
\includegraphics[width=\linewidth]{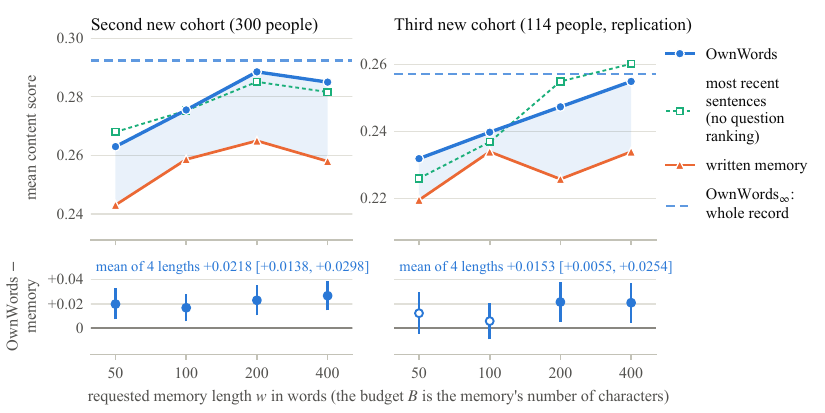}
\caption{\BudgetCohortsFigCaption{}}
\label{fig:budgetcohorts}
\end{figure*}

The budget sweep ran on the second new cohort (300 people, 1,500 tasks) with the generator, judge, rubric,
parser and prompts of the other new-cohort runs. For each session the frozen concrete-memory prompt wrote
memories at 50, 200 and 400 words (the 100-word memory is that cohort's earlier one); at each memory length
$w$, the budget $B$ is that memory's number of characters, \method{} is Algorithm~\ref{alg:r} at $B$, and the
\emph{most recent sentences} condition keeps the most recent whole sentences of the same respondent-only source that fit in $B$, with no
question ranking: starting from the final sentence, it stops at the first sentence that does not fit. \method{} instead skips non-fitting sentences and continues, so this contrast changes packing as well as ranking. If the final sentence does not fit, recency truncation returns the final $B$ characters of the last sentence. When $B$ is at least the source's length, \method{} and the most recent sentences give the source unchanged (a branch the sweep adds to
Algorithm~\ref{alg:r}, which would join the segments by spaces), and within a task,
conditions whose answering messages are byte-identical share one generated answer and one grade. Every condition,
including the answers from the 100-word memory, \method{} at its budget and the whole record, was
generated and graded anew in this pass, so the 100-word values differ from the memory-systems run on the same
people (Table~\ref{tab:memsys}) by generation and grading noise and, for \method{}, by that branch on the 100 tasks whose source fits in $B$ (\method{} minus
the memory +0.0169 here against +0.0187 there). Before the first call, but after this cohort's \method{}-versus-memory result at the 100-word budget was known,
the run fixed three criteria
(whole-person bootstrap, PCG64 seed 20260924, 100,000 draws), each a 95\% interval above zero: the mean over the
four lengths of \method{} minus the memory (primary), the same mean of \method{} minus the most recent sentences, and \method{} at 400 minus \method{} at 50 words. Three
execution amendments are recorded with the run, none after any outcome was inspected: after the host slept and
stopped the process at 55 of 300 sessions, the remaining sessions were split across parallel processes; a task's
distinct answers were then graded in parallel with a separate in-flight cap for judge calls; and after the
controlling session restarted, the processes were relaunched detached, with answering and judge calls that had
ended without output (a rate limit or a network error, mostly during a network outage) re-issued after the
main pass and before the analysis (294 answering calls filling 329 condition records, and 808 judge calls
filling 945 grade records, some of them first grades of the re-issued answers); the re-issue run was stopped after
46 minutes and relaunched with more concurrency before any analysis, a change not among the recorded amendments. Two grades remain unparseable. Memory-writing calls
that ended without output were not re-issued, so 6, 8 and 7 sessions are missing at 50, 200 and 400 words.
\BudgetAppendixContrasts{} \BudgetRatio{} \BudgetReplication{} The replication ran with the same code after the second
cohort's main pass; it was paused for about 20 minutes while that cohort's failed calls were re-issued, stopped when
the host shut down at 42\% of its grades and resumed from its saved records, with 11 answering and 48 judge calls
re-issued, and the 11 re-issued answers graded, before its analysis; one session lacks the 50-word memory.

\BudgetTable{}

\subsection{Detail tables for Section~\ref{sec:retrieval}}
All intervals in Section~\ref{sec:retrieval} and Table~\ref{tab:architectures} are 95\% whole-person bootstrap intervals with 100,000 NumPy PCG64 draws in blocks of 2,000 and linear percentiles; the seed is 20260920 for the held-out and first-cohort analyses (including the second judge) and the development-half factorial, question control and question-free pilot, 20260919 for the development-half pilot of \method{} and the notes and pruning attempts of Appendix~\ref{app:attempts}, 20260918 for the common-experiment contrast that opens Section~\ref{sec:retrieval}, 20260922 for the other analyses added after the first confirmation, 20260924 for the budget sweep (Figure~\ref{fig:budget}, Table~\ref{tab:budget}) and its replication, and 20260923 for the third cohort's other contrasts, the memory-plus-whole-record contrasts and the Twin-2K-500 intervals (the first Twin-2K run without a screener, 20260922).

\VerbatimConfirmationTable{}

%% file: newexp_appendix.tex
\section{Memory Systems, a Trained Reranker, a Cross-Family Judge, and a Public Benchmark}
\label{app:newexp}
This appendix covers the external memory-system baselines (Mem0, MemGPT-style), variants that add the memory to \method{}'s evidence, the second judge and the trained reranker (checks on the fusion result), and the survey port of \method{} to Twin-2K-500 with its superseded development versions. \method{} is the extractive procedure; \method{}$_\infty$ is its whole-record reference (Table~\ref{tab:roles}).
Every comparison in this appendix was added after the results of Section~\ref{sec:joint} and of the
first-cohort confirmation were known; each has its own protocol, hashed before its first request.

\paragraph{Memory systems.} \MemSysAppendix{}

\MemSysTable{}

\RPlusAppendix{}

\paragraph{Trained reranker.} PairRM \citep{jiang2023llmblender} (the \texttt{llm-blender/PairRM}
checkpoint, a pairwise ranker trained on human preference data, run locally with its default truncation)
ranked the three candidates $b_0,b_1,b_2$ of every common-experiment pool. Its instruction field held the
question, then the memory and the screener, so that truncation could not remove the question. The
top-ranked candidate is delivered unchanged and inherits its existing grade, so no new answer or grade is
produced and the comparison with the one-shot, the prompted selector and the candidate mean is contemporaneous by construction. PairRM ranked
1,768 concrete-memory and 1,767 trait-memory pools (one trait-memory pool lacks a candidate) and picked $b_0$, $b_1$ and $b_2$ in 619, 596 and
553 concrete-memory pools. The ranker is trained for general response quality, not for agreement with a particular
person; no trained ranker for that target exists for this data. \PairRMAppendixResult{} We did not test
multi-agent debate, in which several model instances critique and revise one another's answers; like
fusion it spends further calls on the same evidence, and it remains untested here.

\paragraph{Cross-family judge for fusion.} Grok~4.5 (xAI; reasoning effort high, 4,096-token cap) re-graded
every concrete-memory fusion and one-shot answer of the common experiment with the rubric prompt and parser of the primary judge;
no answer was regenerated. \FusionCrossJudgeAppendixResult{}

\paragraph{Twin-2K-500.} The public Twin-2K-500 panel \citep{toubia2025twin} (CC-BY-4.0) holds the answers
of 2,058 respondents to four survey waves. The dataset designates a set of questions (heuristics-and-biases
experiments, a ten-row policy-support matrix, and 40 purchase decisions whose product and price vary across
respondents) that respondents answered in waves 1--3 and again in wave~4, and it releases a wave 1--3
persona from which those questions and the answers to them are removed. We predict, for each respondent,
every closed-form single-answer wave-4 item (80 or 84 per respondent, 108 distinct items) from that persona,
on average about 95,000 characters of structured question--answer text covering demographics, personality
and other psychological scales, cognitive tests and economic preferences; we verified per respondent that it
contains none of the target questions, none of the earlier answers to them, and no question from the
targets' survey blocks. The respondents' own wave 1--3 answers to the same items give a human test--retest
reference (Table~\ref{tab:twin}). As in the interviews, every condition receives a screener: the persona's 14 demographic answer
lines (region, sex, age, education, race, citizenship, marital status, religion and attendance, party,
income, ideology, household size, employment) under the interview header [SCREENER Q\&A]. The conditions
are the screener alone; the written memory (the frozen concrete-memory prompt applied to the whole persona,
one further call); \method{}, which ranks the answer lines of the rest of the persona with the BM25 of
Algorithm~\ref{alg:r}, one ranking per target survey block with that block's question texts as query,
filled round-robin to the memory's length $B$ (adopted after three two-person smoke tests: in the first, Algorithm~\ref{alg:r}'s sentence-level selection picked no answer lines, and the development variants that followed, with other query units and larger fills, are summarized with the earlier runs below); the memory and \method{} together; and the whole persona. One answer
call per respondent and condition returns all of that respondent's answers as JSON, with the generator and
settings of our experiments. Scores are exact-option accuracy and, on ordinal items, absolute deviation
divided by the number of options minus one. The design and both criteria were fixed on a development sample
of 200 respondents (those of the earlier configuration mentioned below), before any model call on the
reported sample: 400 other respondents from the same seeded order, all conditions generated in one pass. \TwinContrasts{} \TwinEarlier{} We did not evaluate on LaMP
\citep{salemi2024lamp} for lack of time and API throughput, and the interviews of
\citet{park2024generative1000} are available only through an access application, which we did not
obtain; Twin-2K-500 is our only external benchmark.

\TwinTable{}

\TwinSecondSample{}

\TwinSecondTable{}

%% file: robustness_appendix.tex
\section{Post Hoc Robustness Analyses}
\label{app:robust}
Every analysis in this appendix is post hoc and was added after the results it examines were known. Unless stated
otherwise, intervals are pointwise 95\% whole-person bootstrap intervals from one separate stream
(NumPy PCG64, seed 20260922, 100{,}000 draws in blocks of 2{,}000) and are not adjusted for
multiplicity. Because this stream differs from those of the main analyses, an interval for the full
sample can differ from the main text in the fourth decimal (for example [0.0046, 0.0272] here against
the prespecified [0.0044, 0.0271] for concrete minus trait memory in Section~\ref{sec:memory}). Each result was
recomputed by a separately written implementation, which reproduced every reported number.

\subsection{Ordinal scale}
\label{app:ordinal}
For integer grades, $a-b=\sum_{k=1}^{3}(\mathbf{1}[a\ge k]-\mathbf{1}[b\ge k])$, so every normalized
difference equals $\tfrac13\sum_{k}\Delta P_{\ge k}$, where $\Delta P_{\ge k}$ is the paired difference in
the share of tasks graded at least $k$; we verified the identity numerically for every contrast in
Table~\ref{tab:ordinal-thresholds} (largest discrepancy below $10^{-17}$ for the point estimates and for
each bootstrap replicate). Any other monotone scoring of the grades with the same endpoints replaces the
equal weights by non-negative weights that sum to one, so its difference lies between the smallest and
largest $\Delta P_{\ge k}$, and its sign is fixed only when all three thresholds agree.

The fusion decrease (fusion minus one-shot with the concrete memory) is negative at every threshold, so its sign does not depend on the
spacing of the grades; only the lowest threshold has an interval excluding zero: the one-shot received at least
grade~1 while fusion received 0 on 112 tasks, against 62 in the opposite direction. On the first new cohort,
\method{} minus the memory (the prespecified confirmation) and the whole record minus the memory
are positive at every threshold; for \method{} minus the memory only the top-grade
interval excludes zero individually (46 against 14 grade-3 answers), and the smallest threshold
difference, which bounds the difference under any monotone scoring, has interval [\mns0.0076, 0.0156].
The prespecified criterion remains the one on the linear score. In the memory experiment, the concrete memory
exceeds the trait memory at the two lower thresholds but not at the top grade (34 against 38 grade-3 answers);
the positive sign holds for every monotone scoring that gives less than $11/12$ of the score range to the
step from grade~2 to grade~3 (the linear scale gives $1/3$). For concrete minus trait memory under one-shot generation in the common experiment, the corresponding cutoff
is 79.5\%. The negative point estimate of $\gamma$ comes from the boundary between grade~0 and grades
1--3 and would change sign only under a scoring that gives more than $23/30$ of the range to the top
step; these pointwise intervals do not alter our decision, based on the family interval, not to claim
an interaction. Selection minus one-shot with the concrete memory has threshold differences of mixed sign, so no direction that holds under
every spacing is supported. Table~\ref{tab:ordinal-transitions} gives the full transition counts.

\begin{table*}[t]
\centering
\scriptsize
\setlength{\tabcolsep}{2pt}
\begin{tabular}{lrccccc}
\toprule
Contrast & Tasks & Normalized & $\Delta P_{\ge 1}$ & $\Delta P_{\ge 2}$ & $\Delta P_{\ge 3}$ & Same sign \\
\midrule
\multicolumn{7}{l}{\emph{Memory experiment (Section~\ref{sec:memory}); 188 people}} \\
\quad Concrete $-$ trait memory & 1,768 & +0.0158 & +0.0249 & +0.0249 & \mns0.0023 & No \\
 & & [0.0046, 0.0272] & [0.0023, 0.0471] & [0.0068, 0.0435] & [\mns0.0092, 0.0045] & \\
\addlinespace
\multicolumn{7}{l}{\emph{Common experiment (Section~\ref{sec:joint}); 188 people}} \\
\quad Concrete: fusion $-$ one-shot & 1,768 & \mns0.0123 & \mns0.0283 & \mns0.0068 & \mns0.0017 & Yes \\
 & & [\mns0.0189, \mns0.0057] & [\mns0.0428, \mns0.0136] & [\mns0.0192, 0.0056] & [\mns0.0044, 0.0006] & \\
\quad One-shot: concrete $-$ trait & 1,768 & +0.0140 & +0.0271 & +0.0198 & \mns0.0051 & No \\
 & & [0.0036, 0.0243] & [0.0057, 0.0487] & [0.0028, 0.0368] & [\mns0.0113, 0.0006] & \\
\quad Concrete: selection $-$ one-shot & 1,768 & \mns0.0038 & \mns0.0130 & +0.0034 & \mns0.0017 & No \\
 & & [\mns0.0101, 0.0026] & [\mns0.0256, \mns0.0006] & [\mns0.0081, 0.0149] & [\mns0.0046, 0.0011] & \\
\quad interaction $\gamma$ & 1,767 & \mns0.0100 & \mns0.0209 & \mns0.0130 & +0.0040 & No \\
 & & [\mns0.0192, \mns0.0009] & [\mns0.0406, \mns0.0011] & [\mns0.0295, 0.0034] & [\mns0.0012, 0.0093] & \\
\addlinespace
\multicolumn{7}{l}{\emph{First new cohort (Section~\ref{sec:retrieval}); 500 people}} \\
\quad \method{} $-$ memory & 2,500 & +0.0127 & +0.0168 & +0.0084 & +0.0128 & Yes \\
 & & [0.0037, 0.0216] & [\mns0.0004, 0.0336] & [\mns0.0072, 0.0240] & [0.0076, 0.0184] & \\
\quad Whole record $-$ memory & 2,500 & +0.0295 & +0.0460 & +0.0300 & +0.0124 & Yes \\
 & & [0.0207, 0.0383] & [0.0292, 0.0628] & [0.0148, 0.0452] & [0.0072, 0.0180] & \\
\bottomrule
\end{tabular}
\caption{Threshold decomposition of the normalized differences. $\Delta P_{\ge k}$ is the paired
difference in the share of tasks graded at least $k$ (for $\gamma$, the difference between the two
workflow differences); each normalized difference equals $\tfrac13\sum_{k=1}^{3}\Delta P_{\ge k}$
exactly. Any monotone rescoring with the same endpoints yields a difference between the smallest and
largest $\Delta P_{\ge k}$, so the sign is independent of spacing only when all three agree (last
column; point estimates). For the new cohort, \method{} is Algorithm~\ref{alg:r} at the written memory's length, and the whole record
is one call on the whole permitted record, \method{} with unlimited budget (Section~\ref{sec:retrieval}).}
\label{tab:ordinal-thresholds}
\end{table*}

\begin{table*}[t]
\centering
\scriptsize
\setlength{\tabcolsep}{4pt}
\begin{tabular}{l*{4}{r}@{\hspace{8pt}}l*{4}{r}}
\toprule
\multicolumn{5}{c}{Concrete vs.\ trait memory} & \multicolumn{5}{c}{Concrete: fusion vs.\ one-shot} \\
\multicolumn{5}{c}{(307 / 1,214 / 247)} & \multicolumn{5}{c}{(119 / 1,472 / 177)} \\
2nd$\backslash$1st & 0 & 1 & 2 & 3 & 2nd$\backslash$1st & 0 & 1 & 2 & 3 \\
0 & 608 & 148 & 44 & 1 & 0 & 688 & 57 & 5 & 0 \\
1 & 129 & 427 & 99 & 3 & 1 & 101 & 525 & 56 & 0 \\
2 & 17 & 81 & 161 & 12 & 2 & 10 & 62 & 234 & 1 \\
3 & 3 & 2 & 15 & 18 & 3 & 1 & 0 & 3 & 25 \\
\midrule
\multicolumn{5}{c}{One-shot: concrete vs.\ trait} & \multicolumn{5}{c}{Concrete: selection vs.\ one-shot} \\
\multicolumn{5}{c}{(290 / 1,236 / 242)} & \multicolumn{5}{c}{(117 / 1,520 / 131)} \\
2nd$\backslash$1st & 0 & 1 & 2 & 3 & 2nd$\backslash$1st & 0 & 1 & 2 & 3 \\
0 & 609 & 143 & 45 & 1 & 0 & 689 & 54 & 7 & 0 \\
1 & 123 & 452 & 94 & 0 & 1 & 71 & 556 & 54 & 1 \\
2 & 18 & 84 & 154 & 7 & 2 & 12 & 43 & 251 & 1 \\
3 & 0 & 3 & 14 & 21 & 3 & 1 & 0 & 4 & 24 \\
\midrule
\multicolumn{5}{c}{Trait: fusion vs.\ one-shot (1{,}767 tasks)} & \multicolumn{5}{c}{\method{} vs.\ memory (first new cohort)} \\
\multicolumn{5}{c}{(114 / 1,527 / 126)} & \multicolumn{5}{c}{(415 / 1,753 / 332)} \\
2nd$\backslash$1st & 0 & 1 & 2 & 3 & 2nd$\backslash$1st & 0 & 1 & 2 & 3 \\
0 & 729 & 60 & 7 & 1 & 0 & 818 & 189 & 51 & 6 \\
1 & 72 & 553 & 44 & 0 & 1 & 150 & 614 & 136 & 4 \\
2 & 9 & 32 & 220 & 2 & 2 & 53 & 122 & 314 & 29 \\
3 & 0 & 0 & 13 & 25 & 3 & 1 & 0 & 6 & 7 \\
\bottomrule
\end{tabular}
\caption{Paired grade transitions. In each block, rows are the grade of the second-named condition and
columns the grade of the first-named condition (0--3); entries are task counts, and the parenthesized
line gives the numbers of tasks on which the first condition scored higher, tied, and lower.}
\label{tab:ordinal-transitions}
\end{table*}

\subsection{Memory length and the rubric's specificity clause}
\label{app:lengthspec}
\paragraph{Length.} Let $\Delta n$ be a session's concrete-profile minus trait-profile word count (median 10, range
\mns15 to 107; 35 of 194 sessions have $\Delta n\le0$). On sessions with $|\Delta n|\le10$ the memory experiment's
concrete-minus-trait difference is +0.0129 ([\mns0.0031, 0.0287]; 881 tasks, 96 people), and with $|\Delta n|\le15$ it is
+0.0148 ([0.0006, 0.0288]; 1,126 tasks, 120 people). A task-weighted regression of the per-task difference
on $\Delta n$ has slope +0.0005 per 10 words ([\mns0.0048, 0.0063]) and predicts +0.0149 ([0.0008, 0.0288])
at $\Delta n=0$. The one-shot concrete-minus-trait contrast of the common experiment, a second generation draw from
the same profiles, gives +0.0125 ([\mns0.0019, 0.0265]) for $|\Delta n|\le10$, +0.0163 ([0.0033, 0.0290]) for
$|\Delta n|\le15$, a slope of +0.0004 ([\mns0.0043, 0.0049]) and an intercept of +0.0132 ([0.0002, 0.0264]).
The difference shows no clear tendency to grow with the length gap (Table~\ref{tab:lengthtertile}).
However, $\Delta n$ was not randomized, the $|\Delta n|\le10$ intervals include zero, and at the mean
task-level gap of 17.2 words the slope interval spans \mns0.0083 to +0.0109, so this is not a
length-controlled estimate.

\begin{table*}[t]\centering\small
\begin{tabular}{@{}lrrrll@{}}\toprule
$\Delta n$ & Sessions & Tasks & People & Memory experiment & Common experiment (one-shot) \\\midrule
\mns15 to 6 & 69 & 630 & 68 & +0.0063 [\mns0.0105, 0.0228] & +0.0090 [\mns0.0084, 0.0261] \\
7 to 17 & 61 & 548 & 59 & +0.0280 [0.0062, 0.0494] & +0.0249 [0.0072, 0.0419] \\
18 to 107 & 64 & 590 & 63 & +0.0147 [\mns0.0062, 0.0362] & +0.0090 [\mns0.0085, 0.0268] \\\midrule
All & 194 & 1,768 & 188 & +0.0158 [0.0046, 0.0272] & +0.0140 [0.0036, 0.0243] \\\bottomrule
\end{tabular}
\caption{Concrete minus trait memory by tertile of the session length gap $\Delta n$ (concrete minus trait profile words). Task-weighted
differences on the 0--1 scale.}
\label{tab:lengthtertile}
\end{table*}

\paragraph{Rubric alignment.} The trait prompt forbids concrete details, while rubric level~3 asks for
comparable specificity. We marked a task as \emph{specific} when the participant's reference answer
contains a digit, or a capitalized word that is neither sentence-initial nor a form of \emph{I}; the rule
marks 527 of 1,768 tasks (29.8\%, 165 people), 523 of them through capitalization alone. In the memory
experiment, concrete minus trait is +0.0253 ([0.0055, 0.0452]) on specific tasks and +0.0118 ([\mns0.0018, 0.0256]) on the others.
The difference between classes, +0.0135 ([\mns0.0104, 0.0374]), is not resolved; in the common experiment's
one-shot concrete-minus-trait contrast, a second draw from the same profiles, it is \mns0.0001 ([\mns0.0209, 0.0206]), with +0.0139
([\mns0.0044, 0.0322]) on specific and +0.0140 ([0.0021, 0.0257]) on the other tasks, and the one resolved
top-level difference ($\Delta P(\ge3)=\mns0.0095$ [\mns0.0186, \mns0.0019] on specific tasks) favours the trait memory. The two lower thresholds carry all of the concrete-minus-trait
difference (Table~\ref{tab:thresholds}); grade~3 is rare (1.9\% of concrete-memory and 2.1\% of trait-memory predictions), at its
upper bound the $\ge3$ threshold contributes at most 0.0015 of the 0.0158, and on specific tasks
$\Delta P(\ge3)$ is \mns0.0038 ([\mns0.0159, 0.0077]). The concrete-memory advantage therefore does not
come from the top level's specificity clause. This check cannot exclude that the judge also rewards
concrete facts at level~2.

\begin{table*}[t]\centering\small
\resizebox{\linewidth}{!}{%
\begin{tabular}{@{}lrllll@{}}\toprule
Tasks & Tasks / people & Concrete $-$ trait (0--1) & $\Delta P(\ge1)$ & $\Delta P(\ge2)$ & $\Delta P(\ge3)$ \\\midrule
All & 1,768 / 188 & +0.0158 [0.0046, 0.0272] & +0.0249 [0.0023, 0.0471] & +0.0249 [0.0068, 0.0435] & \mns0.0023 [\mns0.0092, 0.0045] \\
Specific & 527 / 165 & +0.0253 [0.0055, 0.0452] & +0.0455 [0.0000, 0.0906] & +0.0342 [0.0057, 0.0626] & \mns0.0038 [\mns0.0159, 0.0077] \\
Non-specific & 1,241 / 188 & +0.0118 [\mns0.0018, 0.0256] & +0.0161 [\mns0.0093, 0.0416] & +0.0210 [\mns0.0016, 0.0443] & \mns0.0016 [\mns0.0100, 0.0067] \\
Difference & 1,768 / 188 & +0.0135 [\mns0.0104, 0.0374] & +0.0294 [\mns0.0219, 0.0814] & +0.0132 [\mns0.0233, 0.0491] & \mns0.0022 [\mns0.0169, 0.0121] \\\bottomrule
\end{tabular}}
\caption{Threshold decomposition of the memory experiment's concrete-minus-trait difference by task class. \emph{Specific}: the
reference answer contains a digit or a non-sentence-initial capitalized word other than forms of \emph{I}.}
\label{tab:thresholds}
\end{table*}

\subsection{Fusion and selection diagnostics}
\label{app:fusiondiag}
\paragraph{Length and copying.} Fusion did not lengthen answers. Mean (median) whitespace word counts were 135.2 (137), 133.1 (134) and 139.2 (140) for the one-shot, fusion and selection with the concrete memory, 127.7 (128), 127.3 (127) and 130.4 (130) with the trait memory, and the whole-record and current-dialogue answers averaged 89.1 and 55.8 words. Paired on the same tasks, concrete-memory fusion answers were 2.12 words shorter than the one-shot (difference \mns2.12, [\mns3.21, \mns1.03]; 1,768 tasks, 188 people), and trait-memory fusion differed from its one-shot by \mns0.35 words ([\mns1.45, +0.75]; 1,767 tasks). Split into tertiles of the fusion-to-one-shot word-count ratio (the middle tertile contains the 266 tasks with ratio exactly 1), fusion minus one-shot with the concrete memory was \mns0.0124 ([\mns0.0247, 0.0000]), \mns0.0045 ([\mns0.0134, +0.0042]) and \mns0.0198 ([\mns0.0320, \mns0.0076]); the top-minus-bottom contrast was \mns0.0074 ([\mns0.0243, +0.0094]), so we find no monotone relation between relative length and the fusion loss. Fusion reproduced one of its three candidates byte for byte in 670 of 1,768 concrete-memory tasks (37.9\%) and 916 of 1,767 trait-memory tasks (51.8\%); Table~\ref{tab:copystatus} splits the loss by copy status. Fusion chooses its copy status after seeing the candidates, so the split is descriptive, not causal.

\begin{table*}[t]\centering\small
\begin{tabular}{llrrr}\toprule
Memory & Fusion output & Tasks & Fusion $-$ one-shot & Fusion $-$ candidate mean\\\midrule
Concrete & copies $b_0$ & 223 & 0 (by construction) & +0.0020 [\mns0.0088, +0.0128]\\
Concrete & copies $b_1$ or $b_2$ & 447 & \mns0.0060 [\mns0.0188, +0.0066] & \mns0.0042 [\mns0.0124, +0.0037]\\
Concrete & new text & 1,098 & \mns0.0173 [\mns0.0271, \mns0.0076] & \mns0.0132 [\mns0.0205, \mns0.0059]\\
Trait & copies $b_0$ & 297 & 0 (by construction) & \mns0.0045 [\mns0.0153, +0.0063]\\
Trait & copies $b_1$ or $b_2$ & 619 & \mns0.0075 [\mns0.0186, +0.0036] & \mns0.0045 [\mns0.0111, +0.0022]\\
Trait & new text$^{a}$ & 850 & +0.0008 [\mns0.0095, +0.0114] & +0.0030 [\mns0.0043, +0.0106]\\\bottomrule
\end{tabular}\\[2pt]
{\footnotesize $^{a}$850 of the 851 new-text trait-memory tasks; one has a $b_1$ candidate without a grade, so its candidate mean is undefined. Including it leaves fusion $-$ one-shot at +0.0008.}
\caption{Fusion by whether its output is a byte copy of a candidate (self-selected groups; descriptive).}
\label{tab:copystatus}
\end{table*}

\paragraph{Selection prompt.} Fusion and selection calls used the shared generation system prompt and the unchanged memory answering prompt, including its closing line ``Answer as this participant would. Write ONLY their reply.'' (Appendix~\ref{app:prompts}), so the selection instruction to return only a candidate ID contradicts the system prompt's request for a participant reply. Even so, all 1,768 concrete-memory and 1,767 trait-memory selection responses parsed to a valid ID under the frozen parser (four concrete-memory responses were fenced JSON, and none contained extra fields); the one remaining trait-memory task had no selection call because a candidate was missing.

\paragraph{Display position.} The selector preferred the first displayed candidate: it chose the first, second and third display position in 833/425/510 concrete-memory tasks (47.1\%, 24.0\%, 28.8\%; $\chi^2_2=157.25$) and 732/446/589 trait-memory tasks (41.4\%, 25.2\%, 33.3\%; $\chi^2_2=69.4$). Display order was hash-permuted, so this did not favour the one-shot: $b_0$ appeared in the first, second and third position in 554/581/633 concrete-memory and 585/577/606 trait-memory tasks, and the selector picked $b_0$ in 32.98\% of concrete-memory tasks ([30.85, 35.11]) and 34.75\% of trait-memory tasks ([32.72, 36.77]), consistent with one in three (binomial $p=0.76$ and $0.21$). With the concrete memory, selection minus one-shot was \mns0.0042, \mns0.0023 and \mns0.0047 when $b_0$ was shown first, second or third, and every interval and pairwise contrast includes zero. With the trait memory it was \mns0.0068 ([\mns0.0166, +0.0028]), +0.0087 ([\mns0.0030, +0.0203]) and \mns0.0099 ([\mns0.0204, 0.0000]); two of the three unadjusted pairwise contrasts excluded zero ($b_0$ shown first minus second, \mns0.0155, [\mns0.0306, \mns0.0005]; second minus third, +0.0186, [+0.0029, +0.0346]), with no monotone order and no counterpart with the concrete memory. The selector also tended to pick the longest of three candidates with distinct lengths (720 of 1,669 concrete-memory and 709 of 1,660 trait-memory pools), and its choices were 3.94 (concrete) and 2.78 (trait) words longer than the one-shot. Fusion most often copied the last-displayed candidate: among fusion outputs identical to exactly one candidate, 150/210/307 (concrete) and 176/314/426 (trait) copied the candidate shown first/second/third.

\paragraph{Oracle matches.} No selection produced an invalid ID (0 of 1,768 concrete-memory, 0 of 1,767 trait-memory decisions) and none was an empty-answer zero, so the selector--oracle score-match rates of 1,474/1,768 (83.4\%) for the concrete and 1,503/1,766 (85.1\%) for the trait memory contain no operational zeros; all 622 concrete-memory and 648 trait-memory all-zero-pool matches are valid selections of zero-graded candidates. In pools with unequal candidate grades, the selector matched the oracle in 266/560 (47.5\%) concrete-memory and 243/506 (48.0\%) trait-memory tasks, compared with 47.7\% and 48.7\% expected from a uniformly random pick; the differences were \mns0.0018 ([\mns0.0420, +0.0378]) and \mns0.0072 ([\mns0.0499, +0.0348]).

\subsection{Raw references by target position}
\label{app:firsttarget}
The current dialogue differs from the whole record in its cut, its speakers and its truncation. At each session's first
target, where both raw references share the profile's temporal cut, they do not differ detectably
(\mns0.0017, [\mns0.0274, +0.0241]; 194 tasks, 188 people), and concrete-memory fusion trails them by similar amounts
(Table~\ref{tab:firsttarget}). On the 1,574 later targets (1,573 for contrasts with the whole record),
the current dialogue exceeds the whole record by +0.0426
([+0.0298, +0.0556]), and the fusion shortfall against the current dialogue widens to \mns0.0803, whereas the shortfall
against the whole record stays at \mns0.0377. The fusion versus current-dialogue gap on later targets therefore also
contains the difference between the two raw references; concrete-memory fusion minus the whole record is the source-controlled
contrast.

\begin{table*}[t]\centering\small
\resizebox{\linewidth}{!}{%
\begin{tabular}{lccc}
\toprule
Contrast & First targets (194 tasks) & Later targets (1,574 tasks) & Later minus first \\
\midrule
Concrete fusion $-$ current dialogue & \mns0.0344 [\mns0.0653, \mns0.0051] & \mns0.0803 [\mns0.0942, \mns0.0664] & \mns0.0459 [\mns0.0794, \mns0.0113] \\
Concrete one-shot $-$ current dialogue & \mns0.0258 [\mns0.0590, +0.0069] & \mns0.0676 [\mns0.0816, \mns0.0536] & \mns0.0418 [\mns0.0794, \mns0.0041] \\
Concrete fusion $-$ whole record & \mns0.0361 [\mns0.0653, \mns0.0069] & \mns0.0377 [\mns0.0480, \mns0.0275]$^{a}$ & \mns0.0016 [\mns0.0328, +0.0301] \\
Concrete one-shot $-$ whole record & \mns0.0275 [\mns0.0590, +0.0051] & \mns0.0250 [\mns0.0358, \mns0.0144]$^{a}$ & +0.0025 [\mns0.0320, +0.0368] \\
Current dialogue $-$ whole record & \mns0.0017 [\mns0.0274, +0.0241] & +0.0426 [+0.0298, +0.0556]$^{a}$ & +0.0443 [+0.0145, +0.0741] \\
\bottomrule
\end{tabular}}\\[2pt]
{\footnotesize $^{a}$1,573 tasks; one later target lacks a whole-record grade.}
\caption{Raw-evidence contrasts by target position. A session's first target is its smallest numeric
turn index, where the current dialogue and the whole record share the profile's temporal cut. All 188 people contribute
to both groups.}
\label{tab:firsttarget}
\end{table*}

\subsection{Multi-session participants in the memory experiment}
Removing the five people with more than one session gives concrete minus trait = +0.0144 ([0.0026, 0.0264]; 1,662
tasks, 183 people), and removing the eight risk people of Table~\ref{tab:sensitivity} gives +0.0147
([0.0027, 0.0267]; 1,635 tasks, 180 people), against +0.0158 ([0.0046, 0.0272]) on all tasks under this
stream. The three people added in the second exclusion were flagged for historical direct-excerpt profiles; the rerun
built new earliest-prefix profiles, so this exclusion is a person-level robustness check, not a
provenance correction.

\subsection{Judges and generator configurations}
\label{app:judgeconfig}
\paragraph{History benchmark.} The predictions of Table~\ref{tab:generation} were generated between July 9
and 13, 2026, by claude-fable-5 (native Messages API, effort low), gemini-3.1-pro-preview (native
generateContent, thinking level low), and gpt-5.5 and gpt-5.6-sol (reasoning effort none), each with an
8,000-token output cap; the Anthropic and Gemini arms used model-default sampling and the GPT arms
temperature 0. Every nonempty prediction was graded by Grok 4.5 through the xAI API with the frozen rubric,
temperature 0, reasoning effort high and a 2,000-token cap. The per-task grades behind
Table~\ref{tab:generation} equal these stored judgments on every judged pair, with the question-alone
column taken from the no-persona condition; Claude's 3 history and 20 question-alone empty outputs have no
judgment and score 0 under the historical rule.

\paragraph{Historical ladder.} The stored records of the historical representations (Appendix~\ref{app:lineage}) name neither the
generating model nor the judge. The pipeline module, last modified before these runs, fixes the flash
generator as gemini-3.5-flash; the runner as it exists now, last modified after the runs, defaults to that
model with minimal thinking, 3,000-token profile and 1,600-token answer caps, and service-default
temperature. An earlier description of the same runs names Gemini 3.6 Flash, and local records
cannot resolve this discrepancy. The judging function in the pre-run code uses gemini-3.1-pro-preview with
the frozen rubric and a 2,000-token cap; it requests high reasoning, but the client's native Gemini path did
not forward that setting, so these grades used the service-default thinking level and are not directly
comparable to the explicitly high-reasoning, 4,096-token judge of the memory experiment and the common
experiment. This is one reason the absolute levels of Tables~\ref{tab:oldmemory} and~\ref{tab:strictresult}
differ.

\paragraph{Why two judges.} The two judges were fixed at different times. The history benchmark was
graded by Grok 4.5 in July 2026, under a protocol dated July 9. Gemini 3.1 Pro was first used as a second
judge on 2,000 predictions sampled from that July run, including conditions and one generator not reported
here: half of the sample comes from the cross-interview benchmark (July 14--16; quadratic-weighted
$\kappa=0.891$ overall, 0.914 on the 1,000 cross-interview and 0.837 on the 1,000 history-benchmark
predictions, and 0.834 on the 501 predictions from the exact cells of Table~\ref{tab:generation}). On those 501 predictions (250 with history and 251 question-only, from different tasks, pooled over the four
configurations), history minus question is +0.057 ([0.008, 0.105]) under Grok 4.5 and +0.052 ([\mns0.001,
0.104]) under Gemini 3.1 Pro (10,000 whole-person draws over 178 people, seed 20260917): similar gains,
which this small unpaired sample does not resolve under the second judge. These records are outside the
public package. It
then graded the historical representation runs (August 19--20). The memory experiment's protocol fixing
Gemini 3.1 Pro is dated September 17 and the common experiment's September 18. Grok 4.5 was used again
only for the second-judge checks of Sections~\ref{sec:joint} and~\ref{sec:retrieval}.

\subsection{Answer length in Section~\ref{sec:retrieval}}
\label{app:anslength}
\AnswerLengthSecEight{} \AnswerLengthSweep{}

\subsection{Person-weighted estimates}
\label{app:personweighted}
Weighting every person equally, instead of every task, gives the following (10,000 whole-person NumPy
PCG64 draws, seed 20260923): history minus question +0.0861 ([0.0742, 0.0981]) for Claude Fable 5, +0.0802
([0.0671, 0.0933]) for Gemini 3.1 Pro, +0.0653 ([0.0536, 0.0771]) for GPT-5.5 and +0.0614 ([0.0508, 0.0721])
for GPT-5.6 Sol; memory-experiment concrete minus trait +0.0127 ([0.0014, 0.0242]), smaller than the task-weighted +0.0158;
common-experiment fusion minus one-shot (concrete memory) \mns0.0120 ([\mns0.0186, \mns0.0055]) and concrete minus trait under one-shot generation +0.0118 ([0.0018,
0.0220]). Each new cohort has one session of five tasks per person, so its two weightings coincide except in contrasts where one grade is missing, where they differ by at most 0.0002.

%% file: historical_appendix.tex
\section{Evidence Ledger and Benchmark Boundaries}
\label{app:benchmark}
\begin{table*}[t]
\centering\small
\begin{tabular}{@{}lrrrl@{}}\toprule
Configuration & Question & History & Difference & 95\% interval \\\midrule
Claude Fable 5 & 0.221342 & 0.308069 & +0.086727 & [0.075009, 0.098303] \\
Gemini 3.1 Pro & 0.221719 & 0.304676 & +0.082956 & [0.070146, 0.095699] \\
GPT-5.5 & 0.249246 & 0.314480 & +0.065234 & [0.053780, 0.076965] \\
GPT-5.6 Sol & 0.236991 & 0.299397 & +0.062406 & [0.052310, 0.072731] \\\bottomrule
\end{tabular}
\caption{Available history versus question alone. Historical predictions on 1,768 tasks from 188 people; new 95\% paired respondent-cluster intervals with 10,000 draws. The four intervals are pointwise and share participants. Vendor configurations are not assumed to use equal compute.}
\label{tab:generation}
\end{table*}

The history benchmark scores verified empty successful outputs as zero. Claude has three such history outputs and twenty question-only outputs, with one task empty in both conditions; restricting to the 1,746 pairs with actual judge records yields 0.0846 [0.0727, 0.0963]. The other configurations have grades for all 1,768 pairs.

The original preprocessing retains completed English-language sessions with screeners and transcripts and at least 12 substantive question--answer pairs. A substantive question contains a question mark and at least 20 characters, and its answer has at least five whitespace-delimited words. The producer reports 1,817 eligible interviews. With seed 42, it shuffles interviews within study IDs, iterates sorted study IDs with a cap of three sessions per study, and targets at most 200 sessions. It then requires at least five substantive pairs after the pair midpoint and samples at most ten targets from that second half. The resulting fixed file contains 194 sessions and 1,768 targets; sessions have 5--10 tasks (median 10), and people 5--26 (median 10). This is a bounded platform sample, not a probability sample of individuals or topics. Normalized whole-answer overlap flags are diagnostic and do not by themselves prove that every source is uncontaminated. The strict new source audit separately checks temporal and request bindings.

The evidence ledger distinguishes historical benchmark predictions, historical memory outputs, workflow development batches, the source-controlled memory rerun, and the common memory-by-workflow experiment. The latter two have separate frozen protocols and source-bound execution records. The common experiment generates contemporaneous baselines; it does not reuse the earlier memory rerun's answers as its one-shot conditions.

The historical representation study of Appendix~\ref{app:lineage} reuses these tasks: its six representations and the screener reference share 1,766 tasks, and its broader comparison of 15 conditions plus the screener reference, which includes the wrong-person control, shares 1,753 (13 error and two unscored records).

For the new minimum-effort person-cluster reanalysis, task and person identifiers were projected from the fixed task metadata, and stored integer grades were matched by model, task, and condition. Claude's two prediction logs contain 866 historical error rows and exactly 3,536 unique successful task--condition records. Errors are not additional tasks and do not replace subsequent successful outputs. Only verified empty successful outputs without judge records receive the original policy's zero score; nonempty missing judgments would remain missing. Whole-person bootstrap resampling uses 10,000 draws, seed 20260917, and linearly interpolated percentile endpoints. The four intervals are pointwise, and all models share the same people.

The benchmark also ran higher native-effort settings, which gave no consistent gain (not equivalence; not recomputed here), and GPT-5.5 exceeded GPT-5.6 Sol at minimum effort by 0.0151 ([0.0068, 0.0233], inherited); Section~\ref{sec:history} reports the minimum-effort configurations.

A cross-interview benchmark of the same period (100 earliest/latest session pairs, 782 tasks) enters these analyses only through the judge-agreement sample of Appendix~\ref{app:judgeconfig}; 38 of its 97 cross-study pairs are duplicate records of one session, leaving 59 clean pairs (458 tasks), and scores on duplicates (0.70--0.89) measure retrieval, not human retest reliability or a prediction ceiling.

\section{Historical Representation Results and Source Audit}
\label{app:lineage}
The earlier representation study asked what a short description must preserve for a model to predict a person's answer. Its six constructions span verbatim evidence, concrete paraphrase, free-form traits, a hybrid of traits and excerpts, a closed tag vocabulary, and a combined psychological profile (Table~\ref{tab:oldmemory}). All are memory constructions in Equation~\ref{eq:task}: they replace the available record before the answering workflow begins. These stored outputs motivate the concrete--trait comparison, but their unresolved source provenance prevents their use as confirmatory evidence; none is \method{}. Generator and judge configurations are in Appendix~\ref{app:judgeconfig}.

\paragraph{Schema-constrained memory.} The combined profile contains 28 model-imputed entries: five Big Five domains, one MBTI type, ten Schwartz values, three self-determination needs, five moral foundations, and four dispositional attributes (attachment style, locus of control, regulatory focus, and need for cognition). The model inferred them from interview text; no questionnaire was administered for this representation. Thus the experiment tests an imputation-and-encoding procedure, not the validity of the underlying instruments. Some schemas describe dimensions and others types; the common constraint is a predefined vocabulary. The tag arm instead selects 6--10 labels from the 40 most frequent labels among 511 distinct labels generated from the 188 respondents. Unlike these fixed-vocabulary memories, the trait arm uses unrestricted prose while excluding concrete facts. The labels below map the historical conditions to the current memory terminology.
Table~\ref{tab:oldmemory} uses the independent numerical reanalysis of stored outputs. The common set has 1,766 tasks and 188 people. All intervals use task-weighted means and 10,000 paired respondent-cluster draws with seed 20260917 and linear percentile interpolation. We do not reuse the historical empirical $p$-value formula, which incorrectly returned zero for an all-zero difference vector. Recalculation verifies numbers, not whether the source used to generate each profile was temporally permitted. Representations are named by content: in this appendix \emph{concrete} and \emph{trait} denote the historical outputs of the concrete-paraphrase and trait prompts of Section~\ref{sec:memory}; in Appendix~\ref{app:rerun} they denote the memory experiment's newly generated memories.

\begin{table*}[t]
\centering\small
\begin{tabular}{@{}lrrl@{}}\toprule
Representation & Approx. words & Gain & Recomputed 95\% interval \\\midrule
Direct excerpts (L1) & 96 & +0.024726 & [0.013386, 0.036588] \\
Concrete paraphrase (L2) & 107 & +0.024538 & [0.012842, 0.036320] \\
Traits plus excerpts (L3q) & 104 & +0.016233 & [0.005588, 0.027079] \\
Traits (L3) & 94 & +0.008683 & [\mns0.001322, 0.018560] \\
Closed tags (L4) & 10 & +0.005096 & [\mns0.004419, 0.014648] \\
Imputed scales (L5) & 85 & +0.001321 & [\mns0.007683, 0.010517] \\\bottomrule
\end{tabular}
\caption{Recomputed historical memory results, still limited by source provenance. Gains are relative to the screener-only reference, 0.225179. Approximate word counts describe the old outputs, not exact budget matching.}
\label{tab:oldmemory}
\end{table*}

The constructions differ in length and information content, so their ranking does not isolate abstraction alone. Direct excerpts minus concrete paraphrase is 0.000189 with interval [\mns0.011048, 0.011792], which does not prove equivalence. A separate inherited tight-budget comparison favors a 15-word Big Five representation over a 19-word free description by 0.0170, so neither the ranking in Table~\ref{tab:oldmemory} nor the concrete-over-trait result of Section~\ref{sec:memory} implies a universal ordering of free text over fixed psychological schemas. The combined-scale result also cannot establish the relative value of its individual instrument families. Section~\ref{sec:memory} reran concrete and free-form trait memories only; it did not rerun the tag or scale conditions.

On the 1,753-task control intersection, concrete-paraphrase own-profile quality is 0.249477 versus 0.193002 for a wrong person's profile and 0.224187 for the screener reference. The own-minus-wrong-person difference is 0.056475 [0.042830, 0.070116]. The control reflects both own-person information and harm from irrelevant narratives. Traits and scales also have positive own-minus-wrong-person differences of 0.026051 [0.014956, 0.037390] and 0.016543 [0.006050, 0.027094]. They cannot be described as having no individual signal.

\paragraph{Why these outputs are not used.} Five of the six files hold several profile versions for 81 to 145 of the 188 people (up to six per person), and the recovered logs bind no profile to its request, source or cut, so the original inputs cannot be reconstructed. Content matching found 13 tasks from three people whose only text-compatible source lies later in the same session, and three tasks from two people (two tasks from one person at a 12-word threshold) with target-answer fragments absent from the permitted material; wider, unproven flags cover 84 tasks from 11 people and 60 tasks from five, and an abstractive profile without such fragments is not thereby certified. Two exclusion rules fixed before their scores were examined leave the concrete-minus-trait difference positive (Table~\ref{tab:sensitivity}), which does not certify the remaining provenance. Section~\ref{sec:memory} therefore regenerated both memories from verified prefixes.

\begin{table*}[t]
\centering\small
\begin{tabular}{@{}lrrrl@{}}\toprule
Analysis & Tasks & People & Difference & 95\% person-cluster interval \\\midrule
Original common set & 1,766 & 188 & +0.015855 & [0.005264, 0.026470] \\
Remove 13 flagged tasks & 1,753 & 188 & +0.015592 & [0.005053, 0.026144] \\
Remove eight risk people & 1,633 & 180 & +0.016942 & [0.006324, 0.027847] \\\bottomrule
\end{tabular}
\caption{Historical concrete minus trait memory, with fixed source-risk exclusions. These are reanalyses of old predictions, not permitted-prefix regeneration.}
\label{tab:sensitivity}
\end{table*}

\FloatBarrier
\section{Memory Experiment: Protocol and Completed Results}
\label{app:rerun}
The frozen protocol uses all 1,768 targets and explicitly chooses per-session earliest-prefix memory. Five people have multiple sessions, accounting for six additional sessions. The planner uses numeric cuts rather than lexicographic task ordering; all 194 source bindings and 1,768 target eligibility bindings passed the metadata preparation checks. Each source identity binds person, session, cut, original history hash, parser, capped model-visible text hash, and the complete planned target set. Cache keys also bind representation and model configuration. Old profiles have no import path into this run.

The original protocol allowed 388 profile calls and 3,536 prediction calls. Infrastructure failures required three disclosed execution amendments before any judging: one identical-payload recovery for each of five existing response-free reservations, a bounded network-recovery policy after a further read timeout, and an extension to verified output-free service failures after HTTP 429 admission throttling. The original protocols, requests, errors, and successful outputs remain byte-preserved. Across the latter two generation policies, at most 32 additional recoveries are shared, and every generation slot can receive at most one recovery across the generation amendments.

The final ceilings are 3,961 generation HTTP requests and 3,568 judging HTTP requests, totaling 7,529 main-experiment requests plus two participant-free interface probes. These are ceilings, not completed counts or currency costs. Eligible recovery requires an exact response-free network exception or a strict output-free error envelope with HTTP 429, 502, 503, or 504. The final policy uses the identical payload after a 30-second cooldown; the preceding network-only amendment used two seconds and remains recorded as such. Generation concurrency falls from eight to four; judging starts with eight and resumes with four after the format/service amendment described below. Model content, parse errors, invalid grades, model-version mismatches, and truncated responses cannot trigger a replacement draw.

A request identity is reserved durably before every attempt. Unknown network attempts and received service-error responses are reported separately, and unavailable usage is never entered as zero. Completed artifacts are reused only after verification. The retained response is the first verifiable successful one. Restricted research storage contains request bodies; public summaries contain aggregate usage and verification results. These infrastructure amendments preserve the source boundary, target set, models, prompts, and analysis plan; they do not change the scientific endpoint. Generation completed with 3,931 HTTP reservations: 3,924 successful responses, six original response-free reservations, and one output-free HTTP 429. Seven identical-payload recoveries produced the retained successful outputs without repeating recovery for any slot. The successful generation receipts report 1,745,849 prompt tokens and 631,932 candidate tokens, totaling 2,377,781; no thinking-token field was reported. Usage for unsuccessful attempts remains unknown.

Response parsing required two disclosed implementation amendments before investigators inspected scores or effects. After 127 valid judgments, a saved response had an extra string explanation; after 311, another had an extra structured explanation. Both contained the required rubric fields. The final parser strictly rejects duplicate or non-finite JSON and preserves the original required-field rules: claims as a string list, an integer score in 0--3, and a textual rationale. Additional JSON is retained in the original receipt and ignored when extracting these fields. All earlier scores are verified unchanged, and received model responses are reused without another model call.

The first stopped judging phase retained 131 reservations: 128 successful model responses, two HTTP 429 errors, and one response-free retry. One preidentified slot whose HTTP 429 recovery timed out received a disclosed final identical-payload attempt within the unchanged total HTTP ceiling and shared recovery allowance. The subsequent phase reduced concurrency to four. At the final parser amendment, 311 valid scores and 312 received model outputs were reusable from 316 reservations, with four recovery attempts already consumed. Every earlier protocol, receipt, and failed checkpoint remains preserved.

The final collection continues past isolated invalid core grades, wrong-model responses, incomplete model outputs, or exhausted allowed transport recovery. Such slots remain explicit missing outcomes, never zero scores or grounds for a replacement model draw. Credentials, non-transient service errors, source or payload mismatches, and exhausted global budgets stop submission. Completion therefore means that every planned slot has a recorded terminal outcome; valid paired coverage is reported separately. This execution amendment does not relax the primary requirement of 1,768 valid pairs for complete-plan confirmation.

Final judging used 3,548 HTTP reservations: 3,536 successful responses, ten response-free attempts, and two output-free HTTP 429 errors. Its 12 recoveries include the disclosed third attempt for one slot and remain within the original shared allowance. Received model content was never redrawn. All successful judging receipts report usage: 1,635,563 prompt, 489,297 candidate, and 2,200,098 thinking tokens, totaling 4,324,958. Generation and judging together used 7,479 main-experiment reservations and report 6,702,739 tokens across successful receipts, excluding the two participant-free interface probes. Usage for 19 unsuccessful or unknown attempts is unavailable; generation thinking-token usage was not separately reported. Per-call latency was not independently logged. Execution checks verified source, profile, prediction, first-response, and grade bindings, as well as preservation of earlier files. These internal checks rely on sealed artifacts and frozen control flow, rather than independent service-side timestamps or human validation.

For the valid paired set $\mathcal{P}$, the primary estimator is
\begin{equation}
\widehat\Delta=\frac{\sum_{t\in \mathcal{P}}(s_{t,\mathrm{con}}-s_{t,\mathrm{tra}})}{3|\mathcal{P}|},
\end{equation}
with integer grades $s_{t,r}\in\{0,1,2,3\}$.
The denominator is tasks, not people. The respondent bootstrap samples the paired-set respondents with replacement, carries all of each sampled person's sessions and paired tasks, and recomputes the task-weighted estimator. We use 10,000 draws from Python's \texttt{random.Random} with seed 20260917 and linearly interpolated 2.5th/97.5th percentiles; the same generator and seed are used for every 10,000-draw interval in Section~\ref{sec:history} and Appendices~\ref{app:benchmark}--\ref{app:rerun}. With 10,000 draws the fifth and sixth decimals of interval endpoints are reported for reproducibility with that seed, not as Monte Carlo precision. No interval is reported with fewer than two paired people, and an empty paired set is non-estimable rather than zero.

The full-plan deterministic missingness bound gives a paired task its known difference; a concrete-only integer grade $a$ a difference range $[a-3,a]$; a trait-only grade $b$ the range $[-b,3-b]$; and a wholly missing task $[-3,3]$. Sum lower and upper endpoints over all planned tasks and divide by $3\times1768$. This is not a confidence interval. With no missing grades, it collapses to the observed difference. Judge failures remain missing, rather than being scored zero. A valid rubric grade of zero remains a valid outcome.

\begin{table*}[t]
\centering\small
\begin{tabularx}{\linewidth}{@{}lX@{}}\toprule
Quantity & Completed result \\\midrule
Planned tasks / people / sessions & 1,768 / 188 / 194 \\
Valid concrete grades / valid trait grades / valid pairs & 1,768 / 1,768 / 1,768 \\
Failures / unresolved / missing, each arm & 0 / 0 / 0 \\
Mean concrete / mean trait normalized content score & 0.263575 / 0.247738 \\
Primary concrete $-$ trait difference and 95\% interval & +0.015837 [0.004429, 0.027063] \\
All-task deterministic missingness bounds & [0.015837, 0.015837] \\
Complete-plan positive-contrast decision & Satisfied \\
Profile words: mean (range), 194 per arm & concrete: 111.85 (82--207); trait: 94.94 (80--116) \\
Generation / judging HTTP reservations & 3,931 / 3,548 \\
Reported tokens, successful receipts & 6,702,739; failed-attempt usage unavailable \\
Execution checks / independent numerical check & Passed / passed \\\bottomrule
\end{tabularx}
\caption{Completed memory experiment. All planned pairs are present. The deterministic missingness range collapses to the point estimate because there are no missing grades.}
\label{tab:strictresult}
\end{table*}

\paragraph{Secondary first-versus-later-target diagnostic.}
This analysis was added and frozen during blind judging, before investigators inspected any new scores. Each session's first target uses the prefix at its own cut; later targets reuse that same memory. The first-target group has 194 tasks from all 188 people: concrete 0.261168, trait 0.235395, difference 0.025773 with descriptive 95\% person-cluster interval [\mns0.005208, 0.057823]. The later-target group contains the other 1,574 tasks from the same people: concrete 0.263871, trait 0.249259, difference 0.014612 [0.002433, 0.026551]. All pairs are present. Each interval uses 10,000 whole-person resamples, seed 20260917, and linear percentiles. The later-minus-first difference in contrasts is \mns0.011161, reported descriptively without an interaction test. Question content, difficulty, position, and distance change together; this grouping cannot identify causal memory decay. Neither group replaces the full-sample primary contrast.

\FloatBarrier
\section{Additional Workflow Evidence and Fairness Boundaries}
\label{app:workflow}
The batches here tested earlier workflows (fusion, adaptive access) and demonstration formats (gist), before the common experiment; they are not \method{}, and Section~\ref{sec:workflow} quotes them only as the motivation for the common experiment.
This appendix retains the small development experiments that preceded the common experiment (Tables~\ref{tab:workflow} and~\ref{tab:gist}). They used each target person's full record plus a bank of question--answer demonstrations from other people. ``Native'' and ``raw JSON'' name two input formats for those demonstrations; ``gist'' is a compressed representation of the demonstrations; ``matched'' and ``legacy'' raw are two raw-demonstration references; and ``adaptive access'' lets the model request parts of the record instead of receiving it in one prompt. In these batches a study identifier can group several participants and is the resampling unit, unlike the per-session identifier of the 1,768-task cohort (Section~\ref{sec:info}). With 7 to 17 people in 4 to 9 study clusters, percentile bootstrap intervals are coarse and can have endpoints exactly at zero; they are descriptive only.

\begin{table*}[t]
\centering\small
\begin{tabularx}{\linewidth}{@{}Xlrl@{}}\toprule
Workflow minus contemporaneous reference & People/repeats & Difference & 95\% interval \\\midrule
Adaptive access $-$ full-evidence one-shot & 9 / 1 & +0.037037 & [0, 0.111111] \\
Three-answer fusion $-$ native one-shot & 7 / 2 & +0.047619 & [0, 0.095238] \\
Same fusion $-$ raw JSON one-shot, post hoc & 7 / 2 & +0.071429 & [\mns0.071429, 0.214286] \\
Same fusion $-$ native one-shot, expansion & 10 / 2 & \mns0.016667 & [\mns0.045455, 0] \\
Same fusion $-$ raw JSON one-shot, expansion & 10 / 2 & \mns0.066667 & [\mns0.151515, 0] \\\midrule
Fusion $-$ native one-shot, pooled post hoc & 17 / 2 & +0.009804 & [\mns0.019638, 0.041667] \\
Fusion $-$ raw JSON one-shot, pooled post hoc & 17 / 2 & \mns0.009804 & [\mns0.107843, 0.088889] \\\bottomrule
\end{tabularx}
\caption{Completed same-evidence generation comparisons. Differences are normalized content scores. Batch rows retain original descriptive intervals except the explicitly post hoc seven-person raw comparison. That row and the pooled rows use 10,000 whole-study resamples. Repeats are combined within people, and all respondents are development cases. Study clusters: 9 for the adaptive row (one person per study) and 7 for every other row.}
\label{tab:workflow}
\end{table*}
Person-level differences are mostly ties (positive/tied/negative: adaptive 1/8/0; seven-person fusion 2/5/0; expansion 0/9/1 against native and 0/7/3 against raw JSON demonstrations; pooled 2/14/1 and 3/9/5), and post hoc exact two-sided sign tests give $p\ge0.25$ throughout. The expansion respondents are distinct from the initial seven but not established as historically unseen. The pooled rows were specified after the batch results were known; they keep equal person weights, average each person's two repeats and resample the seven studies 10,000 times (seed 20260917). Both batches share the fusion subgraph (native $b_0$, two further samples with the identical prompt, hash ordering and the fusion prompt; Gemini 3.1 Pro, high reasoning, temperature 1, two repeats), but banks, batch and compute differ, so the pooled rows identify no separate effect of these. None of these comparisons passes its original continuation gate (a 0.05-point gain for adaptive access, a strictly positive lower bound otherwise; the seven-person fusion was a mechanism control), and two further processes were negative: a three-step primary graph (\mns0.0833, [\mns0.1667, \mns0.0208]) and cross-person predictive correction (\mns0.0208, [\mns0.0625, 0]).

\begin{table*}[t]
\centering\small
\begin{tabularx}{\linewidth}{@{}Xlrl@{}}\toprule
Contrast & Estimator/sample & Difference & 95\% interval \\\midrule
Gist + original Q/A $-$ matched raw & Person; 17 people, 7 studies & +0.068627 & [0.031250, 0.105263] \\
Gist + original Q/A $-$ legacy raw & Person; same 17, 7 studies & +0.068627 & [\mns0.010417, 0.137255] \\
Same gist $-$ matched raw, expansion & Study; 11 people, 4 studies & \mns0.013889 & [\mns0.041667, 0] \\
Same gist $-$ legacy raw, expansion & Study; same 11, 4 studies & +0.027778 & [0, 0.055556] \\
Full raw + gist $-$ matched raw & Person; 17 people, 7 studies & \mns0.029412 & [\mns0.088889, 0.022222] \\\bottomrule
\end{tabularx}
\caption{Source-demonstration compression and expansion. These experiments preserve the target's full record but change other people's demonstration representation. They are not the concrete-versus-trait memory comparison. Weights and references must be kept with their own intervals. Studies are the resampling clusters.}
\label{tab:gist}
\end{table*}

Table~\ref{tab:gist} changes how other people's demonstrations are presented, not the target's record. The two raw references have equal means but different person-level vectors, so their intervals differ. The development gain against the matched raw bank is unadjusted for the wider method search and did not hold on the expansion, whose rows are study-weighted (person-weighted \mns0.0152 and +0.0303, not to be paired with those intervals). The one-call format comparison of the two one-shot references (raw JSON minus native) is +0.0196 ([\mns0.0313, 0.0729]; 17 people). No final condition uses other people's demonstrations, bank availability before each interview was not established, and no batch has human validation.

%% file: main.bbl
\begin{thebibliography}{38}
\providecommand{\natexlab}[1]{#1}

\bibitem[{An(2026)}]{an2026fidelity}
Tao An. 2026.
\newblock \href {https://arxiv.org/abs/2601.00821v4} {Fidelity before
  structure: Verbatim chunks beat lossy artifact extraction in
  long-conversation {LLM} memory}.
\newblock \emph{arXiv preprint arXiv:2601.00821v4}.

\bibitem[{Argyle et~al.(2023)Argyle, Busby, Fulda, Gubler, Rytting, and
  Wingate}]{argyle2023outofone}
Lisa~P. Argyle, Ethan~C. Busby, Nancy Fulda, Joshua~R. Gubler, Christopher
  Rytting, and David Wingate. 2023.
\newblock \href {https://doi.org/10.1017/pan.2023.2} {Out of one, many: Using
  language models to simulate human samples}.
\newblock \emph{Political Analysis}, 31(3):337--351.

\bibitem[{Bisbee et~al.(2024)Bisbee, Clinton, Dorff, Kenkel, and
  Larson}]{bisbee2024synthetic}
James Bisbee, Joshua~D. Clinton, Cassy Dorff, Brenton Kenkel, and Jennifer~M.
  Larson. 2024.
\newblock \href {https://doi.org/10.1017/pan.2024.5} {Synthetic replacements
  for human survey data? {The} perils of large language models}.
\newblock \emph{Political Analysis}, 32(4):401--416.

\bibitem[{Brown et~al.(2024)Brown, Juravsky, Ehrlich, Clark, Le, R{\'e}, and
  Mirhoseini}]{brown2024monkeys}
Bradley Brown, Jordan Juravsky, Ryan Ehrlich, Ronald Clark, Quoc~V. Le,
  Christopher R{\'e}, and Azalia Mirhoseini. 2024.
\newblock \href {https://arxiv.org/abs/2407.21787} {Large language monkeys:
  Scaling inference compute with repeated sampling}.
\newblock \emph{arXiv preprint arXiv:2407.21787}.

\bibitem[{Cameron et~al.(2008)Cameron, Gelbach, and
  Miller}]{cameron2008bootstrap}
A.~Colin Cameron, Jonah~B. Gelbach, and Douglas~L. Miller. 2008.
\newblock \href {https://doi.org/10.1162/rest.90.3.414} {Bootstrap-based
  improvements for inference with clustered errors}.
\newblock \emph{Review of Economics and Statistics}, 90(3):414--427.

\bibitem[{Chen et~al.(2023)Chen, Aksitov, Alon, Ren, Xiao, Yin, Prakash,
  Sutton, Wang, and Zhou}]{chen2023usc}
Xinyun Chen, Renat Aksitov, Uri Alon, Jie Ren, Kefan Xiao, Pengcheng Yin,
  Sushant Prakash, Charles Sutton, Xuezhi Wang, and Denny Zhou. 2023.
\newblock \href {https://arxiv.org/abs/2311.17311} {Universal self-consistency
  for large language model generation}.
\newblock \emph{arXiv preprint arXiv:2311.17311}.

\bibitem[{Chhikara et~al.(2025)Chhikara, Khant, Aryan, Singh, and
  Yadav}]{chhikara2025mem0}
Prateek Chhikara, Dev Khant, Saket Aryan, Taranjeet Singh, and Deshraj Yadav.
  2025.
\newblock \href {https://arxiv.org/abs/2504.19413} {{Mem0}: Building
  production-ready {AI} agents with scalable long-term memory}.
\newblock \emph{arXiv preprint arXiv:2504.19413}.

\bibitem[{Dey and Viradecha(2026)}]{dey2026mempalace}
Robin Dey and Panyanon Viradecha. 2026.
\newblock \href {https://arxiv.org/abs/2604.21284} {Spatial metaphors for {LLM}
  memory: A critical analysis of the {MemPalace} architecture}.
\newblock \emph{arXiv preprint arXiv:2604.21284}.

\bibitem[{Dunn(1961)}]{dunn1961multiple}
Olive~Jean Dunn. 1961.
\newblock \href {https://doi.org/10.1080/01621459.1961.10482090} {Multiple
  comparisons among means}.
\newblock \emph{Journal of the American Statistical Association},
  56(293):52--64.

\bibitem[{Efron(1979)}]{efron1979bootstrap}
Bradley Efron. 1979.
\newblock \href {https://doi.org/10.1214/aos/1176344552} {Bootstrap methods:
  Another look at the jackknife}.
\newblock \emph{The Annals of Statistics}, 7(1):1--26.

\bibitem[{Fang et~al.(2026)Fang, Deng, Xu, Jiang, Tang, Xu, Deng, Yao, Wang,
  Qiao, Chen, and Zhang}]{fang2026lightmem}
Jizhan Fang, Xinle Deng, Haoming Xu, Ziyan Jiang, Yuqi Tang, Ziwen Xu, Shumin
  Deng, Yunzhi Yao, Mengru Wang, Shuofei Qiao, Huajun Chen, and Ningyu Zhang.
  2026.
\newblock \href {https://arxiv.org/abs/2510.18866} {{LightMem}: Lightweight and
  efficient memory-augmented generation}.
\newblock In \emph{The Fourteenth International Conference on Learning
  Representations}.

\bibitem[{Harris et~al.(2020)Harris, Millman, van~der Walt, Gommers, Virtanen,
  Cournapeau, Wieser, Taylor, Berg, Smith, Kern, Picus, Hoyer, van Kerkwijk,
  Brett, Haldane, del R{\'i}o, Wiebe, Peterson, G{\'e}rard-Marchant, Sheppard,
  Reddy, Weckesser, Abbasi, Gohlke, and Oliphant}]{harris2020numpy}
Charles~R. Harris, K.~Jarrod Millman, St{\'e}fan~J. van~der Walt, Ralf Gommers,
  Pauli Virtanen, David Cournapeau, Eric Wieser, Julian Taylor, Sebastian Berg,
  Nathaniel~J. Smith, Robert Kern, Matti Picus, Stephan Hoyer, Marten~H. van
  Kerkwijk, Matthew Brett, Allan Haldane, Jaime~Fern{\'a}ndez del R{\'i}o, Mark
  Wiebe, Pearu Peterson, and 7 others. 2020.
\newblock \href {https://doi.org/10.1038/s41586-020-2649-2} {Array programming
  with {NumPy}}.
\newblock \emph{Nature}, 585(7825):357--362.

\bibitem[{Hwang et~al.(2023)Hwang, Majumder, and Tandon}]{hwang2023aligning}
EunJeong Hwang, Bodhisattwa Majumder, and Niket Tandon. 2023.
\newblock \href {https://aclanthology.org/2023.findings-emnlp.393/} {Aligning
  language models to user opinions}.
\newblock In \emph{Findings of the Association for Computational Linguistics:
  EMNLP 2023}, pages 5906--5919, Singapore. Association for Computational
  Linguistics.

\bibitem[{Jiang et~al.(2023)Jiang, Ren, and Lin}]{jiang2023llmblender}
Dongfu Jiang, Xiang Ren, and Bill~Yuchen Lin. 2023.
\newblock \href {https://aclanthology.org/2023.acl-long.792/} {{LLM-Blender}:
  Ensembling large language models with pairwise ranking and generative
  fusion}.
\newblock In \emph{Proceedings of the 61st Annual Meeting of the Association
  for Computational Linguistics (Volume 1: Long Papers)}, pages 14165--14178.
  Association for Computational Linguistics.

\bibitem[{Li et~al.(2025)Li, Lin, Xia, and Jin}]{li2025selfmoa}
Wenzhe Li, Yong Lin, Mengzhou Xia, and Chi Jin. 2025.
\newblock \href {https://arxiv.org/abs/2502.00674} {Rethinking
  {Mixture-of-Agents}: Is mixing different large language models beneficial?}
\newblock \emph{arXiv preprint arXiv:2502.00674}.

\bibitem[{Liu et~al.(2023)Liu, Iter, Xu, Wang, Xu, and Zhu}]{liu2023geval}
Yang Liu, Dan Iter, Yichong Xu, Shuohang Wang, Ruochen Xu, and Chenguang Zhu.
  2023.
\newblock \href {https://aclanthology.org/2023.emnlp-main.153/} {{G-Eval}:
  {NLG} evaluation using {GPT-4} with better human alignment}.
\newblock In \emph{Proceedings of the 2023 Conference on Empirical Methods in
  Natural Language Processing}, pages 2511--2522, Singapore. Association for
  Computational Linguistics.

\bibitem[{Ma et~al.(2025)Ma, Yoztyurk, Haensch, Wang, Herklotz, Kreuter, Plank,
  and A{\ss}enmacher}]{ma2025algorithmic}
Bolei Ma, Berk Yoztyurk, Anna-Carolina Haensch, Xinpeng Wang, Markus Herklotz,
  Frauke Kreuter, Barbara Plank, and Matthias A{\ss}enmacher. 2025.
\newblock \href {https://aclanthology.org/2025.acl-long.90/} {Algorithmic
  fidelity of large language models in generating synthetic {German} public
  opinions: A case study}.
\newblock In \emph{Proceedings of the 63rd Annual Meeting of the Association
  for Computational Linguistics (Volume 1: Long Papers)}, pages 1785--1809.
  Association for Computational Linguistics.

\bibitem[{Maharana et~al.(2024)Maharana, Lee, Tulyakov, Bansal, Barbieri, and
  Fang}]{maharana2024locomo}
Adyasha Maharana, Dong-Ho Lee, Sergey Tulyakov, Mohit Bansal, Francesco
  Barbieri, and Yuwei Fang. 2024.
\newblock \href {https://aclanthology.org/2024.acl-long.747/} {Evaluating very
  long-term conversational memory of {LLM} agents}.
\newblock In \emph{Proceedings of the 62nd Annual Meeting of the Association
  for Computational Linguistics (Volume 1: Long Papers)}, pages 13851--13870,
  Bangkok, Thailand. Association for Computational Linguistics.

\bibitem[{Miller(2024)}]{miller2024errorbars}
Evan Miller. 2024.
\newblock \href {https://arxiv.org/abs/2411.00640} {Adding error bars to evals:
  A statistical approach to language model evaluations}.
\newblock \emph{arXiv preprint arXiv:2411.00640}.

\bibitem[{Packer et~al.(2023)Packer, Wooders, Lin, Fang, Patil, Stoica, and
  Gonzalez}]{packer2023memgpt}
Charles Packer, Sarah Wooders, Kevin Lin, Vivian Fang, Shishir~G. Patil, Ion
  Stoica, and Joseph~E. Gonzalez. 2023.
\newblock \href {https://arxiv.org/abs/2310.08560} {{MemGPT}: Towards {LLMs} as
  operating systems}.
\newblock \emph{arXiv preprint arXiv:2310.08560}.

\bibitem[{Pan et~al.(2025)Pan, Wu, Jiang, Luo, Cheng, Li, Yang, Lin, Zhao, Qiu,
  and Gao}]{pan2025secom}
Zhuoshi Pan, Qianhui Wu, Huiqiang Jiang, Xufang Luo, Hao Cheng, Dongsheng Li,
  Yuqing Yang, Chin-Yew Lin, H.~Vicky Zhao, Lili Qiu, and Jianfeng Gao. 2025.
\newblock \href {https://openreview.net/forum?id=xKDZAW0He3} {{SeCom}: On
  memory construction and retrieval for personalized conversational agents}.
\newblock In \emph{The Thirteenth International Conference on Learning
  Representations}.

\bibitem[{Panickssery et~al.(2024)Panickssery, Bowman, and
  Feng}]{panickssery2024selfpreference}
Arjun Panickssery, Samuel~R. Bowman, and Shi Feng. 2024.
\newblock \href
  {https://proceedings.neurips.cc/paper_files/paper/2024/hash/7f1f0218e45f5414c79c0679633e47bc-Abstract-Conference.html}
  {{LLM} evaluators recognize and favor their own generations}.
\newblock In \emph{Advances in Neural Information Processing Systems},
  volume~37, pages 68772--68802.

\bibitem[{Park et~al.(2023)Park, O'Brien, Cai, Morris, Liang, and
  Bernstein}]{park2023generativeagents}
Joon~Sung Park, Joseph~C. O'Brien, Carrie~J. Cai, Meredith~Ringel Morris, Percy
  Liang, and Michael~S. Bernstein. 2023.
\newblock \href {https://doi.org/10.1145/3586183.3606763} {Generative agents:
  Interactive simulacra of human behavior}.
\newblock In \emph{Proceedings of the 36th Annual ACM Symposium on User
  Interface Software and Technology}.

\bibitem[{Park et~al.(2026)Park, Zou, Kamphorst, Egan, Shaw, Hill, Cai, Morris,
  Liang, Willer, and Bernstein}]{park2024generative1000}
Joon~Sung Park, Carolyn~Q. Zou, Jonne Kamphorst, Niles Egan, Aaron Shaw,
  Benjamin~Mako Hill, Carrie Cai, Meredith~Ringel Morris, Percy Liang, Robb
  Willer, and Michael~S. Bernstein. 2026.
\newblock \href {https://arxiv.org/abs/2411.10109v3} {{LLM} agents grounded in
  self-reports enable general-purpose simulation of individuals}.
\newblock \emph{arXiv preprint arXiv:2411.10109v3}.

\bibitem[{Reimers and Gurevych(2019)}]{reimers2019sbert}
Nils Reimers and Iryna Gurevych. 2019.
\newblock \href {https://doi.org/10.18653/v1/D19-1410} {Sentence-{BERT}:
  Sentence embeddings using {S}iamese {BERT}-networks}.
\newblock In \emph{Proceedings of the 2019 Conference on Empirical Methods in
  Natural Language Processing and the 9th International Joint Conference on
  Natural Language Processing (EMNLP-IJCNLP)}, pages 3982--3992.

\bibitem[{Robertson and Zaragoza(2009)}]{robertson2009bm25}
Stephen Robertson and Hugo Zaragoza. 2009.
\newblock \href {https://doi.org/10.1561/1500000019} {The probabilistic
  relevance framework: {BM25} and beyond}.
\newblock \emph{Foundations and Trends in Information Retrieval},
  3(4):333--389.

\bibitem[{Salemi et~al.(2024)Salemi, Mysore, Bendersky, and
  Zamani}]{salemi2024lamp}
Alireza Salemi, Sheshera Mysore, Michael Bendersky, and Hamed Zamani. 2024.
\newblock \href {https://aclanthology.org/2024.acl-long.399/} {{LaMP}: When
  large language models meet personalization}.
\newblock In \emph{Proceedings of the 62nd Annual Meeting of the Association
  for Computational Linguistics (Volume 1: Long Papers)}, pages 7370--7392,
  Bangkok, Thailand. Association for Computational Linguistics.

\bibitem[{Santurkar et~al.(2023)Santurkar, Durmus, Ladhak, Lee, Liang, and
  Hashimoto}]{santurkar2023opinions}
Shibani Santurkar, Esin Durmus, Faisal Ladhak, Cinoo Lee, Percy Liang, and
  Tatsunori Hashimoto. 2023.
\newblock \href {https://proceedings.mlr.press/v202/santurkar23a.html} {Whose
  opinions do language models reflect?}
\newblock In \emph{Proceedings of the 40th International Conference on Machine
  Learning}, volume 202 of \emph{Proceedings of Machine Learning Research},
  pages 29971--30004. PMLR.

\bibitem[{Snell et~al.(2025)Snell, Lee, Xu, and Kumar}]{snell2025scaling}
Charlie Snell, Jaehoon Lee, Kelvin Xu, and Aviral Kumar. 2025.
\newblock \href {https://openreview.net/forum?id=4FWAwZtd2n} {Scaling {LLM}
  test-time compute optimally can be more effective than scaling parameters for
  reasoning}.
\newblock In \emph{The Thirteenth International Conference on Learning
  Representations}, pages 10131--10165.

\bibitem[{Toubia et~al.(2025)Toubia, Gui, Peng, Merlau, Li, and
  Chen}]{toubia2025twin}
Olivier Toubia, George~Z. Gui, Tianyi Peng, Daniel~J. Merlau, Ang Li, and
  Haozhe Chen. 2025.
\newblock \href {https://doi.org/10.1287/mksc.2025.0262} {Database report:
  {Twin-2K-500}: A data set for building digital twins of over 2,000 people
  based on their answers to over 500 questions}.
\newblock \emph{Marketing Science}, 44(6):1446--1455.
\newblock Preprint available as arXiv:2505.17479.

\bibitem[{Wang et~al.(2025{\natexlab{a}})Wang, Morgenstern, and
  Dickerson}]{wang2025flatten}
Angelina Wang, Jamie Morgenstern, and John~P. Dickerson. 2025{\natexlab{a}}.
\newblock \href {https://doi.org/10.1038/s42256-025-00986-z} {Large language
  models that replace human participants can harmfully misportray and flatten
  identity groups}.
\newblock \emph{Nature Machine Intelligence}, 7(3):400--411.

\bibitem[{Wang et~al.(2025{\natexlab{b}})Wang, Wang, Athiwaratkun, Zhang, and
  Zou}]{wang2025moa}
Junlin Wang, Jue Wang, Ben Athiwaratkun, Ce~Zhang, and James Zou.
  2025{\natexlab{b}}.
\newblock \href {https://openreview.net/forum?id=h0ZfDIrj7T}
  {{Mixture-of-Agents} enhances large language model capabilities}.
\newblock In \emph{The Thirteenth International Conference on Learning
  Representations}, pages 33944--33963.

\bibitem[{Wang et~al.(2023)Wang, Wei, Schuurmans, Le, Chi, Narang, Chowdhery,
  and Zhou}]{wang2023selfconsistency}
Xuezhi Wang, Jason Wei, Dale Schuurmans, Quoc~V. Le, Ed~H. Chi, Sharan Narang,
  Aakanksha Chowdhery, and Denny Zhou. 2023.
\newblock \href {https://openreview.net/forum?id=1PL1NIMMrw} {Self-consistency
  improves chain of thought reasoning in language models}.
\newblock In \emph{The Eleventh International Conference on Learning
  Representations}.

\bibitem[{Wu et~al.(2025)Wu, Wang, Yu, Zhang, Chang, and
  Yu}]{wu2025longmemeval}
Di~Wu, Hongwei Wang, Wenhao Yu, Yuwei Zhang, Kai-Wei Chang, and Dong Yu. 2025.
\newblock \href {https://openreview.net/forum?id=pZiyCaVuti} {{LongMemEval}:
  Benchmarking chat assistants on long-term interactive memory}.
\newblock In \emph{The Thirteenth International Conference on Learning
  Representations}, pages 86809--86836.

\bibitem[{Zhang et~al.(2025)Zhang, Hosseini, Bansal, Kazemi, Kumar, and
  Agarwal}]{zhang2025generativeverifiers}
Lunjun Zhang, Arian Hosseini, Hritik Bansal, Seyed~Mehran Kazemi, Aviral Kumar,
  and Rishabh Agarwal. 2025.
\newblock \href
  {https://proceedings.iclr.cc/paper_files/paper/2025/file/214308a2d5e3f83ef9ad2739e1cbc46d-Paper-Conference.pdf}
  {Generative verifiers: Reward modeling as next-token prediction}.
\newblock In \emph{The Thirteenth International Conference on Learning
  Representations}.

\bibitem[{Zhang et~al.(2018)Zhang, Dinan, Urbanek, Szlam, Kiela, and
  Weston}]{zhang2018personachat}
Saizheng Zhang, Emily Dinan, Jack Urbanek, Arthur Szlam, Douwe Kiela, and Jason
  Weston. 2018.
\newblock \href {https://aclanthology.org/P18-1205/} {Personalizing dialogue
  agents: {I} have a dog, do you have pets too?}
\newblock In \emph{Proceedings of the 56th Annual Meeting of the Association
  for Computational Linguistics (Volume 1: Long Papers)}, pages 2204--2213,
  Melbourne, Australia. Association for Computational Linguistics.

\bibitem[{Zheng et~al.(2023)Zheng, Chiang, Sheng, Zhuang, Wu, Zhuang, Lin, Li,
  Li, Xing, Zhang, Gonzalez, and Stoica}]{zheng2023judging}
Lianmin Zheng, Wei-Lin Chiang, Ying Sheng, Siyuan Zhuang, Zhanghao Wu, Yonghao
  Zhuang, Zi~Lin, Zhuohan Li, Dacheng Li, Eric~P. Xing, Hao Zhang, Joseph~E.
  Gonzalez, and Ion Stoica. 2023.
\newblock \href
  {https://proceedings.neurips.cc/paper_files/paper/2023/hash/91f18a1287b398d378ef22505bf41832-Abstract-Datasets_and_Benchmarks.html}
  {Judging {LLM}-as-a-judge with {MT-Bench} and {Chatbot Arena}}.
\newblock In \emph{Advances in Neural Information Processing Systems},
  volume~36, pages 46595--46623.
\newblock Datasets and Benchmarks Track.

\bibitem[{Zhou et~al.(2026)Zhou, Wang, Koopman, and Zuccon}]{zhou2026lightmem}
Yongjie Zhou, Shuai Wang, Bevan Koopman, and Guido Zuccon. 2026.
\newblock \href {https://arxiv.org/abs/2607.29104} {Reproducing {LightMem}:
  Naive {RAG} is just as good for memory management}.
\newblock \emph{arXiv preprint arXiv:2607.29104}.

\end{thebibliography}
